\documentclass{aa}  

\usepackage{graphicx}
\usepackage[varg]{txfonts}
\newlength{\dlugskr}

\def\figref#1{Fig.\,\ref{#1}}

\def\CCar{\textrm{CC$_{ar}$}}        
\def\CCal{\textrm{CC$_{al}$}}        
\def\CCaral{\textrm{CC$_{ar+al}$}}   

\def\OCcr{\textrm{OC$_{cr}$}}        
\def\OCam{\textrm{OC$_{am}$}}        
\def\OCamcr{\textrm{OC$_{am+cr}$}}   

\def\DCcr{\textrm{DC$_{cr}$}}        
\def\DCamcr{\textrm{DC$_{am+cr}$}}   
\def\F{\textrm{F}}

\def\msun{M$_{\odot}$}

\def\O++/O{O$^{++}$/(O$^{+}$$+$O$^{++}$)}
\def\He++/He{He$^{++}$/(He$^{+}$$+$He$^{++}$)}

\newcommand{\oii}{[O~{\sc ii}]}
\newcommand{\Oiib}{[O~{\sc ii}] $\lambda$3727}
\newcommand{\Oiir}{[O~{\sc ii}] $\lambda$7325}
\newcommand{\oiii}{[O~{\sc iii}]}
\newcommand{\Oiii}{[O~{\sc iii}] $\lambda$5007}
\newcommand{\Oiiit}{[O~{\sc iii}] $\lambda$4363}

\newcommand{\Niit}{[N~{\sc ii}] $\lambda$5755}
\newcommand{\Nii}{[N~{\sc ii}] $\lambda$6584}
\newcommand{\rSii}{[S~{\sc ii}] $\lambda$6731/6717}

\begin{document} 

   \title{Chemical abundances in Galactic planetary nebulae with Spitzer
   spectra}

   \titlerunning{Chemical abundances in Galactic PNe}

   \author{D. A. Garc\'{\i}a-Hern\'andez\inst{1,2} \and S. K. G\'orny\inst{3}}

   \authorrunning{Garc\'{\i}a-Hern\'andez \& G\'orny}

          
   \institute{Instituto de Astrof\'{\i}sica de Canarias, C/ Via L\'actea s/n, E$-$38205 La Laguna, Spain \email{agarcia@iac.es} 
   \and Departamento de Astrof\'{\i}sica, Universidad de La Laguna (ULL), E$-$38206 La Laguna, Spain
   \and N. Copernicus Astronomical Center, Rabia\'nska 8, 87-100 Toru\'n, Poland
\email{skg@ncac.torun.pl}
	}


\date{Received February 19, 2014; accepted April 21, 2014}

 
\abstract
{We present new low-resolution (R$\sim$800) optical spectra of 22 Galactic
planetary nebulae (PNe) with {\it Spitzer} spectra. These data are combined with
recent optical spectroscopic data available in the literature to construct
representative samples of compact (and presumably young) Galactic disc and bulge
PNe with {\it Spitzer} spectra. Attending to the nature of the dust features -
C-rich, O-rich, and both C- and O-rich dust features (or double chemistry) - 
seen in their {\it Spitzer} spectra, the Galactic disc and bulge PNe are
classified according to four major dust types (oxygen chemistry or OC, carbon
chemistry or CC, double chemistry or DC, featureless or F) and subtypes
(amorphous and crystalline, and aliphatic and aromatic), and their Galactic
distributions are presented. Nebular gas abundances of He, N, O, Ne, S, Cl, and
Ar, as well as plasma parameters (e.g.\ N$_e$, T$_e$) are homogeneously derived
by using the classical empirical method. We study the median chemical abundances
and nebular properties in Galactic disc and bulge PNe depending on their {\it
Spitzer} dust types and subtypes. The differences and similarities between PNe
in the Galactic disc and bulge are reported. In particular, the median
abundances for the major {\it Spitzer} dust types CC and OC are representative
of the dominant dust subtype (which are different in both Galactic
environments), while these values in DC PNe are representative of the two DC
subtypes. A comparison of the derived median abundance patterns with AGB
nucleosynthesis predictions mainly show that i) DC PNe, both with amorphous and
crystalline silicates, display high-metallicity (solar/supra-solar) and the
highest He abundances and N/O abundance ratios, suggesting relatively massive
($\sim$3--5 M$_{\odot}$) hot bottom burning AGB stars as progenitors; ii) PNe
with O-rich and C-rich unevolved dust (amorphous and aliphatic) seem to evolve
from subsolar metallicity (z$\sim$0.008) and lower mass ($<$3 M$_{\odot}$) AGB
stars; iii) a few O-rich PNe and a significant fraction of C-rich PNe with more
evolved dust (crystalline and aromatic, respectively) display chemical
abundances similar to DC PNe, suggesting that they are related objects. A
comparison of the derived nebular properties with predictions from models
combining the theoretical central star evolution with a simple nebular model is
also presented. Finally, a possible link between the {\it Spitzer} dust
properties, chemical abundances, and evolutionary status is discussed.}


   \maketitle
%

\section{Introduction}

Planetary nebulae (PNe) are a short evolutionary phase in the life of low- and
intermediate-mass stars (0.8 $\leq$ M $\leq$ 8 M$_{\odot}$) occurring after they
leave the ssymptotic giant branch (AGB) and before ending their lives as white
dwarfs \citep[e.g.][]{Iben1995}. However, many details of the physical processes
leading to the creation of the nebula and its subsequent evolution remain
unclear. Still, a better knowledge of the PN phase is necessary for
understanding not only the final fate of stars like our Sun, but also the
formation and the chemical evolution of the Milky Way and other galaxies.

At the tip of the AGB phase, stars experience a strong superwind that
efficiently enriches the surrounding interstellar medium with huge amounts of
gas and dust from the outer layers of the star \citep[e.g.][]{Herwig2005}.  When
the strong mass loss stops, they leave the AGB, and the future central star
rapidly evolves towards hotter effective temperatures in the
Hertzsprung--Russell diagram.  Thus, when the ionization of the ejected gas
takes place, a new PN emerges.

Owing to their emission-line nature, PNe can be easily observed at very large
distances, and the chemical composition of the gas and other properties can be
derived.  Some of the abundances (e.g.\ the Ar/H and Cl/H ratios) may remain
practically unchanged in these objects, reflecting the primordial composition of
the interstellar matter where their central stars were born.  But there are
other abundance ratios (e.g.\ N/O or C/O) that are strongly modified during the
life of low- and intermediate-mass stars.  The products of the hydrogen burning
and shell helium burning are brought to the stars' outer layers in dredge-up
episodes (the third dredge-up, TDU) taking place during the thermally pulsing
phase on the AGB, converting originally O-rich stars into C-rich ones. In
addition, the stellar surface can be enriched in products of the so-called hot
bottom burning \citep[HBB, e.g.][]{SackmannBoothroyd1992,Mazzitelli1999} process
for the more massive (M$>$3-4 M$_{\odot}$) AGB stars
\citep[e.g.][]{GH06,GH07,GH09}, preventing formation of C-rich stars.  At solar
metallicity, low-mass ($\sim$1.5$-$3-4 M$_{\odot}$) stars are expected to be
C-rich (C/O$>$1) at the end of the AGB phase, while more massive HBB stars may
remain O-rich (C/O$<$1) during the full AGB evolution.\footnote{Very
low-mass (e.g. $\leq$1.5 M$_{\odot}$) and solar metallicity AGB stars are
expected to be O-rich due to a rather inefficient TDU.}

It has been believed for a long time that post-AGB objects belong to only one of
the two chemical branches mentioned above: either those characterized by an
O-rich chemistry or those surrounded by C-rich material.  PNe with rare
Wolf--Rayet type central stars \citep[e.g.][]{GornyTylenda2000} were the first
and the only ones in the Galactic disc to simultaneously show the presence of
both carbon-based (e.g.\ polycyclic aromatic hydrocarbons; PAHs) and
oxygen-based dust (e.g.\ crystalline silicates) - \cite{Waters1998a}.  However,
a sample of Galactic bulge PNe (GBPNe) observed with {\it Spitzer/IRS} has been
analysed, and a majority of them show such dual-dust chemistry
\citep{Gutenkunst2008,PereaCalderon2009}; the dual-dust chemistry phenomenon is
clearly not restricted to objects with Wolf--Rayet (WR) central stars.  More
recently, \cite{GuzmanRamirez2011} have proposed a scenario that may explain the
simultaneous presence of PAHs and crystalline silicates in circumstellar
disc-like structures around the central stars of GBPNe, but this scenario does
not address the crucial question of why such a phenomenon is obserwed in
Galactic disc PNe only in PNe with [WR] central stars. Apparently, in the
Galactic bulge only a small fraction of PNe do not share that phenomenon and
only have oxygen-based dust. The reason for the apparent difference between
Galactic disc and bulge PNe remains unknown and is studied in this paper.

Various characteristics of GBPNe with peculiar {\it Spitzer} IR spectra were
analysed by \cite{Gorny2010}, who found that PNe characterized by carbon-based
dust (i.e.\ PAHs) and simultaneously oxygen-based dust (in the form of both
crystalline and amorphous silicates) - the so-called DC$_{a+cr}$-type PNe in
this paper - have unusual chemical compositions of the nebular gas.  Oxygen
seems to be under-abundant relative to hydrogen and nitrogen (see the location
of DC$_{a+cr}$-type PNe in their figure 11) but not to other elements. This
cannot be explained in the standard picture of the AGB chemical evolution for
objects with the typical Milky Way metallicity.  On the other hand, PNe
surrounded by only oxygen-rich dust (both in amorphous and crystalline forms) -
the so-called OC$_{a+cr}$-type PNe in this paper - have very low abundances of
nitrogen (see their figure 11).  The latter could be explained if these
OC$_{a+cr}$-type PNe do not come from single stars but from binary systems
because the presence of a companion may change the final abundances of the
nebula \citep[e.g.][]{DeMarco2009}.

Very recently, a large and complete sample ($\sim$150) of compact Galactic disc
PNe have been analysed through {\it Spitzer/IRS} spectroscopy by
\cite{Stanghellini2012}.  These authors find many PNe with peculiar dust
characteristics (the DC$_{a+cr}$- and OC$_{a+cr}$-type PNe described above)
similar to those previously found in the Galactic bulge \citep{Gorny2010}.  In
particular, 28\% (42 PNe) and 30\% (45 PNe) of the sample turned out to be among
the DC$_{a+cr}$- and OC$_{a+cr}$-type PNe, respectively. Owing to sensitivity
limits of the \textit{Infrared Space Observatory} (\textit{ISO}), most of the
disc PNe with IR spectra have [WR] central stars probably because they are
bright sources. It is a paradox of the high sensitivity of the {\it Spitzer} IRS
instrument that, in contrast, only the optically fainter Galactic disc PNe could
be observed \citep{Stanghellini2012}. As a result, proper optical spectra are
not available yet for most of them. New optical spectroscopic observations are
needed to form a proper reference sample of Galactic disc PNe and overcome
selection effect problems. We want to alleviate this problem with the present
paper by forming balanced samples of the Galactic disc and Galactic bulge PNe
for a more extensive investigation.

In this paper, we present new low-resolution spectroscopy of 22 Galactic PNe
with {\it Spitzer} spectra. These spectroscopic data are combined with recent
data to construct representative samples of the Galactic disc and bulge PNe.
Various nebular gas abundances of the Galactic disc and bulge PNe are studied
depending on their dust properties (i.e.\ {\it Spitzer} dust types/subtypes).
PNe in environments with different metallicity and chemical history such, as the
Galactic disc and bulge, are also compared. Our new low-resolution spectroscopic
observations, the nebular chemical abundance analysis, and the optical data
available in the literature are described in Section 2. We give an overview of
the {\it Spitzer} dust types/subtypes and the Galactic distribution of our final
samples of the Galactic disc and bulge PNe in Section 3, while in Section 4 we
report the derived chemical abundances versus the {\it Spitzer} dust types and
subtypes. Section 5 presents the nebular properties of the PNe with different
{\it Spitzer} dust types and subtypes. Our results are discussed in Section 6,
while a final summary of our work is given in Section 7.

\section{Optical spectroscopy of Galactic PNe with {\it Spitzer} spectra}

This section is devoted to the optical data available for PNe with {\it Spitzer}
spectra.  First, we present our own low-resolution spectroscopic observations. 
We determine their quality, present measured line intensities, and describe the
way they were used to derive nebular plasma parameters and chemical abundances. 
At the end of the section we also describe the literature data collected for the
same purpose. The six PNe with emission-line central stars discovered with our
new spectroscopic observations and the three suspected symbiotic stars we
observed are presented in Appendixes~A and B, respectively.

\subsection{New optical spectroscopic observations}

We conducted low-resolution (R$\sim$800) optical spectroscopic observations of
19 PNe and three suspected symbiotic stars with the DOLORES spectrograph at the
3.6 m Telescopio Nazionale Galileo (TNG) located on the island of La Palma
(Spain).  The PNe were observed during two nights in August 2011 and two nights
in July 2012. The detailed log of our TNG observations is presented in Table 1.
By selecting the targets a strong preference was given to PNe not present in
papers published after 1992 and that present large, homogeneous sets of
spectroscopic observations.  For this reason our data allow us to derive their
chemical composition and plasma parameters for the first time for almost all of
them.

\begin{table*}
\caption{ Log of
the TNG observations~~~~~~~~~~~~~~~~~~~~~~~~~~~~~~~~~~~~~~~~~~~~~~~~~
}
\begin{tabular}{ l l l r r l l}
\hline
 PNG        & name      & dust type      & date         & exp.  & -log F(H$\beta$) & diam. \\ 
            &           &                &              & (s)   &               	     & (")	    \\     
\hline
 107.4-02.6 & K 3-87    &  \CCal         & Aug  6, 2011 &  3600 &  ~~~13.4            & ~6.     \\ 
 107.4-00.6 & K 4-57    &  na    / symb. & Aug  7, 2011 &  5400 &  ~~~13.6            & stellar \\ 
 097.6-02.4 & M 2-50    &  \OCam         & Aug  6, 2011 &  5400 &  ~~~12.48           & ~4.5    \\ 
 095.2+00.7 & Bl 2- 1   &  \CCal         & Jul 26, 2012 &  2400 &  ~~~13.07           & ~2.5    \\ 
 079.9+06.4 & K 3-56    &  \OCam         & Jul 25, 2012 &  2400 &  ~~~12.9            & ~3.7    \\ 
 069.2+02.8 & K 3-49    &  \OCam         & Aug  7, 2011 &  3600 &  ~~~13.16           & <0.5    \\ 
 068.7+01.9 & K 4-41    &  \OCcr         & Jul 26, 2012 &  2400 &  ~~~12.98           & ~3.     \\ 
 060.5+01.8 & He 2-440  &  \OCcr         & Jul 25, 2012 &  2400 &  ~~~12.80           & ~2.2    \\ 
 052.9+02.7 & K 3-31    &  \CCal         & Jul 26, 2012 &  2400 &  ~~~13.8            & ~1.5    \\ 
 051.0+02.8 & WhMe 1$^*$&  \OCcr         & Jul 25, 2012 &  1800 &  ~~~13.52           & stellar \\ 
 044.1+05.8 & CTSS 2    &  na    / symb. & Aug  7, 2011 &  2400 &  ~~~13.5            & stellar \\ 
 042.9-06.9 & NGC 6807  &  \OCamcr       & Aug  6, 2011 &  2400 &  ~~~11.48           & ~0.8    \\ 
 041.8+04.4 & K 3-15    &  \CCaral       & Jul 26, 2012 &  2400 &  ~~~12.76           & <0.5    \\ 
 038.7-03.3 & M 1-69    &  \F            & Aug  6, 2011 &  3600 &  ~~~12.25           & stellar \\ 
 027.6-09.6 & IC 4846   &  \OCam         & Aug  6, 2011 &  2400 &  ~~~11.34           & ~2.9    \\ 
 025.3-04.6 & K 4- 8    &  \OCam         & Aug  7, 2011 &  2400 &  ~~~12.35           & stellar \\ 
 011.1+07.0 & Sa 2-237  &  \OCam         & Aug  7, 2011 &  3600 &  ~~~13.3            & stellar \\ 
 008.6-02.6 & MaC 1-11  &  \OCam         & Jul 25, 2012 &  3600 &  ~~~13.45           & ~3.1    \\ 
 007.5+04.3 & Th 4- 1   &  \OCcr / symb. & Aug  7, 2011 &  1800 &  ~~~13.0            & stellar \\ 
 004.3-02.6 & H 1-53    &  \DCamcr       & Jul 26, 2012 &  2400 &  ~~~12.56           & stellar \\ 
 000.6-02.3 & H 2-32$^*$&  \DCamcr       & Jul 25, 2012 &  3600 &  ~~~13.20           & stellar \\ 
 354.9+03.5 & Th 3- 6   &  \DCcr         & Jul 26, 2012 &  2400 &  ~~~13.67           & ~3.9    \\ 
\hline
\end{tabular}\\
(*) - faint or unusual spectra, no chemical abundances could be derived.
\end{table*} 

During the observations, the DOLORES instrument was configured to the
low-resolution long-slit spectroscopic mode.  We used the volume phase
holographic LR-B grism for science data and most of the instrument calibration
frames (bias, dome flats, and arc lamps), plus the LR-R grism  for some additional
dome flats in the low sensitivity blue part. The effective useful range of the
obtained spectra is $\sim$3600$-$8050 \AA.

The on-sky projected width of the slit was 0.7 arcsec for observations of the
programme targets. This resulted in about 4 to 5 \AA~resolution of the secured
spectra from the blue to the red part of the spectrum. Each night the
spectrophotometric standard star Feige\,110 was observed with at least one
additional standard from the list of BD$+$25$^o$3941, BD$+$28$^o$4211, and
G158-100. The standard stars were observed through 2 and 10 arcsec wide slits. 
The position angle of the slit was always aligned to the paralactic angle at the
moment of the observations and positioned at the centre of the PN where its
central star should be located.  The combined light of Ar+Kr+Ne+Hg lamps were
observed for wavelength calibration.

The basic exposure times of PNe ranged from 30 to 90 minutes split into two or
three sub-exposures. During such a long time, high S/N for the weak but
important features could be achieved in most cases, but the strongest nebular
emission lines (H$\alpha$, \Oiii, or \Nii, rarely some other lines) were
usually saturated.  Therefore, additional snapshot spectra with exposures
lasting from several seconds to a few minutes were also taken to allow the
measurement of such prominent lines.

\subsection{Data reduction and quality}

We used the long-slit spectral package of MIDAS\footnote{MIDAS was developed
and maintained by the European Southern Observatory.} to reduce and calibrate
the optical spectra. That involved the standard procedures of bias subtraction,
flat-field correction, atmospheric extinction correction, and wavelength
calibration.  After sky subtraction the one-dimensional spectra of standard
stars and PNe were extracted by summing over the appropriate rows of the
frames. In the case of PNe, however, since they were often located in crowded
areas, we first had to remove contaminating continuum contributions of all
stars in the field. Because this, a good fit to the sky emission could be found
before it was subtracted from the nebular spectrum; see \cite{Gorny2009} for a
more detailed description of this method. 

\begin{figure*}[t]
\resizebox{\hsize}{!}{\includegraphics{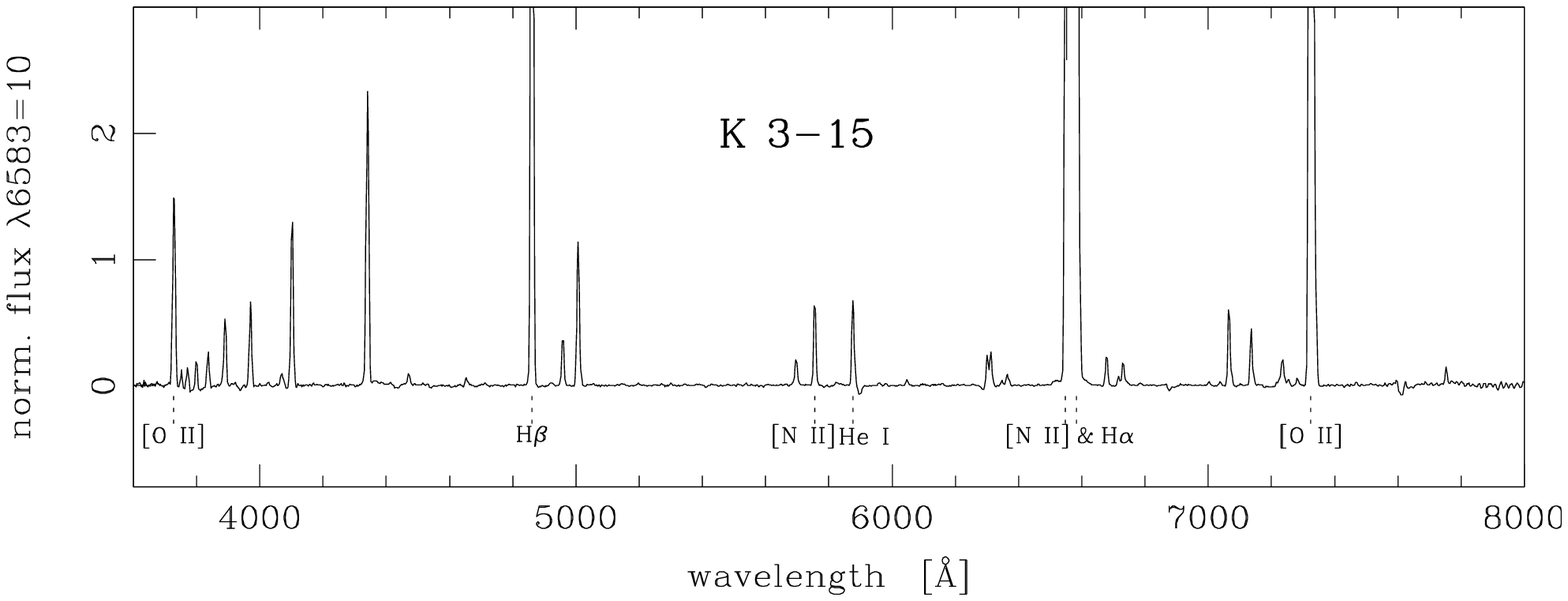}}
\resizebox{\hsize}{!}{\includegraphics{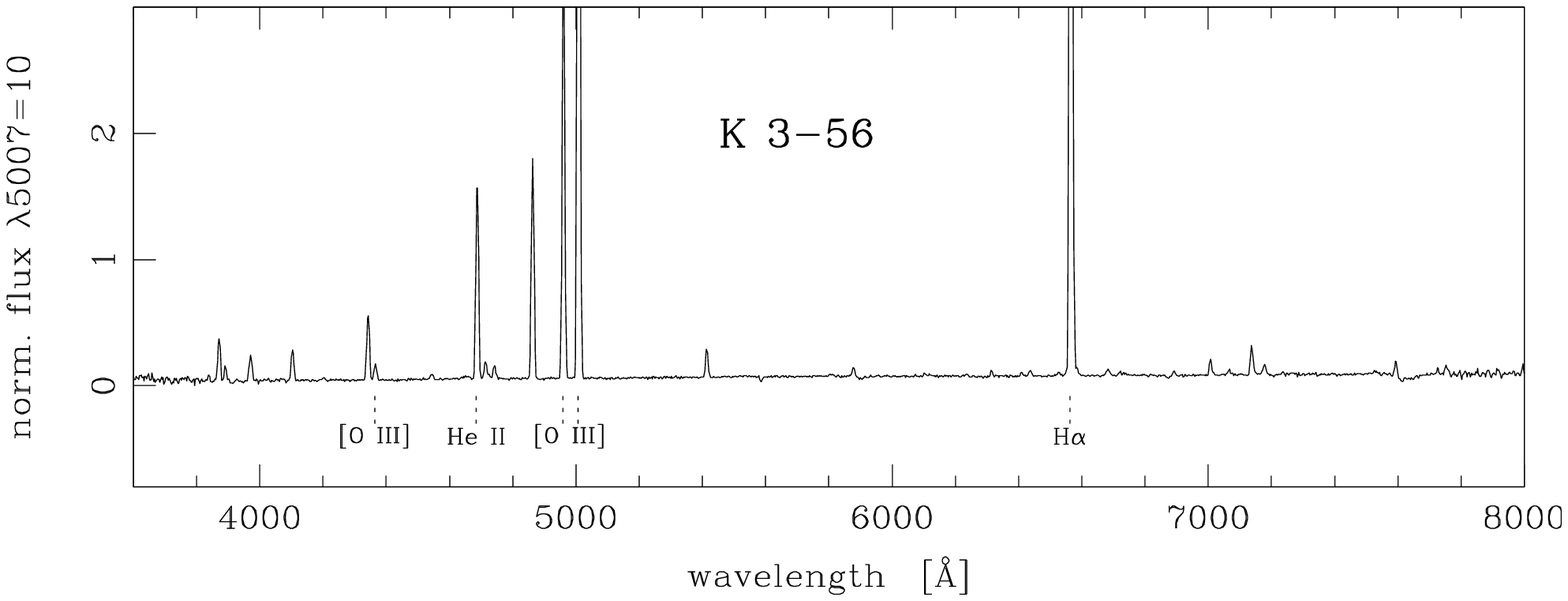}}
\caption[]{Illustrative examples of the acquired low-resolution optical
TNG/DOLORES spectra of a low ionization PN (K\,3-15, top panel) and a high
ionization PN (K\,3-56, bottom panel) in our observing programme. Some spectral
features are indicated.
}
\label{spectra_ex}
\end{figure*}

Two examples of PNe spectra obtained by us with the DOLORES spectrograph are
presented in \figref{spectra_ex}. This figure shows the reduced, integrated, and
flux-calibrated one-dimensional spectra of the low ionization PN K\,3-15 (top
panel) and the higher ionization PN K\,3-56.

The one-dimensional spectra were used to measure the intensities of the nebular
emission lines with the REWIA package developed by J.\,Borkowski from the
Copernicus Astronomical Centre (Torun, Poland). Gaussian profiles were used to
fit the nebular lines and multi-Gaussian fits were performed when necessary. 
Table 2 presents the measured intensities of all important
nebular lines on the scale H($\beta$)=100 for the PNe observed in our TNG
programme.

\begin{figure}[t]
\resizebox{\hsize}{!}{\includegraphics{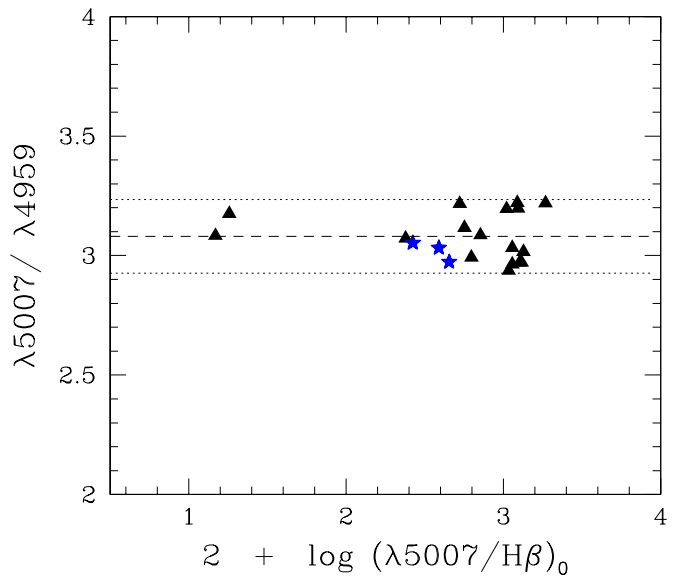}}
\caption[]{ 
  Intensity ratio of the [O~{\sc iii}] $\lambda$5007\AA~to $\lambda$4959\AA~lines
  as a function of the flux of [O~{\sc iii}] $\lambda$5007\AA~(in units of
  H$\beta$). The dashed line represents the mean value for the present TNG
  observations of PNe, and the dotted lines are 5\% deviations from it.  The
  observed PNe are marked by triangles, and the star symbols mark values for
  the suspected symbiotic stars observed.
}
\label{o3}
\end{figure}

In \figref{o3} we present the intensity ratios of the \oiii\ $\lambda$5007\AA~to
$\lambda$4959\AA~lines as derived from our optical spectra. The mean derived
value is marked, along with 5\% deviations from it. From atomic physics the
$\lambda$5007\AA/$\lambda$4959\AA~line ratio should have practically a constant
value for all PNe and can therefore be  used to assess the quality of our
spectra. The error of this line ratio is mostly influenced by the error of the
weakest spectral line that can be measured from our spectra, and this error is
typically found to be less than 5\%. It is to be noted here that, although the
\oiii\ $\lambda$4959\AA~is often one of the strongest spectral features in the
PNe spectra, the $\lambda$5007\AA/$\lambda$4959\AA~line ratio has to be
established using the additional snapshot spectra. This is due to the frequent
saturation of \oiii\ $\lambda$5007\AA~line in our long exposures. We therefore
assume that 5\% is a good error estimate also for the fainter lines measured
from our basic spectra. In the case of some weaker lines or lines contaminated
with sky features, the flux uncertainty is estimated to be around 20\% (marked
with a colon in Table 2) or rarely to be as high as 40\% (marked with a
semicolon in Table 2). In addition, in the case of some strongly blended nebular
lines the uncertainty is assumed to be at the 10\% level.

\begin{figure}[t] 
\resizebox{\hsize}{!}{\includegraphics{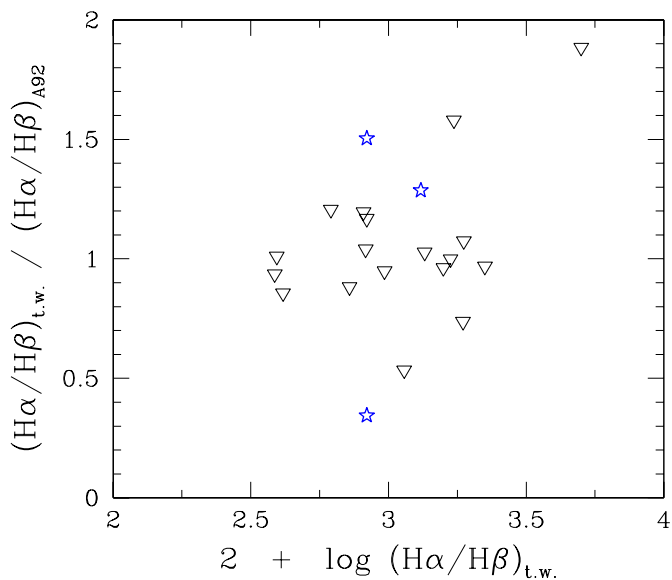}}
\caption[]{
  Comparison of the derived H$\alpha$/H$\beta$ flux ratios in this work
  (t.w.) and the survey spectra of the Strasbourg--ESO Catalogue by
  \cite{Acker1992} (A92). The reversed triangles mark our
  observed PNe and stars the suspected symbiotic stars.
}
\label{Ha}
\end{figure}

As mentioned above, the PNe observed in our TNG programme were deliberately
selected among those PNe with {\it Spitzer} spectra and without existing
high-quality optical spectra.  The only available large set of data for them are
the survey spectra collected during the preparation of the Strasbourg--ESO
Catalogue of Planetary Nebulae by A. Acker and co-workers. However, their
spectra were of considerably lower quality where many important nebular lines
were not detected.  In \figref{Ha} we compare our derived H$\alpha$/H$\beta$
ratios with those from \cite{Acker1992}.  A considerable spread can be observed
in \figref{Ha} that we naturally attribute to the lower quality of the latter
data.  Nevertheless, for half of the observed PNe an agreement with
\cite{Acker1992} is better than 10\%, and no systematic trend can be noticed.

As a final reduction step, we corrected the line intensities measured  from our
observations for the interstellar reddening by adopting the extinction law of
\cite{Seaton1979}. An iterative procedure as described in \cite{Gorny2004} was
used to reproduce the theoretical case B~Balmer H$\alpha$/H$\beta$ line ratios
at the electron temperature and density derived for each object. However, with
such a procedure, the reddening-corrected H$\gamma$/H$\beta$ and
H$\delta$/H$\beta$ ratios were usually different than theoretically predicted. 
Those deviations had no systematic direction for our sample of PNe but were
larger than could be expected from individual random errors of measurements of
these lines.  We attribute them rather to a generally lower accuracy of the flux
calibration by the decreasing sensitivity in the blue part of the spectra. For
the chemical abundance calculations, it is very important to determine the
intensities of lines like \Oiiit\ or \Oiib\ as reliably as possible. Thus, we
used the method described in \cite{Gorny2009} to calculate additional
corrections that would bring the H$\gamma$/H$\beta$ ratio (and if still
necessary the H$\delta$/H$\beta$ ratio) to the theoretical recombination
values.  These corrections were then linearly propagated to the observed fluxes
at all wavelengths shorter than 4681 \AA.

The final extinction-corrected line intensities are listed in Table 2 and the
lines additionally corrected in the way described above are marked with `c' in
this table.

\subsection{Plasma parameters and chemical abundances}

We used the ABELION code developed by G.\,Stasi{\'n}ska to derive plasma
parameters and chemical abundances exactly as in \cite{Gorny2009}.  This
code is based on the classical empirical method.  The electron densities
(N$_e$) are obtained from the \rSii\ ratio and the electron temperatures
(T$_e$) from the [O~{\sc iii}] $\lambda$4363/5007 and [N~{\sc ii}]
$\lambda$5755/6584 ratios. These were first used to refine the inferred
reddening correction as described above.

The chemical abundances of PNe were derived using the electron temperature
derived from [N~{\sc ii}] for ions of low ionization potential and that from
\oiii\ for ions of high ionization potential. Objects for which the electron
temperature could not be determined from either line ratio were discarded for
abundance determinations. This concerned two PNe, H\,2-32 and WhMe\,1, both with
\Oiiit\ below the detection limit and with \Niit\ too weak and uncertain to be
used for electron temperature estimation. On the other hand, if the electron
density (N$_e$) could not be precisely derived because the \rSii\ ratio was
estimated to be outside of the useful range, then the chemical composition was
nevertheless derived by assuming the N$_e$ upper limit as 10$^5$ electrons per
cm$^3$.  This was the case in nine of our targets and seems to be an unusually high
proportion among our PN sample (see below).  

We have attempted to investigate the unusually high electron densities derived
from our observations in more detail.  First, we established that the high
ratios of \rSii\ are not caused by any defect in the CCD and is not an artefact
of our reduction procedure. The two close lines of the [S~{\sc ii}] doublet are
resolved well in our spectra, and inspection of the raw frames already indicate
that the $\lambda$6731 line is usually much stronger than the $\lambda$6717
line. A comparison with other observers is difficult because our observations
are usually the first ones to have the high quality needed for plasma analysis.
Out of the nine unusually high electron density cases mentioned above we have
found literature entries with an \rSii\ ratio for only three PNe.  In two cases
(Th\,3-6 and NGC\,6807), \cite{Acker1992} give N$_{e}$ values of 1.25 and 1.50
(in units of 10$^5$ electrons per cm$^3$), which are lower than our estimated
N$_{e}$ values of 3.10 and 2.33, respectively. In the case of NGC\,6807,
however, our N$_{e}$ estimate is supported by the much closer value of 2.16
found by \cite{AllerKeyes1987}. 

In the third case (H\,1-53), \cite{Acker1992} found an approximate N$_{e}$ value
of 3.0 that is comparable to our value of 2.74 but against the estimate by
\cite{Exter2004}, who derived N$_{e}$=1.87. It is worth mentioning that in
another peculiar case (IC\,4846), although not concerning extremely dense PNe,
we derived an \rSii\ ratio of 2.13 that is comparable to the N$_{e}$ value of
2.35 by \cite{Acker1992} but in disagreement with the N$_{e}$ values of 1.70 and
1.67 by \cite{AllerCzyzak1979} and \cite{Barker1978}, respectively. However, the
latter N$_{e}$ measurements were derived from pre-CCD observations, and it is
also doubtful that the 6731/17 doublet was properly resolved. Finally, for PN
H\,2-32 we found an \rSii\ ratio of 1.47 in raw agreement with the N$_{e}$
values of 1.79 and 1.27 by \cite{Ratag1997} and \cite{Exter2004}, respectively. 
Thus, there is no indication that our electron density estimates is less
reliable than those in earlier studies. We therefore conclude that the very high
electron densities we derive for a considerable number of PNe in our sample are
real and probably reflect the fact that these objects are extremely young PNe
that are characterized by very high densities. This is not surprising if we
consider that most of our observed objects are compact (with a size of $\leq$4")
Galactic disc sources that are expected to be young, unevolved PNe
\citep[see][]{Stanghellini2012}. 

In the applied empirical method, the derived elemental abundances are the sum of
the ionic abundances derived from the measured intensities of the various lines
available in the optical domain. Wherever information on ionic abundances is
missing, this was corrected by the appropriate ionization correction factors
(ICFs) using the same ICFs scheme as in \cite{Gorny2009}. The only noteworthy
point is that similar to \cite{Gorny2009}, we compute the O$^+$ ionic abundance
for all PNe as an average of the values derived by using the \Oiib\ and \Oiir\
lines. However, no additional correction factors have to be applied to the
latter set of lines since both estimates from our TNG/DOLORES observations are,
on average, in satisfactory agreement.

We also note that for the analysis presented in this work and discussed in the
following sections, we have refrained from using the infrared atomic lines from
the {\it Spitzer} spectra (beyond the scope of this paper) to preserve direct
comparability with the results of the earlier studies conducted when such data
were not available for a representative sample of the Galactic PN population.
The use of the {\it Spitzer} IR data in the nebular chemical analysis could
provide some insight into additional ions, mainly O IV. However, ICFs for
optical studies work well, as was shown by \cite{Shaw2010}, who derived
elemental abundances in Small Magellanic Cloud PNe both from optical and {\it
Spitzer} spectra.

\subsection{Nebular abundances from literature data}

Apart from the new optical data presented in this work (Sect. 2.1), we also
search optical spectroscopic literature data for all Galactic (bulge and disc)
PNe with {\it Spitzer} spectra listed in \cite{Stanghellini2012},
\cite{PereaCalderon2009}, and \cite{Gutenkunst2008}. It is to be noted here that
the \cite{Stanghellini2012} PNe sample was selected in a more uniform way (see
below) than the \cite{PereaCalderon2009} and \cite{Gutenkunst2008} PNe samples.
The \cite{Stanghellini2012} PNe sample (157 PNe) is more complete, including PNe
in both the Galactic disc and bulge, while the \cite{PereaCalderon2009} (40 PNe)
and \cite{Gutenkunst2008} PNe (11 PNe) samples mainly contain PNe in the
Galactic bulge.\footnote{We have excluded the three suspected symbiotic stars
and five objects without {\it Spitzer} spectral classification in
\cite{Stanghellini2012} from further consideration, as well as another five
objects from \cite{PereaCalderon2009} because their classification is uncertain
(H\,1-12, M\,3-13, GLMP\,698, NGC\,6644, and 19W32). In addition, we excluded
Wray 16-423 since it is located in the Sagittarius dwarf galaxy. Also, M\,2-10
was observed both by \cite{Stanghellini2012} and \cite{Gutenkunst2008}, and we
adopt its double-chemistry dust classification from the latter authors.} In
addition, all Galactic PNe in \cite{Stanghellini2012} are compact (size $<$ 4")
sources. This means that, in principle, the \cite{Stanghellini2012} PN sample is
expected to be mainly composed of relatively young PNe, where we are sampling
the early dust stages of the PN evolution \citep[see][for more
details]{Stanghellini2012}. However, the \cite{PereaCalderon2009} and
\cite{Gutenkunst2008} Galactic bulge PN samples also contain some more extended
(size $>$ 4") sources.

In selecting literature data we have limited applicable entries to papers based
on modern observations published after 1992 and making use of CCD-equipped
spectrographs. They also had to contain enough objects to allow us to assess
their quality and internal integrity in a way that is similar to our own data
presented here. These requirements were met by the following papers:
\cite{FreitasPacheco1992},    
\cite{KingsburghBarlow1994},  
\cite{Cuisinier1996},         
\cite{Ratag1997},             
\cite{Cuisinier2000},         
\cite{Escudero2001},          
\cite{KwitterHenry2001},      
\cite{Pena2001},              
\cite{Milingo2002},           
\cite{Escudero2004},          
\cite{Exter2004},             
\cite{Gorny2004},             
\cite{Girard2007},            
\cite{WangLiu2007},           
\cite{Gorny2009},             
\cite{Cavichia2010},          
\cite{Henry2010}, and          
\cite{Gorny2014}. In short, we found useful optical spectroscopic literature
data for 76, 30, and nine Galactic PNe in the lists by
\cite{Stanghellini2012}, \cite{PereaCalderon2009}, and \cite{Gutenkunst2008},
respectively. 

The same ABELION code with identical assumptions (see Sect. 2.3) was applied to
these literature data. However, for obvious reasons the dereddening procedure
and additional corrections to the blue nebular lines described above were
calculated and applied only if the original still-reddened data were available. 
The corrections to match the O$^+$ abundances derived from \Oiir\ with those
from \Oiib\ were estimated individually for each set of data. If more than one
reference to optical spectroscopic data was available for a given object, then
we gave preference to papers with data of higher internal integrity, e.g. those 
containing complete measurements ranging from the blue to red spectral regions
and thus also both \oii\ doublets. However, if several literature sources had
similar quality for a given PN, then N$_{e}$, T$_{e}$ and the abundances were
derived separately, and the averaged values were calculated for such objects.

By joining the useful optical spectroscopic literature data mentioned above with
our new DOLORES optical data (17 PNe, by excluding two faint objects and the
three suspected symbiotic stars; see Sect. 2.1), we end up with a final sample
of 131 Galactic (bulge and disc) PNe with available {\it Spitzer} spectra and
nebular abundances. The derived plasma diagnostics and elemental abundances are
listed in Table 3 for the objects observed in our TNG programme and for all
other PNe with optical data in the literature (see above). In Table 3 there are
three rows for each object, and a fourth row used to separate them. The first
row gives the values of parameters computed from the nominal values of the
observational data or averages of estimates calculated from nominal values. The
second and third rows give the upper and lower limits of these parameters,
respectively. If there was only one source of data, these errors were calculated
with Monte Carlo simulations \citep[see][for more details]{Chiappini2009}. Also,
if there was more than one entry for a given object, these values are again
averages of individually determined errors. However, if nominal values of
individual entries were outside such estimated errors, then we replaced them
with actual deviations of these values from the adopted mean value given in the
first row.

Entries in Table 3 are ordered by PNG numbers that are given in column (1);
col.\   (2) gives the usual name of the PN; col.\ (3) indicates whether the
object belongs to the Galactic disc or bulge population of PNe (see Sect.\ 3 for
more details); col.\  (4) informs whether the object was included in the sample
of PNe with {\it Spitzer} spectra analysed by \cite{Stanghellini2012},
\cite{PereaCalderon2009}, or \cite{Gutenkunst2008}; col.\ (5) presents the dust
classification of the object (see Sect.\ 3 for more details); col.\ (6) gives
the references to data used in the calculations; col.\ (7) gives the electron
density (N$_{e}$); cols (8) and (9) give the electron temperature (T$_{e}$)
deduced from \Oiii\ and \Nii\ line ratios, respectively; cols (10) to (16) list
the He/H, N/H, O/H, Ne/H, S/H, Ar/H, Cl/H abundance ratios, respectively.
Finally, col.\ (17) gives the logarithmic extinction C at H$\beta$.

We do not expect the results of our work to be affected by the so-called
abundance discrepancy between collisionally excited lines (CELs) and optical
recombination lines (ORLs) in PNe \citep[see e.g.][and references
therein]{liu2006}. If these discrepancies are produced by temperature
fluctuations \citep{peimbert67} and they are moderate, then they would affect 
most of the elements in a similar way \citep[see e.g.][]{Garcia-Rojas2009,
Garcia-Rojas2013}. On the other hand, if the discrepancy is due to the presence
of chemical inhomogeneities, our analysis would not be affected, since the bulk
of the emission of the PN is coming from CELs \citep{liu2006}.

\section{{\textbf{\emph{ Spitzer}}} dust types/subtypes and Galactic distribution}

Galactic (disc and bulge) PNe in our final sample (131 PNe) can be classified
inro four major dust types that indicate the nature of the dust features in
their {\it Spitzer} infrared spectra \citep{Stanghellini2012}. PNe with C-rich
and O-rich  dust features belong to the carbon chemistry (CC) and oxygen
chemistry (OC) dust types, respectively, while those PNe with both C-rich and
O-rich dust features belong to the double chemistry (DC) dust type. Finally, PNe
with no dust emission features (and generally a very low dust continuum
emission) are within the featureless (F) dust type. These major PN dust types
can be subclassified into several dust subtypes depending on the specific nature
of the C-rich (aromatic and/or aliphatic) and O-rich (crystalline and/or
amorphous) dust features present in the {\it Spitzer} spectra. Thus, we define
the following dust subtypes among the major dust CC, OC, and DC types: i) carbon
chemistry (CC): CC$_{ar}$ (aromatic), CC$_{al}$ (aliphatic), and CC$_{ar+al}$
(aromatic and aliphatic); ii) oxygen chemistry (OC): OC$_{am}$ (amorphous),
OC$_{cr}$ (crystalline), and OC$_{am+cr}$ (amorphous and crystalline); iii)
double chemistry (DC): DC$_{cr}$ (crystalline) and DC$_{am+cr}$ (amorphous and
crystalline)\footnote{DC PNe display C-rich aromatic (e.g. PAH)
features only \citep[see e.g.][]{PereaCalderon2009}.}.

For the sample of the compact Galactic disc and bulge PNe, we adopt the dust
types and subtypes reported in \cite{Stanghellini2012}. We note that no dust
subclasses were defined for the DC (or mixed chemistry dust, MCD) PNe by
\cite{Stanghellini2012}, and we give the DC subtypes for the first time.  For
the \cite{PereaCalderon2009} and \cite{Gutenkunst2008} PNe, we double-checked
the major dust types, and we classified them among our defined dust subtypes by
inspecting the {\it Spitzer} spectra published by these authors.  Table 4 shows
the correspondence of our major {\it Spitzer} dust types and subtypes with the
equivalent dust classes defined by \cite{Stanghellini2012}.  All {\it Spitzer}
dust types/subtypes are listed in Table 3.

\addtocounter{table}{2}

\begin{table}
\caption{Correspondence of our major {\it Spitzer} dust types/subtypes
with the dust classes defined by \cite{Stanghellini2012}.
}
\begin{center}
\begin{tabular}{ l l }
\hline
 this work~~~~~~~~~    &     Stanghellini et al. (2012)  \\
\hline
 ~~\CCar        &    ~~~~~1~~~(CRD)    \\
 ~~\CCal        &    ~~~~~2~~~(CRD)    \\
 ~~\CCaral      &    ~~~~~3~~~(CRD)    \\
              &                 \\
 ~~\OCcr        &    ~~~~~4~~~(ORD)    \\
 ~~\OCam        &    ~~~~~5~~~(ORD)    \\
 ~~\OCamcr      &    ~~~~~6~~~(ORD)    \\
              &                 \\
 ~~\DCcr        &    ~~~~~7~=~MCD~$^*$  \\
 ~~\DCamcr      &    ~~~~~7~=~MCD~$^*$  \\
              &                 \\
 ~~F            &    ~~~~~0~=~F        \\
\hline
\end{tabular}

(*) - no dust subclasses were assigned for the MCD PNe by Stanghellini et al.\ (2012).
\end{center} 
\end{table} 

To study the Galactic distribution of  the Galactic PNe with {\it Spitzer}
spectra, we use the sample of \cite{Stanghellini2012}, which is the largest
sample so was selected in a more uniform way. We divide them into PNe pertaining
to the Galactic disc and those pertaining to the Galactic bulge. To select the
latter PNe we use the following standard criteria: i) they have to be located
within ten degrees of the centre of the Milky Way; ii) their diameters are
smaller than 20\arcsec, and iii) known radio fluxes at 5\,GHz are smaller than
100\,mJy. It has been established from simulations by \cite{Stasinska1991} for a
general population of PNe that the contamination of the bulge PN sample defined
in this way by disc PNe is most probably at the 5\% level.  

The sample of \cite{Stanghellini2012} that we use here, however, is  different
from the general population of PNe in the sense that it has been preselected,
giving preference to small and presumably young objects. For this reason the
second criterion concerning the nebular diameters is in fact fulfilled by all
these PNe. Concerning the third requirement, radio fluxes at 5\,GHz are
available for 26 out of 52 PNe located in the direction of the bulge. Only those
26 PNe were considered by \cite{Stanghellini2012} as true bulge members, whereas
the remainder have been adopted as disc members. In our opinion this may be
rather unlikely, and we prefer to use the 5\,GHz condition in the classical way
to eliminate clear disc members. The 5\,GHz flux has not been measured in
existing radio surveys and can in general be regarded as smaller than the
assumed 100\,mJy limit. It seems that the 5\,GHz flux is a rather poor
discriminative parameter for the specific sample of \cite{Stanghellini2012}.
Indeed, only for six of them undoubtedly located in the Galactic disc, i.e.\ far
away from the Galactic centre, is the 5\.GHz flux  larger than 100\,mJy. We
therefore prefer to adopt that all the 52 PNe in \cite{Stanghellini2012} sample
pertain physically to the Galactic bulge and that the percentage of disc
contaminators in this sample (52 compact Galactic bulge PNe) is low.

\begin{table}
\caption{Statistics of the major {\it Spitzer} dust types among the compact
Galactic bulge and disc PNe found in this work. The number of PNe is given in
parentheses.
}
\begin{center}
\begin{tabular}{ l c r c}
\hline
 dust type    &     bulge     &         &     disc        \\
\hline
 F            &    14\% (~7)  &         &     19\% (18)   \\
 CC           &     8\% (~4)  &         &     33\% (32)   \\
 OC           &    26\% (13)  &         &     31\% (31)   \\
 DC           &    52\% (26)  &         &     18\% (17)   \\
\hline
\end{tabular}
\end{center} 
\end{table} 

We present in Table 5 the distribution of the compact PNe according to their
main {\it Spitzer} dust types among the Galactic bulge and disc populations. By
comparing with Table 6 in \cite{Stanghellini2012}, one can see that with our
criteria for bulge membership, there is little change ($\leq$10\%) in the
distribution into dust types in the Galactic bulge. We found almost identical
percentages of CC and slightly increased ones of DC bulge PNe. The most
prominent differences are a decrease and an increase (both of about 10\%) in OC
and F bulge PNe, respectively. We indeed found that both OC and F PNe seem to
populate the Galactic disc and bulge with a very similar percentage (differences
$\leq$6\%). Taking into account that F-type PNe can be more evolved objects,
hence radio-dimmed, selection criteria based on available 5\,GHz flux
measurements could discriminate artificially against them pertaining to the
bulge population.

Regarding the dust distribution among disc PNe presented in Table 5, we also
found percentages of CC, OC, DC, and F disc PNe similar (differences $\leq$6\%)
to those found by \cite{Stanghellini2012}. The DC disc PNe seem to be slightly
less common (although only by about 6\%) than established by these authors.
\cite{Stanghellini2012} have already noted that the DC PNe are apparently
concentrated towards the inner regions (i.e.\ the Galactic centre) of the Milky
Way. This fact is also reflected in our statistics (see also Fig.\ 4), and DC
PNe are less common in the Galactic disc than CC and OC PNe, but they clearly
dominate the Galactic bulge region.

\begin{table} 
\caption{Statistics of the {\it Spitzer} dust subtypes among the compact
Galactic bulge and disc PNe found in this work. The number of PNe is given in
parentheses.} 
\begin{center} 
\begin{tabular}{ l c r c} 
\hline
 subtype   &     bulge     & &     disc        \\ 
\hline 
\CCar      & ~~   (~2)     & & 31\% (10) \\ 
\CCal      & ~~   (~1)     & & 66\% (21) \\ 
\CCaral    & ~~   (~1)     & & ~3\% ( 1) \\
              &            & &                 \\ 
\OCcr      & 46\% (~6)     & & 29\% (~9) \\ 
\OCam      & 39\% (~5)     & & 61\% (19) \\ 
\OCamcr    & 15\% (~2)     & & 10\% (~3) \\
              &            & &                 \\ 
\DCcr      & 69\% (18)     & & 53\% (~9) \\ 
\DCamcr    & 31\% (~8)     & & 47\% (~8) \\
\hline
\end{tabular}
\end{center} 
\end{table} 

More interesting is the distribution of the compact Galactic disc and bulge PNe
depending on their {\it Spitzer} dust subtypes. This is shown in Table 6 where
we list the corresponding percentages of the {\it Spitzer} dust subtypes among
the compact PNe in the Galactic disc and bulge. Table 6 inmediately suggests
that the several {\it Spitzer} dust subtypes are distributed differently between
the Galactic disc and bulge. CC bulge PNe are very rare, as expected from the
known lack of C-rich AGB stars in the Galactic bulge \citep[see e.g.][and
references therein]{Uttenthaler2007}. In addition, CC PNe with aliphatic dust
(CC$_{al}$ and CC$_{ar+al}$) are extremely rare (only two sources!) in the
Galactic bulge, while they completely dominate in the Galactic disc. The
apparent lack of CC$_{ar}$ disc PNe (i.e.\ with aromatic PAH-like dust features)
may be understood in terms of the dominant dust evolutionary stage around these
compact (and presumably young) Galactic disc PNe. There is no time for efficient
dust processing \citep[e.g.\ by the UV irradiation from the central
star;][]{Kwok2001}  to transform aliphatic groups to aromatic ones. Thus, we
generally see the early stages of the circumtellar dust grains (in the form of
broad aliphatic dust features) or the precursors of the aromatic groups that are
typically seen in much more evolved C-rich PNe (in the form of narrower and
aromatic PAH-like features). Alternatively, CC$_{al}$ PNe in the Galactic disc
may have somewhat low metallicity, favouring the presence of unprocessed dust
grains in their circumstellar envelopes \citep[see][]{Stanghellini2007}.

On the other hand, the distributions of the OC and DC {\it Spitzer} dust
subtypes in the Galactic disc and bulge also show remarkable differences.  Both
OC and DC PNe with O-rich amorphous dust features are more common in the
Galactic disc than in the bulge. One expects the evolution of the O-rich dust
features to proceed from amorphous silicates (in the AGB/post-AGB stage) to
crystalline silicates (in the PN phase)  \citep[][see also Garcia-Hernandez 2012
and references therein]{Garcia-Lario2003}.   Similarly to the CC PNe, the higher
detection rate of amorphous silicates in the compact Galactic disc PNe may be
related to the early dust evolutionary stage around these presumably young
objects and/or with a low metallicity.  In particular, \cite{Stanghellini2012}
have already pointed out that the OC$_{cr}$ disc PNe occur at lower Galactic
latitudes (b) than the OC$_{am}$ disc PNe (see below; Fig.  4). This would be
consistent with the OC$_{cr}$ PNe being more massive (and/or more metal rich)
than the OC$_{am}$ ones.  Curiously, similar fractions of OC$_{am+cr}$ (with
amorphous and crystalline silicate features) are found in both the Galactic disc
and bulge, although they are very rare and uncommon in both environments. 
Finally, the OC and DC PNe with O-rich crystalline silicate features seem to be
more frequent in the Galactic bulge. A higher detection rate of crystalline
silicates in the Galactic bulge may again be related to a more advanced
evolutionary stage or with higher metallicity. For example, higher metallicity
in the Galactic bulge may favour higher mass loss rates at the end of the
previous AGB phase and a more efficient crystallization of silicates (e.g.\
Waters et al.  1996; see also Section 6).

\begin{figure*}[t] 
\resizebox{0.47\hsize}{!}{\includegraphics{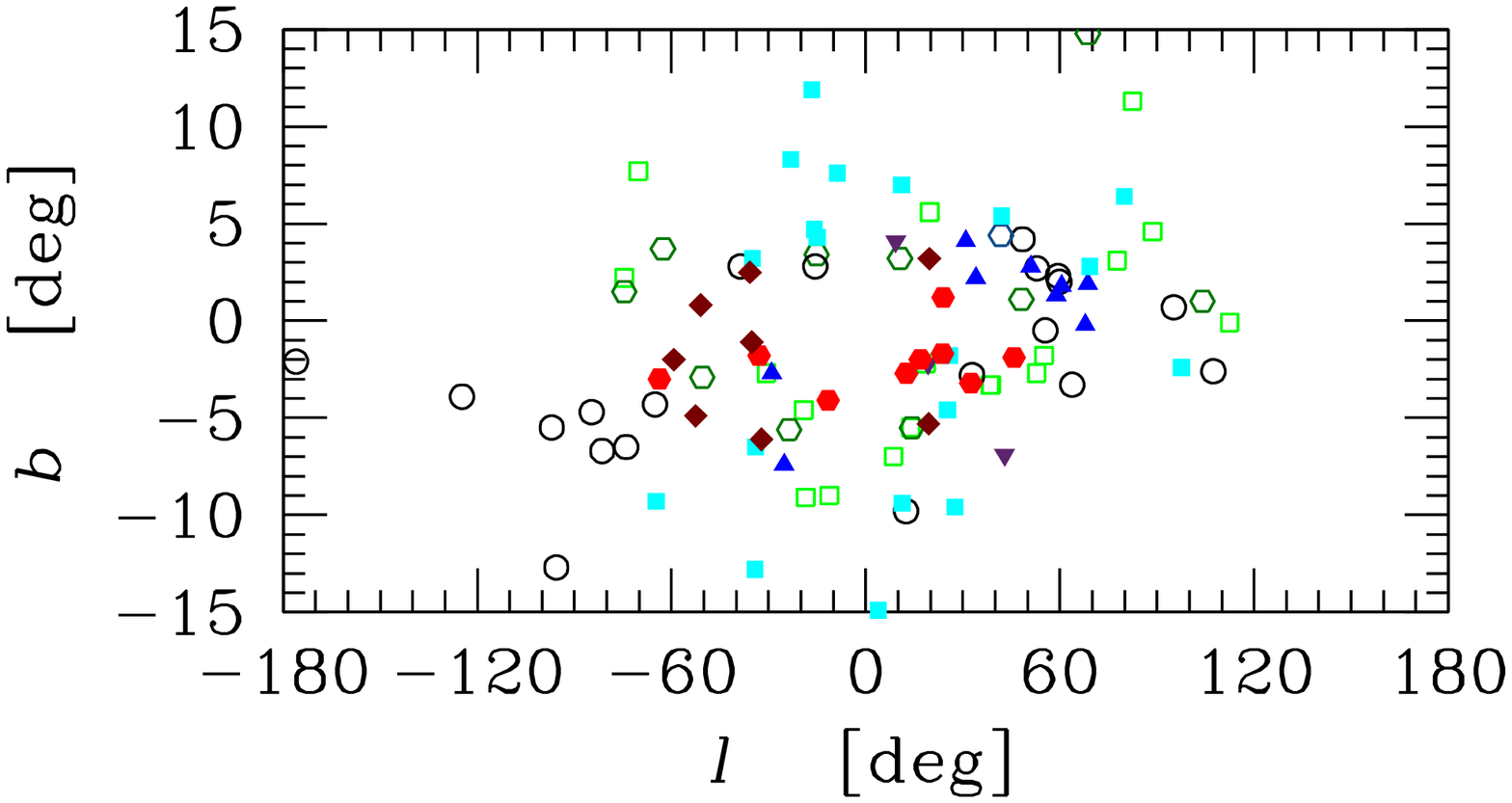}}
\resizebox{0.47\hsize}{!}{\includegraphics{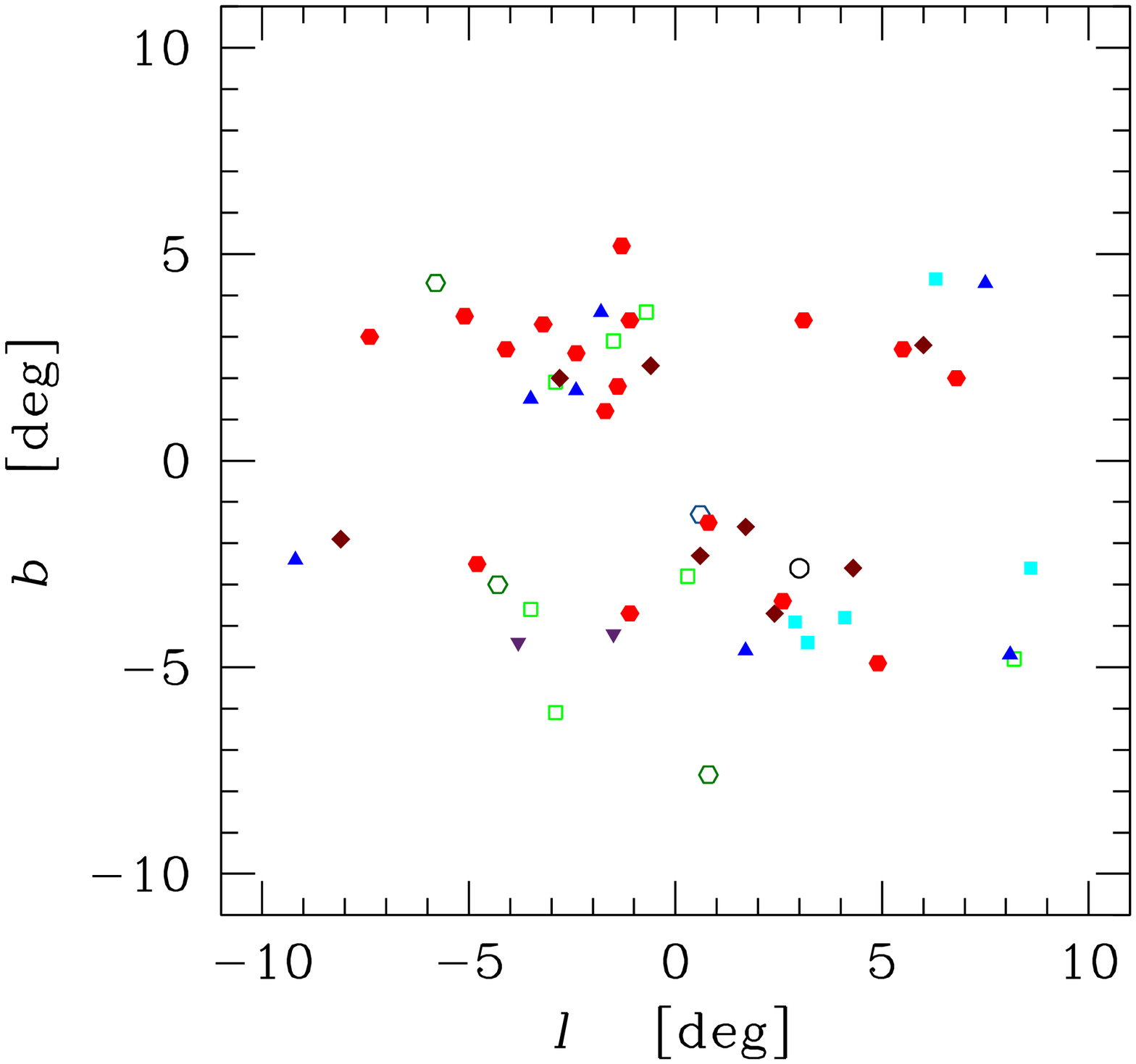}}
\caption[]{
  The distribution of compact Galactic PNe of various subtypes in the Galactic coordinates
  (Galactic disc and bulge PNe in the left and right panels, respectively):
  \CCar - open hexagons; \CCal - open circles; \CCaral - open diamonds; 
  \OCcr - triangles; \OCam - filled squares; \OCamcr - reversed triangles;
  \DCcr - filled hexagons; \DCamcr - filled diamonds; F - open squares.}
\label{fig_lb}
\end{figure*}

In \figref{fig_lb} we present the locations (in Galactic coordinates) of the
compact Galactic disc (left panel) and bulge (right panel) PNe divided into the
various dust subtypes. As mentioned above, DC PNe in the Galactic disc are
mainly located towards the Galactic centre (Fig.\  4, left panel), and there is
no obvious difference between the DC$_{cr}$ and DC$_{am+cr}$ dust subtypes.  CC
and OC Galactic disc PNe are rather uniformly distributed in the Galaxy,
although CC PNe seems to be more concentrated in the Galactic plane than (at
least) the OC PNe with amorphous silicates (see below). In Fig.\  4 (left panel)
there is no significant difference for the Galactic coordinates exhibited by the
several CC dust subtypes (CC$_{am}$, CC$_{al}$ and CC$_{am+al}$).  However, the
OC PNe in Fig.\  4 (left panel) display, on average, different Galactic
latitudes depending on the OC dust subtypes (mainly OC$_{am}$ and OC$_{cr}$), as
previously pointed out by \cite{Stanghellini2012}. The OC$_{cr}$ PNe in the
Galactic disc are more concentrated at lower latitudes than the OC$_{am}$ PNe,
since they are consistent with the former objects being more massive (and/or
more metal-rich) than the latter (see above). In fact, judging only by their
median {\it b}=6\fdg5, OC$_{am}$ PNe seem to be located two to three times
further away from the Galactic plane than almost any other dust subtype in the
disc.  Surprisingly, the other PN group with high Galactic latitudes (with a
median of {\it b}=4\fdg0) are the F PNe with featureless {\it Spitzer} spectra.
The situation is very similar in the Galactic bulge (Fig.\  4, right panel), and
there is no clear distinction between the different {\it Spitzer} dust subtypes,
although it has to be noted that the samples are considerably smaller, and the
number of objects in the central region is naturally depleted due to strong
obscuration by dust.  As in the disc, the OC$_{cr}$ bulge PNe seem to show lower
Galactic latitudes (and longitudes) than their amorphous dust counterparts
(OC$_{am}$); this difference, however, is marginal. Finally, PNe with amorphous
and crystalline O-rich dust \citep[OC$_{am+cr}$; see e.g.][]{Gorny2010} are
rather peculiar objects in both Galactic populations (disc and bulge) of PNe.


\section{PNe nebular abundances versus {\it Spitzer} dust types/subtypes}

In this section we compare nebular gas abundances (He, N, O, Ne, S, Cl, and Ar)
of PNe in our sample with the several major dust types and subtypes as defined
above. We also discuss the differences and similarities between PNe in the
Galactic disc and bulge and compare our results with theoretical nucleosynthesis
predictions for AGB stars. In the previous section, we only considered the
Galactic bulge and disc PNe from the more uniform \cite{Stanghellini2012} sample
of compact PNe. This is because we wanted to check the rate of occurrence of the
different {\it Spitzer} dust types/subtypes in both environments when using
different criteria (more appropriate in our opinion) for selecting bulge
objects. In the following, however, we merge the bulge PNe from
\cite{Stanghellini2012} - according to our criteria for selecting bulge objects
as defined above - with those from \cite{PereaCalderon2009} and
\cite{Gutenkunst2008} in order to construct a unique sample of Galactic bulge
PNe with a higher statistical significance. By comparing the derived abundances
among these different bulge PNe samples, we find that the only apparent
difference is that OC PNe from \cite{PereaCalderon2009} have slightly lower N/O
so we can safely merge the latter bulge PNe samples. Finally, we do not
include the derived abundances (X/H and X/Y) for low ionization PNe in the
abundance diagrams and statistical tests presented in this section. This
concerns the elements He, Ar, and S if the O$^{++}$/(O$^{+}$ $+$ O$^{++}$)
ionization ratio is below 0.4 and Cl if this ratio is less than 0.8.
In principle, also Ne cannot be derived if O$^{++}$ ions are not observed
(e.g.\  they are needed for ICF), but we checked that this is not the case for
our objects.  By inspecting high ionization PNe and assuming similar spectral
energy distributions (SEDs) of the central stars it is clear that the
differences in abundances through the several {\it Spitzer} dust types/subtypes
discussed here are probably real and not due to ICFs.

\begin{table*}
\caption{Kolmogorov--Smirnov (K-S) and Wilcoxon (W) statistical tests for bulge
PNe (OC vs.\ DC) and disc PNe (OC vs.\ DC, OC vs.\ CC, and DC vs.\ CC) (see
text for more details).}
\label{test_b}
\tiny{
\begin{tabular}{lclclll}
\hline\hline
                              & &     {\it bulge}   & &      {\it disc}   &     {\it disc}  &      {\it disc}   \\
Parameter/element             & &        OC vs. DC    & &        OC vs. DC    &        OC vs. CC    &        DC vs. CC    \\
                              & &     K-S / W   & &      K-S / W    &       K-S / W   &      K-S / W    \\
\cline{1-1} \cline{3-3} \cline{5-7}
 C                            & & {\bf 0.01 / 0.01} & &                   &                   &                   \\
 log N$_e$                    & &      0.11 / 0.10  & & {\bf 0.01}/ 0.05  &      0.21 / 0.12  &      0.70 / 0.87  \\
 log T$_e$ (O\,III)           & &      0.22 / 0.30  & & {\bf 0.01 / 0.01} &      0.98 / 0.88  & {\bf 0.01 / 0.00} \\
 log T$_e$ (N\,II)            & & {\bf 0.00 / 0.00} & &      0.60 / 0.43  &      0.83 / 0.60  &      0.80 / 0.49  \\
 O$^{++}$/(O$^+$+O$^{++}$)    & & {\bf 0.01 / 0.00} & &      0.38 / 0.31  &      0.40 / 0.59  &      0.77 / 0.67  \\
 He$^{++}$/(He$^+$+He$^{++}$)$^{*}$& & 0.32 / {\it 0.00}& & 1.00 / {\it 0.00}& 0.90 / {\it 0.01}& 0.96 / {\it 0.00}\\
\cline{1-1} \cline{3-3} \cline{5-7}
$\log\epsilon$(He)            & & {\bf 0.01 / 0.01} & & {\bf 0.00 / 0.00} &      0.45 / 0.49  &      0.82 / 0.02  \\
$\log\epsilon$(O)             & &      0.66 / 0.86  & &      0.17 / 0.31  &      0.50 / 0.72  & {\bf 0.00 / 0.00} \\
$\log\epsilon$(Ar)            & & {\bf 0.00 / 0.00} & & {\bf 0.00 / 0.00} &      0.43 / 0.52  & {\bf 0.00 / 0.00} \\
$\log\epsilon$(Ne)            & &      0.04 / 0.02  & &      0.67 / 0.65  &      0.43 / 0.24  &      0.04 / 0.03  \\
$\log\epsilon$(S)             & &      0.03 / 0.01  & & {\bf 0.00 / 0.00} &      0.17 / 0.08  & {\bf 0.00 / 0.00} \\
$\log\epsilon$(N)             & & {\bf 0.00 / 0.00} & & {\bf 0.00 / 0.00} &      0.92 / 0.87  & {\bf 0.00 / 0.00} \\
$\log\epsilon$(Cl)            & &      0.38 / 0.11  & &      0.21 / 0.13  &      0.54 / 0.39  & {\bf 0.01}/ 0.02  \\
\cline{1-1} \cline{3-3} \cline{5-7}
log(S/O)                      & & {\bf 0.00 / 0.00} & & {\bf 0.01 / 0.01} & {\bf 0.01 / 0.00} & {\bf 0.00 / 0.00} \\
log(Ne/O)                     & & {\bf 0.00 / 0.00} & &      0.55 / 0.53  & {\bf 0.00 / 0.00} & {\bf 0.00 / 0.00} \\
log(Ar/O)                     & & {\bf 0.01 / 0.00} & & {\bf 0.00 / 0.00} &      0.33 / 0.28  & {\bf 0.00 / 0.00} \\
log(Cl/O)                     & &      0.81 / 0.39  & &      0.72 / 0.41  &      0.98 / 0.87  &      0.51 / 0.24  \\
log(N/O)                      & & {\bf 0.00 / 0.00} & & {\bf 0.00 / 0.00} &      0.24 / 0.53  & {\bf 0.00 / 0.00} \\
log(N/Ar)                     & & {\bf 0.00 / 0.00} & &      0.18 / 0.28  &      0.70 / 0.38  &      0.16 / 0.04  \\
log(N/Ne)                     & &      0.04 / 0.04  & & 0.02 / {\bf 0.01} &      0.03 / 0.11  &      0.20 / 0.09  \\
log(N/S)                      & &      0.62 / 0.37  & & {\bf 0.00 / 0.00} & {\bf 0.00 / 0.01} &      0.43 / 0.74  \\
log(N/Cl)                     & & {\bf 0.00 / 0.00} & & {\bf 0.01}/ 0.08  &      0.13 / 0.27  &      0.21 / 0.21  \\
log(Ne/Ar)                    & &      0.96 / 0.87  & & {\bf 0.00 / 0.00} & {\bf 0.00 / 0.01} &      0.85 / 0.50  \\
log(S/Ne)                     & &      0.81 / 0.59  & & 0.02 / {\bf 0.01} &      0.95 / 0.71  & 0.05 / {\bf 0.01} \\
log(S/Ar)                     & &      0.76 / 0.65  & &      0.09 / 0.02  & {\bf 0.00 / 0.00} & 0.04 / {\bf 0.00} \\
\hline
\end{tabular}\\
$^{*}$ Results of Wilcoxon test are not reliable in this case.}
\end{table*}

\begin{table*}
\caption{Kolmogorov--Smirnov (K-S) and Wilcoxon (W) statistical tests for OC and
DC PNe (disc vs.\ bulge) (see text for more details).}
\label{test_b}
\tiny{
\begin{tabular}{lll}
\hline\hline
                              &           OC      &          DC       \\

Parameter/element             &  {\it disc vs. bulge}      &   {\it disc vs. bulge}       \\
                              &          K-S / W      &         K-S / W       \\

\hline
 log N$_e$                    &      0.02 / 0.03  &      0.69 / 0.74  \\
 log T$_e$ (O\,III)           &      0.43 / 0.29  &      0.91 / 0.88  \\
 log T$_e$ (N\,II)            &      0.30 / 0.14  &      0.10 / 0.15  \\
 O$^{++}$/(O$^+$+O$^{++}$)    &      0.23 / 0.89  &      0.32 / 0.03  \\
 He$^{++}$/(He$^+$+He$^{++}$) & 0.96 / {\it 0.00} & 0.29 / {\it 0.00} \\
\hline
$\log\epsilon$(He)            &      0.14 / 0.10  &      0.98 / 0.87  \\
$\log\epsilon$(O)             &      0.74 / 0.84  &      0.16 / 0.49  \\
$\log\epsilon$(Ar)            &      0.03 / 0.05  &      0.54 / 0.36  \\
$\log\epsilon$(Ne)            &      0.71 / 0.44  &      0.95 / 0.63  \\
$\log\epsilon$(S)             &      0.95 / 0.75  &      0.03 / 0.04  \\
$\log\epsilon$(N)             &      0.39 / 0.49  &      0.86 / 0.32  \\
$\log\epsilon$(Cl)            &      0.21 / 0.32  &      0.90 / 0.93  \\
\hline\hline
log(S/O)                      &      0.19 / 0.08  &      0.13 / 0.05  \\
log(Ne/O)                     & {\bf 0.01}/ 0.03  &      0.88 / 0.47  \\
log(Ar/O)                     &      0.33 / 0.18  &      0.73 / 0.68  \\
log(Cl/O)                     &      0.70 / 0.66  &      0.38 / 0.72  \\
log(N/O)                      &      0.39 / 0.57  &      0.68 / 0.97  \\
log(N/Ar)                     &      0.21 / 0.04  &      0.63 / 0.34  \\
log(N/Ne)                     &      0.58 / 0.64  &      0.56 / 0.27  \\
log(N/S)                      &      0.42 / 0.44  &      0.21 / 0.21  \\
log(N/Cl)                     &      0.16 / 0.28  &      0.90 / 1.00  \\
log(Ne/Ar)                    &      0.15 / 0.04  &      0.05 / 0.08  \\
log(S/Ne)                     &      0.68 / 0.66  &      0.23 / 0.12  \\
log(S/Ar)                     & {\bf 0.01 / 0.00} &      0.25 / 0.11  \\
\hline
\end{tabular}\\
}
\end{table*}

\begin{table*}
\caption{Median plasma parameters and abundance values for disc PNe.
The 25 and 75 percentiles are given in brackets and the number of objects in parentheses.}
\label{test_b}
\tiny{
\begin{tabular}{lrrrr}
\hline\hline
Parameter/element             &        OC                  &        DC                  &        CC                  &        F                      \\
\hline
 log N$_e$                    &  4.09 [ 3.70,  5.00] (16)  &  3.76 [ 3.60,  3.89] (21)  &  3.83 [ 3.47,  4.00] (25)  &  3.32 [ 3.03,  3.44] ( 6)     \\
 log T$_e$ (O\,III)           &  4.05 [ 4.01,  4.10] (16)  &  3.98 [ 3.96,  4.01] (16)  &  4.05 [ 4.00,  4.08] (22)  &  4.05 [ 3.93,  4.08] ( 6)     \\ 
 log T$_e$ (N\,II)            &  4.07 [ 3.95,  4.13] (12)  &  4.00 [ 3.91,  4.07] (21)  &  4.05 [ 3.93,  4.09] (21)  &   -   [  -  ,   -  ] ( 2)     \\
 O$^{++}$/(O$^+$+O$^{++}$)    &  0.92 [ 0.43,  0.96] (18)  &  0.84 [ 0.19,  0.94] (21)  &  0.83 [ 0.51,  0.93] (27)  &  0.95 [ 0.88,  0.96] ( 6)     \\
 He$^{++}$/(He$^+$+He$^{++}$) &  0.00 [ 0.00,  0.02] (18)  &  0.00 [ 0.00,  0.10] (20)  &  0.00 [ 0.00,  0.07] (27)  &  0.33 [ 0.03,  0.38] ( 5)     \\
\hline
$\log\epsilon$(He)            & 11.02 [10.99, 11.03] (15)  & 11.10 [11.05, 11.13] (16)  & 11.03 [10.99, 11.07] (21)  & 11.07 [11.01, 11.09] ( 5)     \\
$\log\epsilon$(O)             &  8.57 [ 8.17,  8.72] (17)  &  8.66 [ 8.54,  8.71] (21)  &  8.52 [ 8.35,  8.59] (27)  &  8.52 [ 7.94,  8.67] ( 6)     \\
$\log\epsilon$(Ar)            &  6.04 [ 5.91,  6.12] (14)  &  6.56 [ 6.48,  6.71] (16)  &  6.07 [ 5.92,  6.42] (21)  &  6.21 [ 5.84,  6.40] ( 6)     \\
$\log\epsilon$(Ne)            &  8.00 [ 7.61,  8.19] (10)  &  8.10 [ 7.95,  8.22] (16)  &  7.75 [ 7.55,  7.96] (15)  &  7.83 [ 7.33,  7.98] ( 5)     \\ 
$\log\epsilon$(S)             &  6.66 [ 6.44,  6.83] (13)  &  7.02 [ 6.94,  7.17] (16)  &  6.47 [ 6.23,  6.71] (20)  &  6.43 [ 6.14,  6.52] ( 5)     \\  
$\log\epsilon$(N)             &  7.82 [ 7.44,  8.02] (17)  &  8.38 [ 8.14,  8.56] (21)  &  7.76 [ 7.61,  8.07] (25)  &  7.71 [ 7.15,  7.74] ( 5)     \\  
$\log\epsilon$(Cl)            &  6.08 [ 5.92,  6.13] ( 6)  &  6.21 [ 6.04,  6.57] (10)  &  5.99 [ 5.83,  6.01] ( 7)  &   -   [  -  ,   -  ] ( 2)     \\ 
\hline
log(S/O)                      & -1.84 [-1.95, -1.68] (13)  & -1.62 [-1.69, -1.52] (16)  & -1.99 [-2.14, -1.94] (19)  & -1.86 [-2.14, -1.71] ( 5)     \\
log(Ne/O)                     & -0.60 [-0.64, -0.58] (10)  & -0.57 [-0.66, -0.52] (15)  & -0.69 [-0.82, -0.66] (15)  & -0.57 [-0.70, -0.52] ( 5)     \\ 
log(Ar/O)                     & -2.40 [-2.53, -2.27] (14)  & -2.09 [-2.16, -1.99] (16)  & -2.39 [-2.49, -2.17] (21)  & -2.26 [-2.38, -2.38] ( 6)     \\  
log(Cl/O)                     & -2.53 [-2.83, -2.33] ( 6)  & -2.45 [-2.66, -2.10] (10)  & -2.63 [-2.75, -2.49] ( 7)  &   -   [  -  ,   -  ] ( 2)     \\  
log(N/O)                      & -0.81 [-0.95, -0.53] (17)  & -0.27 [-0.40, -.10 ] (21)  & -0.70 [-0.81, -0.55] (25)  & -0.65 [-1.35, -0.40] ( 5)     \\ 
\hline\hline
\end{tabular}\\
}
\end{table*}

\begin{table*}
\caption{Median plasma parameters and abundance values for bulge PNe.}
\label{test_b}
\tiny{
\begin{tabular}{lrr}
\hline\hline
Parameter/element             &        OC                  &        DC                     \\
\hline
 C                            &  1.33 [~1.15, ~1.55] (15)  &  1.99 [~1.44, ~2.36] (30)    \\
 log N$_e$                    &  3.57 [~3.32, ~3.80] (17)  &  3.73 [~3.61, ~3.96] (38)     \\
 log T$_e$ (O\,III)           &  4.01 [~3.97, ~4.05] (15)  &  3.98 [~3.96, ~4.03] (16)     \\ 
 log T$_e$ (N\,II)            &  4.14 [~4.03, ~4.16] (13)  &  3.94 [~3.83, ~4.07] (37)     \\
 O$^{++}$/(O$^+$+O$^{++}$)    &  0.90 [~0.71, ~0.94] (17)  &  0.55 [~0.00, ~0.88] (38)     \\
 He$^{++}$/(He$^+$+He$^{++}$) &  0.00 [~0.00, ~0.04] (16)  &  0.00 [~0.00, ~0.04] (38)     \\
\hline
$\log\epsilon$(He)            & 11.04 [11.01, 11.06] (17)  & 11.09 [11.05, 11.14] (22)     \\
$\log\epsilon$(O)             &  8.61 [ 8.38,  8.70] (17)  &  8.61 [ 8.21,  8.71] (38)     \\
$\log\epsilon$(Ar)            &  6.29 [ 6.05,  6.40] (17)  &  6.51 [ 6.38,  6.63] (22)     \\
$\log\epsilon$(Ne)            &  7.77 [ 7.67,  8.05] (14)  &  8.12 [ 7.86,  8.23] (17)     \\ 
$\log\epsilon$(S)             &  6.64 [ 6.41,  6.79] (17)  &  6.87 [ 6.66,  7.03] (22)     \\  
$\log\epsilon$(N)             &  7.74 [ 7.38,  7.98] (17)  &  8.30 [ 8.03,  8.49] (38)     \\  
$\log\epsilon$(Cl)            &  6.17 [ 5.84,  6.28] (10)  &  6.32 [ 6.14,  6.57] (~8)     \\ 
\hline
log(S/O)                      & -1.98 [-2.03, -1.80] (17)  & -1.71 [-1.78, -1.61] (22)     \\
log(Ne/O)                     & -0.66 [-0.69, -0.64] (14)  & -0.55 [-0.60, -0.44] (17)     \\
log(Ar/O)                     & -2.31 [-2.39, -2.19] (17)  & -2.03 [-2.27, -1.96] (22)     \\
log(Cl/O)                     & -2.44 [-2.63, -2.24] (10)  & -2.37 [-2.47, -2.20] (~8)     \\
log(N/O)                      & -0.80 [-1.04, -0.53] (17)  & -0.23 [-0.48, -0.11] (38)     \\
\hline\hline
\end{tabular}
}
\end{table*}

\begin{table*}
\caption{Nucleosynthesis predictions for AGB stars from Karakas (2010).}
\label{test_b}
\tiny{
\begin{tabular}{ccccccc}
\hline\hline
Mass & Z & PMZ & $\log\epsilon$(He) & $\log$(N/O) &  $\log$(Ne/O)& $\log$(S/O) \\
\hline
1.0 & 0.02  & no & 11.065 &  -0.742    & -0.839  &  -1.685 \\
1.5 & 0.02  & no & 11.056 &  -0.604    & -0.839  &  -1.685 \\
1.9 & 0.02  & no & 11.045 &  -0.579    & -0.840  &  -1.684 \\
2.5 & 0.02  & no & 11.056 &  -0.545    & -0.774  &  -1.676 \\
3.0 & 0.02  & no & 11.078 &  -0.515    & -0.668  &  -1.663 \\
3.0 & 0.02  & yes& 11.078 &  -0.519    & -0.570  &  -1.668 \\
4.0 & 0.02  & no & 11.062 &  -0.470    & -0.742  &  -1.661 \\
5.0 & 0.02  & no & 11.113 &  -0.282    & -0.751  &  -1.644 \\
6.0 & 0.02  & no & 11.160 &   0.070    & -0.723  &  -1.609 \\
\hline
1.0 & 0.008 & no & 10.982 &  -0.717 &  -0.838 &  -1.683 \\
1.5 & 0.008 & no & 10.982 &  -0.573 &  -0.834 &  -1.683 \\
1.9 & 0.008 & no & 10.984 &  -0.555 &  -0.766 &  -1.691 \\
2.5 & 0.008 & no & 11.032 &  -0.496 &  -0.479 &  -1.673 \\
3.0 & 0.008 & no & 11.065 &  -0.467 &  -0.273 &  -1.659 \\
3.0 & 0.008 & yes& 11.064 &  -0.489 &  -0.103 &  -1.682 \\
4.0 & 0.008 & no & 10.990 &  -0.469 &  -0.684 &  -1.671 \\
5.0 & 0.008 & no & 11.086 &   0.724 &  -0.488 &  -1.572 \\
6.0 & 0.008 & no & 11.111 &   0.691 &  -0.516 &  -1.467 \\
\hline
1.0 & 0.004 & no  &  10.964 &  -0.682 &   -0.837 &  -1.683 \\
1.5 & 0.004 & no  &  10.970 &  -0.574 &   -0.803 &  -1.692 \\
1.9 & 0.004 & no  &  10.992 &  -0.544 &   -0.644 &  -1.708 \\
2.5 & 0.004 & no  &  11.064 &  -0.504 &   -0.065 &  -1.700 \\
3.0 & 0.004 & no  &  11.044 &  -0.496 &   -0.133 &  -1.692 \\
3.0 & 0.008 & yes &  11.042 &  -0.539 &    0.079 &  -1.736 \\
4.0 & 0.004 & no  &  11.016 &   0.649 &   -0.580 &  -1.693 \\
5.0 & 0.004 & no  &  11.114 &   1.165 &   -0.254 &  -1.509 \\
6.0 & 0.004 & no  &  11.135 &   1.209 &   -0.268 &  -1.290 \\
\hline
\end{tabular}\\
The abundances listed here (on the usual scale $\log\epsilon$(X)~=~$\log$(X/H) +
12) have been calculated by us from the average abundances (mass fraction) given
by Karakas (2010). We only consider the most abundant isotopes (i.e.\ 
$^{4}$He, $^{14}$N, $^{16}$O, $^{20,22}$Ne, and $^{32}$S).
}
\end{table*}

\begin{table*}
\caption{Kolmogorov--Smirnov (K-S) and Wilcoxon (W) statistical tests for 
the {\it Spitzer} dust subtypes DC (DC$_{cr}$ vs.\ DC$_{am+cr}$), CC (CC$_{ar}$
vs.\ CC$_{al}$), and OC (OC$_{cr}$ vs.\ OC$_{am}$, OC$_{cr}$ vs.\ OC$_{am+cr}$, and 
OC$_{am}$ vs.\ OC$_{am+cr}$) in disc and bulge PNe.}
\label{test_b}
\tiny{
\begin{tabular}{lclclclll}
\hline\hline
                              & & {\it disc $+$ bulge}       & & {\it disc $+$ bulge}    & &                         & {\it disc $+$ bulge}       &  \\
                              & & DC subtypes                       & & CC subtypes      & &                         & OC subtypes      &  \\
Parameter/element             & & DC$_{cr}$ vs. DC$_{am+cr}$ & & CC$_{ar}$ vs. CC$_{al}$ & & OC$_{cr}$ vs. OC$_{am}$ & OC$_{cr}$ vs. OC$_{am+cr}$ & OC$_{am}$ vs. OC$_{am+cr}$ \\
                              & &  K-S / W                   & & K-S / W                 & &  K-S / W                &  K-S / W                   & K-S / W  \\
\cline{1-1} \cline{3-3} \cline{5-5} \cline{7-9}
 log N$_e$                    & &      0.05 / 0.03  & &      0.05 / 0.07  & &      0.56 / 0.65  &      0.08 / 0.12  &      0.25 / 0.40  \\
 log T$_e$ (O\,III)           & &      0.22 / 0.46  & &                   & &      0.21 / 0.56  &      0.19 / 0.83  &      0.05 / 0.03  \\
 log T$_e$ (N\,II)            & &      0.22 / 0.09  & &      0.97 / 0.76  & &      0.23 / 0.11  &      0.08 / 0.05  &      0.33 / 0.32  \\
 O$^{++}$/(O$^+$+O$^{++}$)    & &      0.31 / 0.07  & &      0.33 / 0.22  & &      0.10 / 0.14  &      0.95 / 0.93  &      0.24 / 0.21  \\
 He$^{++}$/(He$^+$+He$^{++}$) & & {\bf 0.00} / {\it 0.00} & &      0.71 / {\it 0.10}  & &      0.63 / {\it 0.02}  &      0.09 /  -          & 0.14 / {\it 0.00} \\
\cline{1-1} \cline{3-3} \cline{5-5} \cline{7-9}
$\log\epsilon$(He)            & &      0.46 / 0.18  & &      0.04 / 0.08  & &      0.02 / 0.24  &      0.09 / 0.29  &      0.06 / 0.05  \\
$\log\epsilon$(O)             & &      0.96 / 0.97  & &      0.08 / 0.06  & &      0.36 / 0.46  &      0.70 / 0.93  &      0.24 / 0.19  \\
$\log\epsilon$(Ar)            & &      0.03 / 0.05  & &      0.05 / 0.06  & & {\bf 0.00 / 0.01} & {\bf 0.01}/ 0.02  &      0.15 / 0.08  \\
$\log\epsilon$(Ne)            & &      0.30 / 0.20  & &                   & &      0.19 / 0.30  &      0.32 / 0.26  &      1.00 / 0.93  \\
$\log\epsilon$(S)             & &      0.38 / 0.29  & &      0.02 / 0.03  & &      0.07 / 0.10  &      0.52 / 0.40  &      0.19 / 0.14  \\
$\log\epsilon$(N)             & &      0.05 / 0.16  & &      0.07 / 0.02  & &      0.22 / 0.42  &      0.34 / 0.35  &      0.96 / 0.83  \\
$\log\epsilon$(Cl)            & &      0.38 / 0.78  & &                   & &                   &                   &                   \\
\cline{1-1} \cline{3-3} \cline{5-5} \cline{7-9}
log(S/O)                      & &      0.33 / 0.52  & &      0.07 / 0.03  & &      0.86 / 1.00  &      0.52 / 0.56  &      0.44 / 1.00  \\
log(Ne/O)                     & &      0.49 / 0.24  & &                   & &      1.00 / 0.85  &      0.32 / 0.26  &      0.19 / 0.11  \\
log(Ar/O)                     & &      0.71 / 0.57  & &      0.05 / 0.10  & & 0.04 / {\bf 0.01} & {\bf 0.01 / 0.00} &      0.48 / 0.52  \\
log(Cl/O)                     & &      0.19 / 0.57  & &                   & &                   &                   &                   \\
log(N/O)                      & &      0.25 / 0.12  & &      0.05 / 0.02  & &      0.72 / 0.62  &      0.34 / 0.25  &      0.75 / 0.27  \\
log(N/Ar)                     & &      0.94 / 0.49  & &      0.23 / 0.11  & &      0.83 / 0.42  &      0.96 / 0.89  &      0.56 / 0.57  \\
log(N/Ne)                     & &      0.26 / 0.17  & &                   & &      0.97 / 1.00  &      0.81 / 0.75  &      0.47 / 0.48  \\
log(N/S)                      & &      0.99 / 0.95  & &      0.94 / 0.75  & &      0.09 / 0.09  &      0.08 / 0.02  &      0.47 / 0.31  \\
log(N/Cl)                     & &      0.93 / 0.71  & &                   & &                   &                   &                   \\
log(Ne/Ar)                    & &      0.19 / 0.19  & &                   & &      0.19 / 0.16  &      1.00 / 0.75  &      0.67 / 0.40  \\
log(S/Ne)                     & &      0.30 / 0.13  & &                   & &      0.74 / 0.88  &      0.32 / 0.15  &      0.11 / 0.06  \\
log(S/Ar)                     & &      0.95 / 0.77  & &      0.35 / 0.14  & &      0.07 / 0.03  & {\bf 0.01}/ 0.02  &      0.91 / 0.76  \\
\hline
\end{tabular}\\
}
\end{table*}

\begin{table*}
\caption{Median plasma parameters and abundance values for 
 bulge + disc OC subtypes PNe.}
\label{test_b}
\tiny{
\begin{tabular}{lrrrr}
\hline\hline
Parameter/element             &       \OCcr                &       \OCam                &       \OCamcr              &                               \\
\hline
 log N$_e$                    &  3.47 [ 3.15,  3.79] ( 9)  &  3.87 [ 3.23,  4.15] (15)  &  3.92 [ 3.65,  4.20] ( 9)  &                               \\
 log T$_e$ (O\,III)           &  4.00 [ 3.92,  4.28] ( 8)  &  4.04 [ 4.01,  4.10] (15)  &  4.00 [ 4.00,  4.04] ( 8)  &                               \\ 
 log T$_e$ (N\,II)            &  3.99 [ 3.87,  4.06] ( 6)  &  4.10 [ 4.04,  4.15] (11)  &  4.17 [ 4.04,  4.15] ( 8)  &                               \\
 O$^{++}$/(O$^+$+O$^{++}$)    &  0.84 [ 0.59,  0.92] (10)  &  0.94 [ 0.67,  0.98] (16)  &  0.87 [ 0.57,  0.91] ( 9)  &                               \\
 He$^{++}$/(He$^+$+He$^{++}$) &  0.01 [ 0.00,  0.03] ( 9)  &  0.01 [ 0.00,  0.17] (16)  &  0.00 [ 0.00,  0.00] ( 9)  &                               \\
\hline
$\log\epsilon$(He)            & 11.06 [10.86, 11.09] ( 9)  & 11.02 [11.00, 11.03] (15)  & 11.03 [11.03, 11.05] ( 8)  &                               \\
$\log\epsilon$(O)             &  8.63 [ 7.60,  8.75] ( 9)  &  8.42 [ 8.29,  8.64] (16)  &  8.61 [ 8.43,  8.73] ( 9)  &                               \\
$\log\epsilon$(Ar)            &  6.44 [ 6.32,  6.58] ( 8)  &  6.03 [ 5.90,  6.11] (15)  &  6.13 [ 6.05,  6.25] ( 8)  &                               \\
$\log\epsilon$(Ne)            &  8.18 [ 7.26,  8.28] ( 6)  &  7.78 [ 7.67,  8.00] (12)  &  7.87 [ 7.63,  8.05] ( 6)  &                               \\ 
$\log\epsilon$(S)             &  6.82 [ 6.57,  6.96] ( 8)  &  6.50 [ 6.40,  6.66] (14)  &  6.68 [ 6.60,  6.83] ( 8)  &                               \\  
$\log\epsilon$(N)             &  8.01 [ 7.10,  8.43] (10)  &  7.78 [ 7.34,  7.91] (15)  &  7.75 [ 7.47,  7.81] ( 9)  &                               \\  
$\log\epsilon$(Cl)            &   -   [  -  ,   -  ] (  )  &  6.15 [ 6.00,  6.22] ( 8)  &  5.84 [ 5.61,  6.11] ( 5)  &                               \\ 
\hline
log(S/O)                      & -1.96 [-2.01, -1.65] ( 8)  & -1.87 [-2.01, -1.70] (14)  & -1.87 [-1.94, -1.81] ( 8)  &                               \\
log(Ne/O)                     & -0.63 [-0.78, -0.60] ( 6)  & -0.62 [-0.73, -0.58] (12)  & -0.61 [-0.83, -0.64] ( 6)  &                               \\ 
log(Ar/O)                     & -2.12 [-2.28, -2.03] ( 8)  & -2.38 [-2.50, -2.28] (15)  & -2.38 [-2.52, -2.35] ( 8)  &                               \\  
log(Cl/O)                     &   -   [  -  ,   -  ] (  )  & -2.23 [-2.52, -2.17] ( 8)  & -2.49 [-2.87, -2.42] ( 5)  &                               \\  
log(N/O)                      & -0.65 [-1.05, -0.22] (10)  & -0.80 [-0.96, -0.52] (15)  & -0.88 [-1.04, -0.82] ( 9)  &                               \\ 
\hline\hline
\end{tabular}
}
\end{table*}

\begin{table*}
\caption{Median plasma parameters and abundance values for 
 bulge + disc DC and CC subtypes PNe.}
\label{test_b}
\tiny{
\begin{tabular}{lrrrr}
\hline\hline
Parameter/element             &       \DCcr                &       \DCamcr              &       \CCar                &       \CCal                   \\
\hline
 log N$_e$                    &  3.71 [ 3.85,  3.84] (42)  &  3.89 [ 3.65,  4.05] (17)  &  3.48 [ 3.31,  3.74] ( 8)  &  3.90 [ 3.50,  4.11] (15)     \\
 log T$_e$ (O\,III)           &  3.99 [ 3.95,  4.03] (23)  &  3.97 [ 3.96,  3.98] ( 9)  &  4.07 [ 3.95,  4.08] ( 5)  &  4.05 [ 4.01,  4.08] (17)     \\ 
 log T$_e$ (N\,II)            &  3.96 [ 3.84,  4.06] (40)  &  4.05 [ 3.90,  4.11] (17)  &  4.05 [ 3.92,  4.09] ( 8)  &  4.06 [ 3.97,  4.11] (12)     \\
 O$^{++}$/(O$^+$+O$^{++}$)    &  0.58 [ 0.04,  0.89] (42)  &  0.86 [ 0.10,  0.93] (17)  &  0.82 [ 0.45,  0.89] ( 8)  &  0.91 [ 0.62,  0.98] (18)     \\
 He$^{++}$/(He$^+$+He$^{++}$) &  0.00 [ 0.00,  0.03] (41)  &  0.00 [ 0.00,  0.00] (17)  &  0.01 [ 0.00,  0.06] ( 8)  &  0.01 [ 0.00,  0.10] (18)     \\
\hline
$\log\epsilon$(He)            & 11.10 [11.06, 11.14] (26)  & 11.08 [11.04, 11.13] (12)  & 11.12 [11.02, 11.12] ( 6)  & 11.03 [10.99, 11.06] (15)     \\
$\log\epsilon$(O)             &  8.64 [ 8.49,  8.72] (42)  &  8.63 [ 8.38,  8.74] (17)  &  8.59 [ 8.54,  8.64] ( 8)  &  8.44 [ 8.32,  8.58] (18)     \\
$\log\epsilon$(Ar)            &  6.56 [ 6.48,  6.69] (26)  &  6.44 [ 6.31,  6.60] (12)  &  6.44 [ 6.14,  6.46] ( 6)  &  5.97 [ 5.88,  6.18] (14)     \\
$\log\epsilon$(Ne)            &  8.09 [ 7.94,  8.16] (21)  &  8.19 [ 7.94,  8.24] (12)  &   -   [  -  ,   -  ] (  )  &  7.66 [ 7.35,  7.84] (11)     \\ 
$\log\epsilon$(S)             &  6.99 [ 6.84,  7.11] (26)  &  6.89 [ 6.66,  7.06] (12)  &  6.71 [ 6.45,  6.87] ( 6)  &  6.35 [ 6.06,  6.50] (13)     \\  
$\log\epsilon$(N)             &  8.39 [ 8.12,  8.52] (42)  &  8.24 [ 7.98,  8.35] (17)  &  8.20 [ 7.82,  8.46] ( 8)  &  7.66 [ 7.53,  8.02] (15)     \\  
$\log\epsilon$(Cl)            &  6.26 [ 6.11,  6.49] (12)  &  6.37 [ 5.95,  6.67] ( 6)  &   -   [  -  ,   -  ] (  )  &  6.00 [ 5.94,  6.12] ( 4)     \\ 
\hline
log(S/O)                      & -1.65 [-1.75, -1.60] (26)  & -1.72 [-1.81, -1.54] (12)  & -1.88 [-2.06, -1.72] ( 6)  & -2.07 [-2.28, -1.98] (13)     \\
log(Ne/O)                     & -0.57 [-0.63, -0.48] (21)  & -0.53 [-0.59, -0.37] (12)  &   -   [  -  ,   -  ] (  )  & -0.73 [-0.84, -0.67] (11)     \\ 
log(Ar/O)                     & -2.06 [-2.22, -1.96] (26)  & -2.11 [-2.24, -1.97] (12)  & -2.17 [-2.42, -2.15] ( 6)  & -2.40 [-2.49, -2.32] (14)     \\  
log(Cl/O)                     & -2.43 [-2.51, -2.24] (12)  & -2.24 [-2.65, -1.98] ( 6)  &   -   [  -  ,   -  ] (  )  & -2.47 [-2.70, -2.20] ( 4)     \\  
log(N/O)                      & -0.21 [-0.41, -0.07] (42)  & -0.36 [-0.62, -0.17] (17)  & -0.38 [-0.71, -0.18] ( 8)  & -0.71 [-0.96, -0.68] (15)     \\ 
\hline\hline
\end{tabular}
}
\end{table*}

\begin{figure*}[t]
\resizebox{0.47\hsize}{!}{\includegraphics{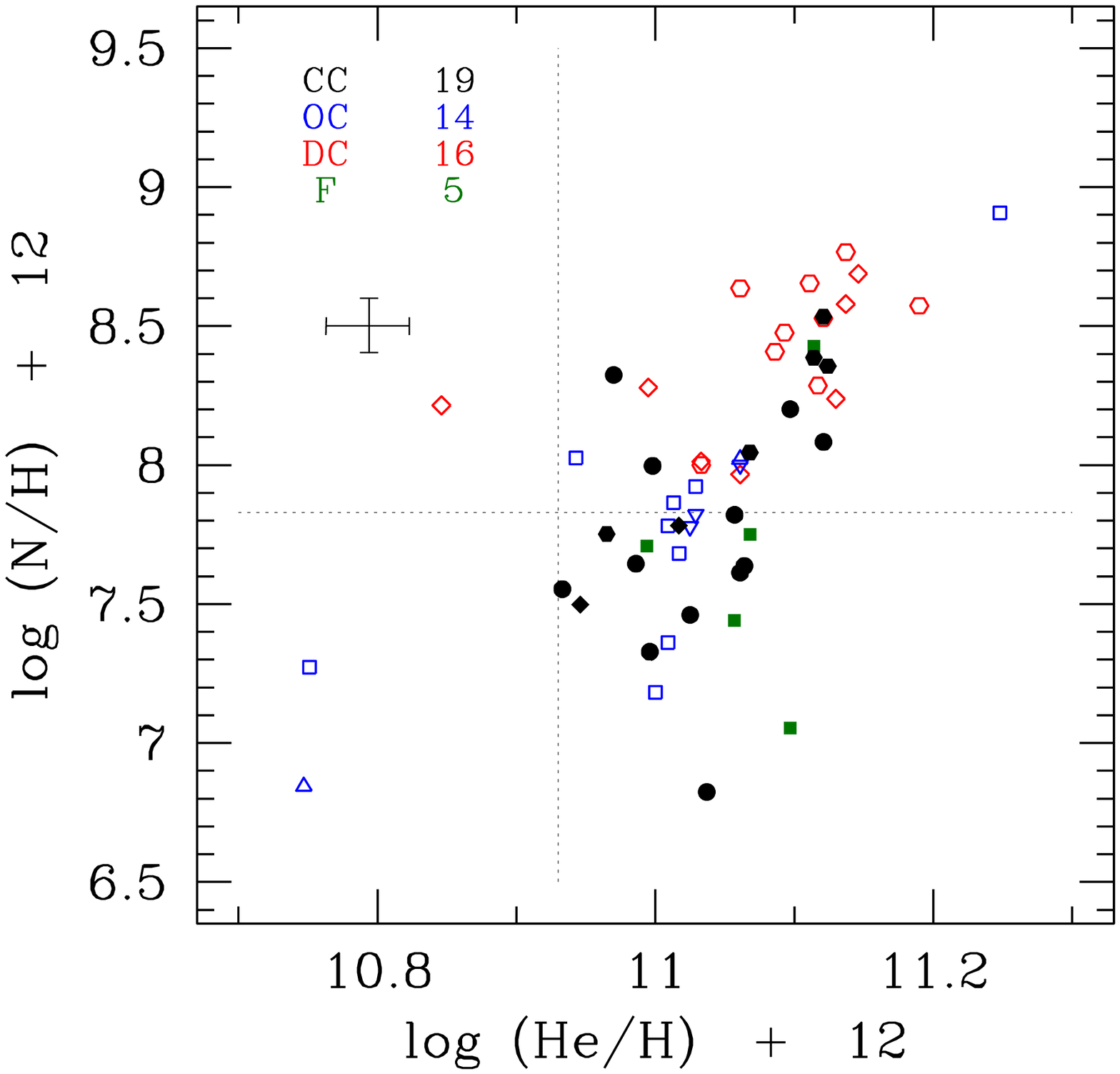}}
\resizebox{0.47\hsize}{!}{\includegraphics{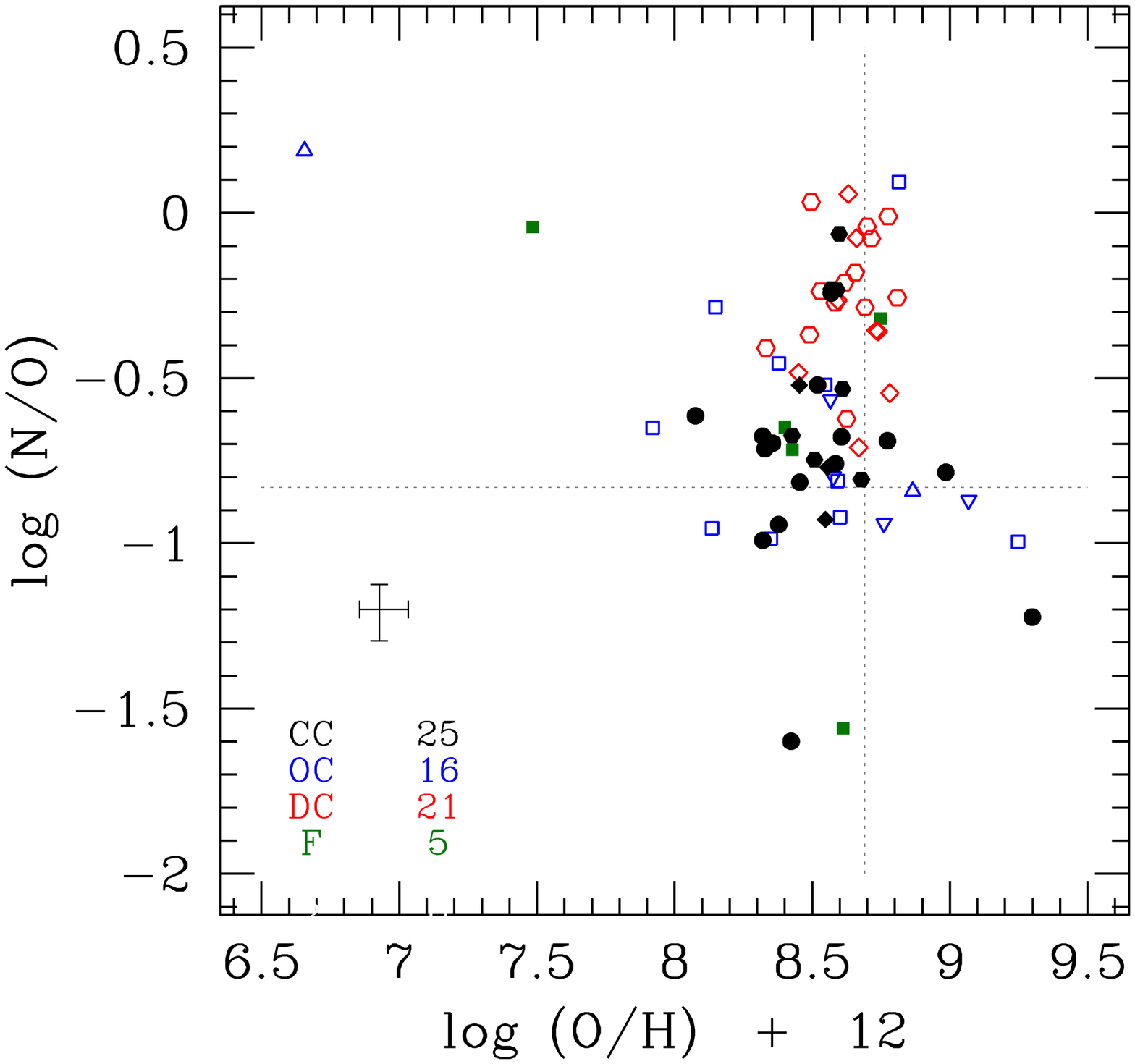}}
\caption[]{
 Diagrams of nebular abundance ratios He/H vs.\ N/H (left panel) and O/H vs.\ N/O
 (right panel) for Galactic disc PNe. Blue symbols mark OC~PNe (\OCcr\ - open
 triangles; \OCam\ - open squares; \OCamcr\ - open reversed triangles).  Red
 symbols mark CD~PNe (\DCcr\ - open hexagons; \DCamcr\ - open diamonds); Black
 symbols mark CC~PNe (\CCar\ - filled hexagon; \CCal\ - filled circles; \CCaral\
 - filled diamonds); Green filled squares mark F type PNe. The horizontal/vertical lines mark the solar abundance
 values.
}
\label{abund_d1}
\end{figure*}

\begin{figure*}[t]
\resizebox{0.47\hsize}{!}{\includegraphics{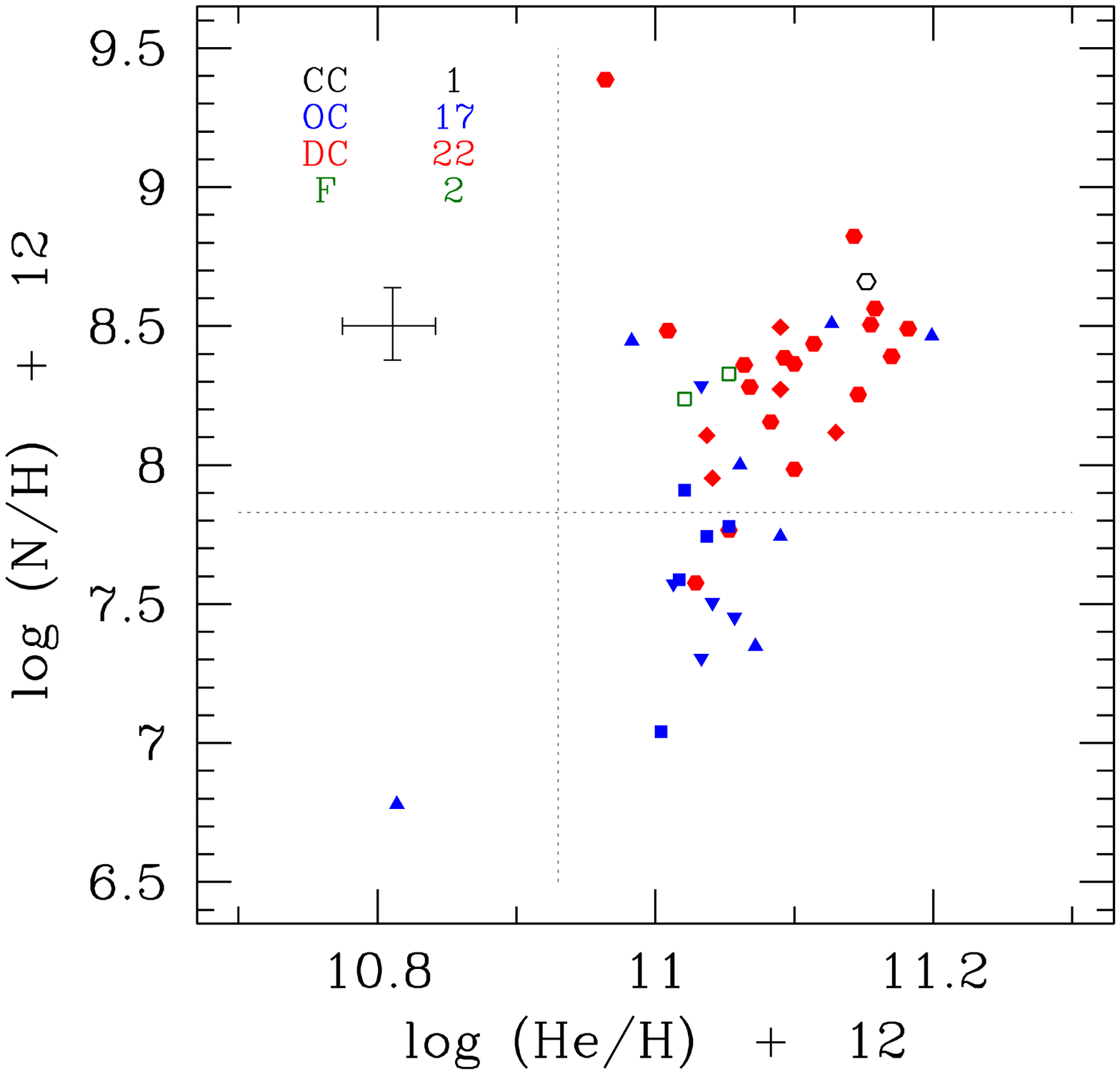}}
\resizebox{0.47\hsize}{!}{\includegraphics{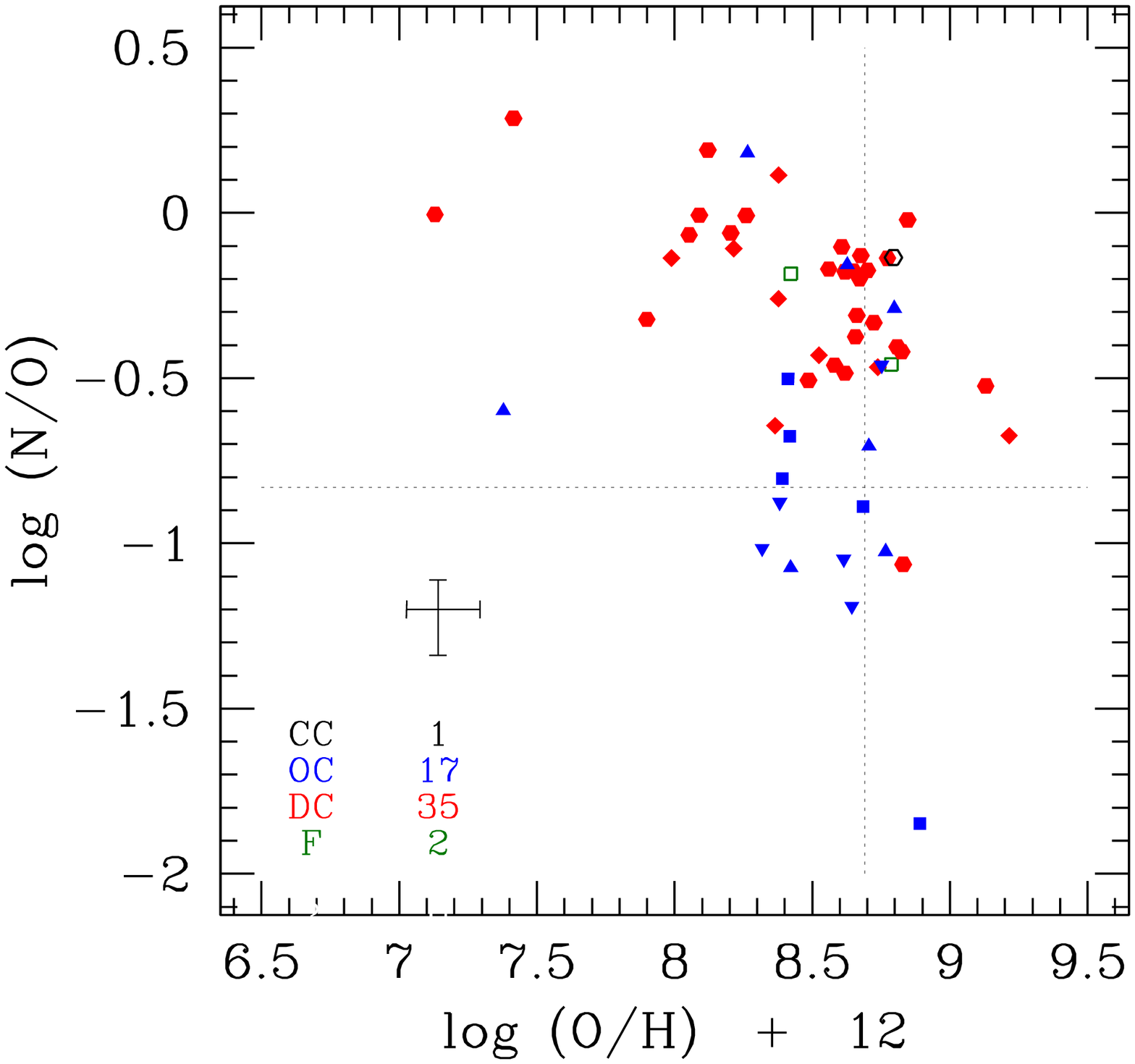}}
\caption[]{
 Diagrams of nebular abundance ratios He/H vs.\ N/H (left panel) and O/H vs.\ N/O
 (right panel) for Galactic bulge PNe.  Blue symbols mark OC~PNe (\OCcr\ - filled
 triangles; \OCam\ - filled squares; \OCamcr\ - filled reversed triangles).  Red
 symbols mark CD~PNe (\DCcr\ - filled hexagons; \DCamcr\ - filled diamonds);
 Black symbols mark CC~PNe (\CCar\ - open hexagon); Green open squares mark F
 type PNe. The horizontal/vertical lines mark the solar abundance
 values.
}
\label{abund_b1}
\end{figure*}

\begin{figure*}[t]
\resizebox{0.47\hsize}{!}{\includegraphics{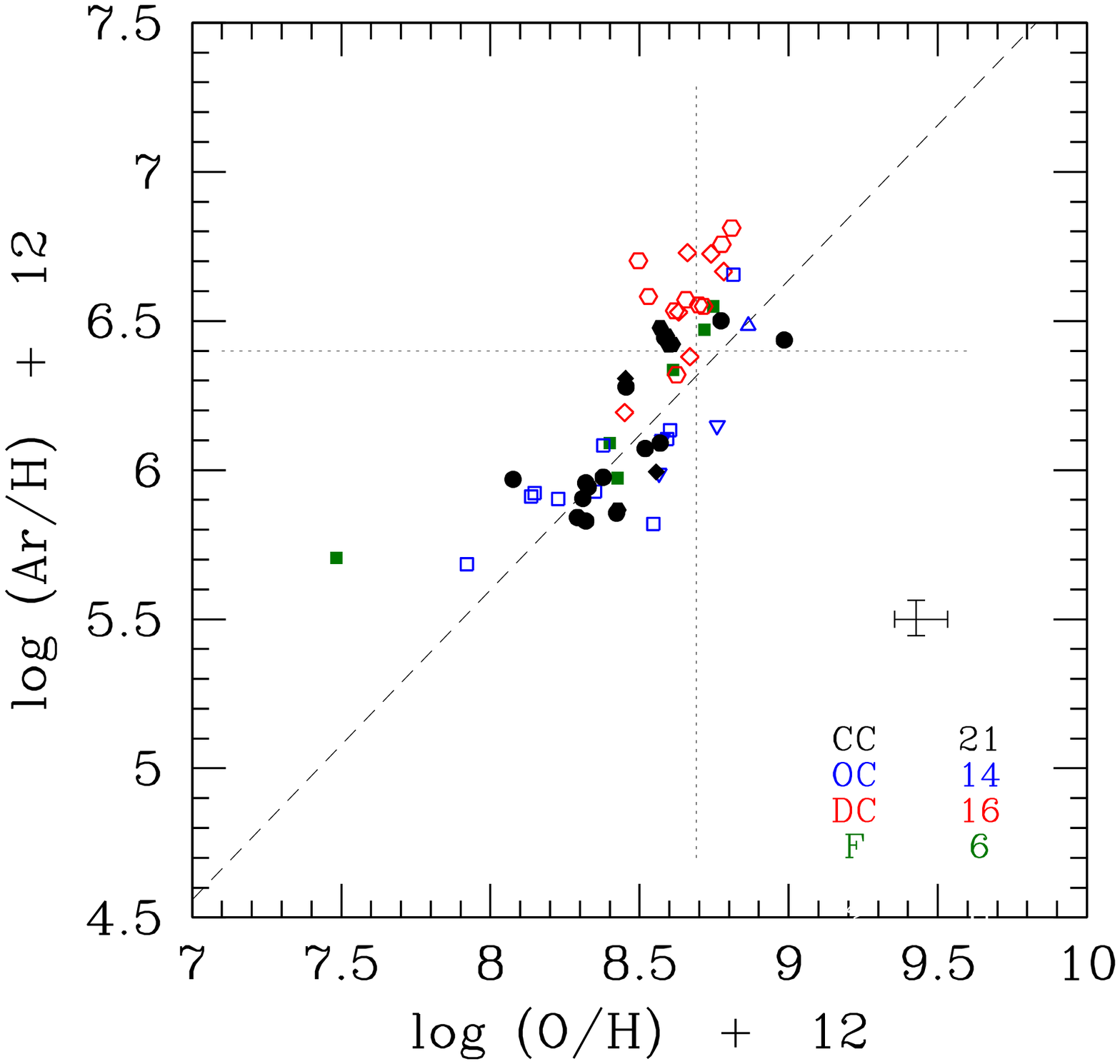}}
\resizebox{0.47\hsize}{!}{\includegraphics{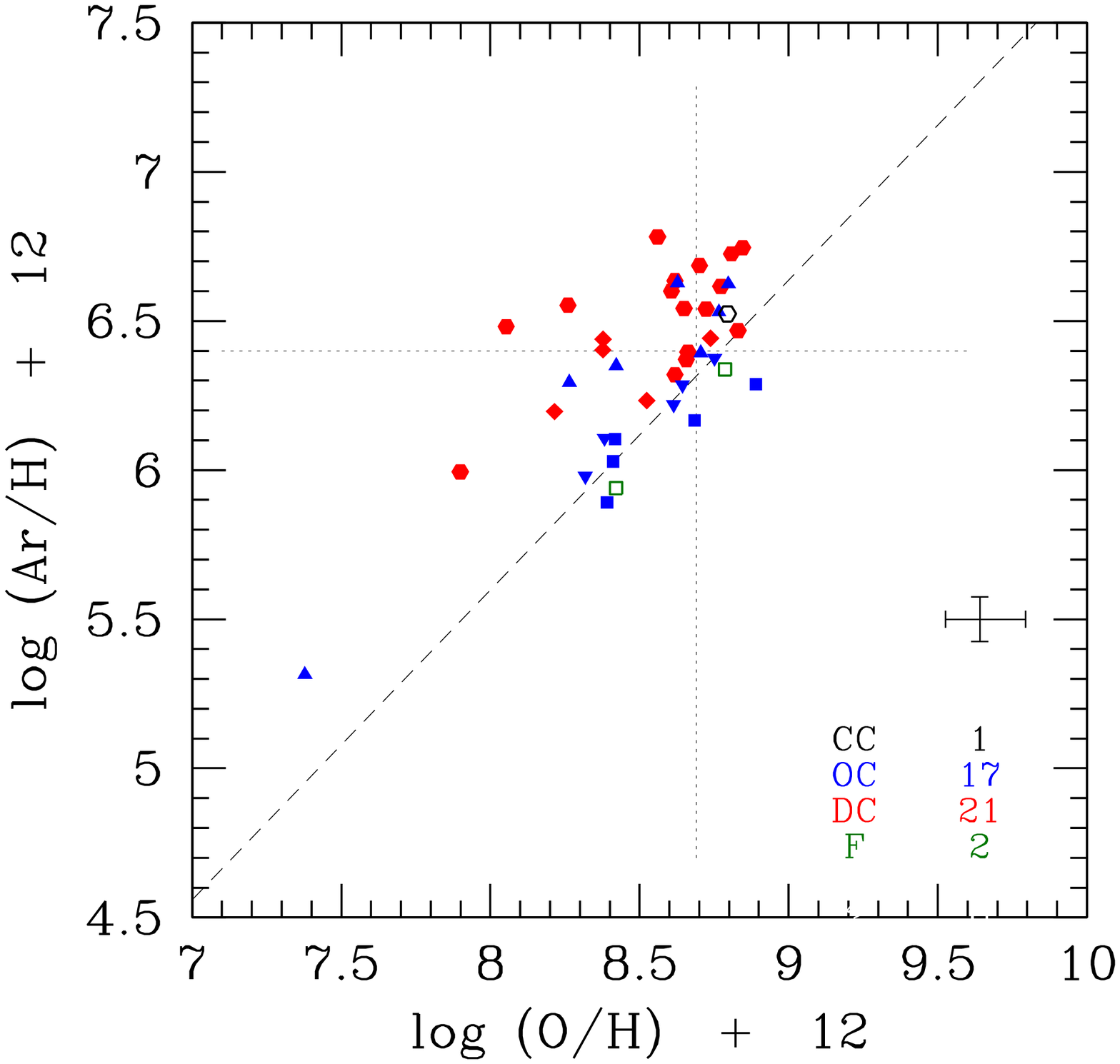}}
\caption[]{
 Diagrams of nebular abundance ratio O/H vs.\ Ar/H for Galactic disc PNe (left
 panel) and Galactic bulge PNe (right panel). The same meaning of symbols as in
 Figs. 5 and 6. The horizontal/vertical lines mark the solar abundance values,
 while the dashed line marks the relation derived by \citet{Izotov06}.
}
\label{abund_arh_db}
\end{figure*}

\begin{figure*}
\resizebox{0.47\hsize}{!}{\includegraphics{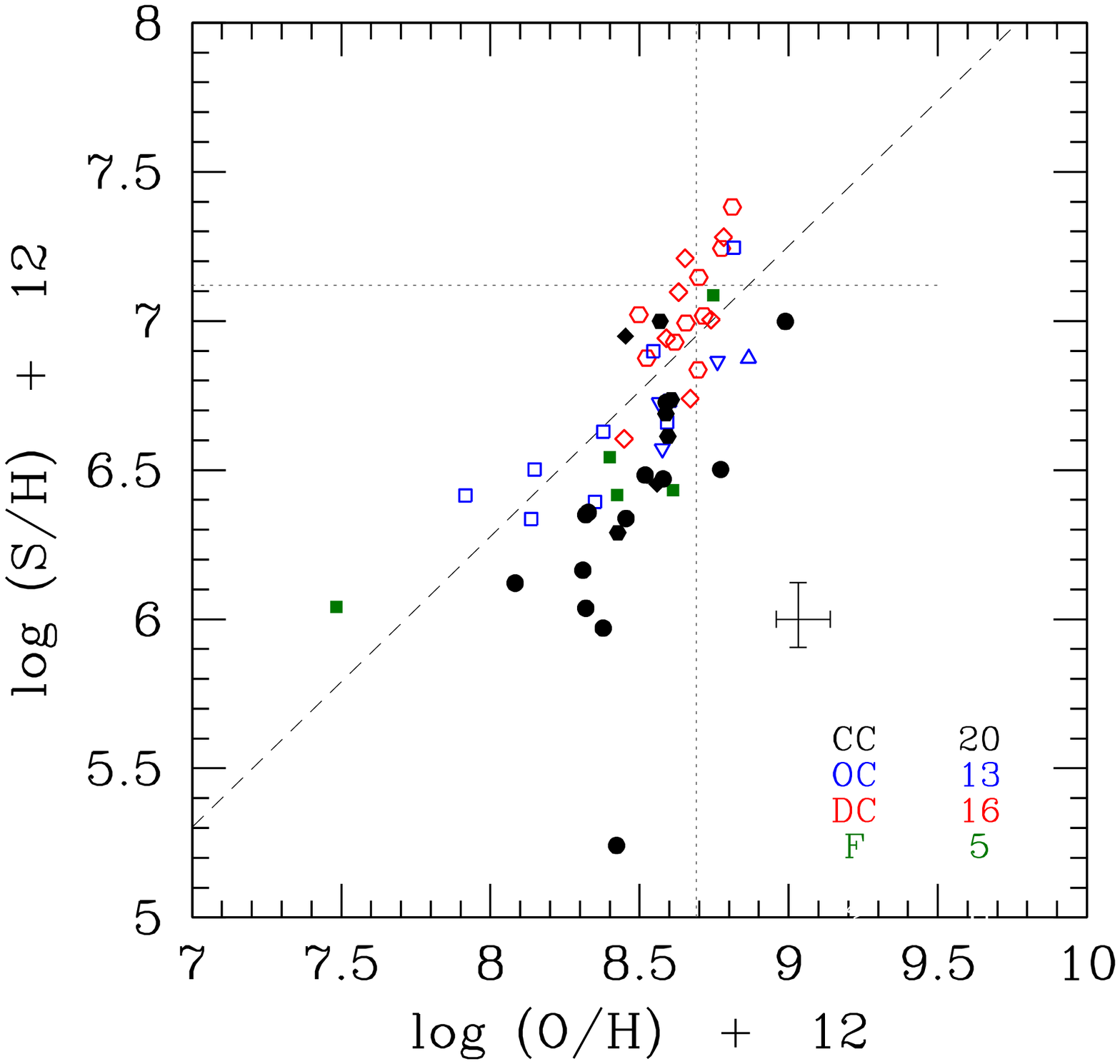}}
\resizebox{0.47\hsize}{!}{\includegraphics{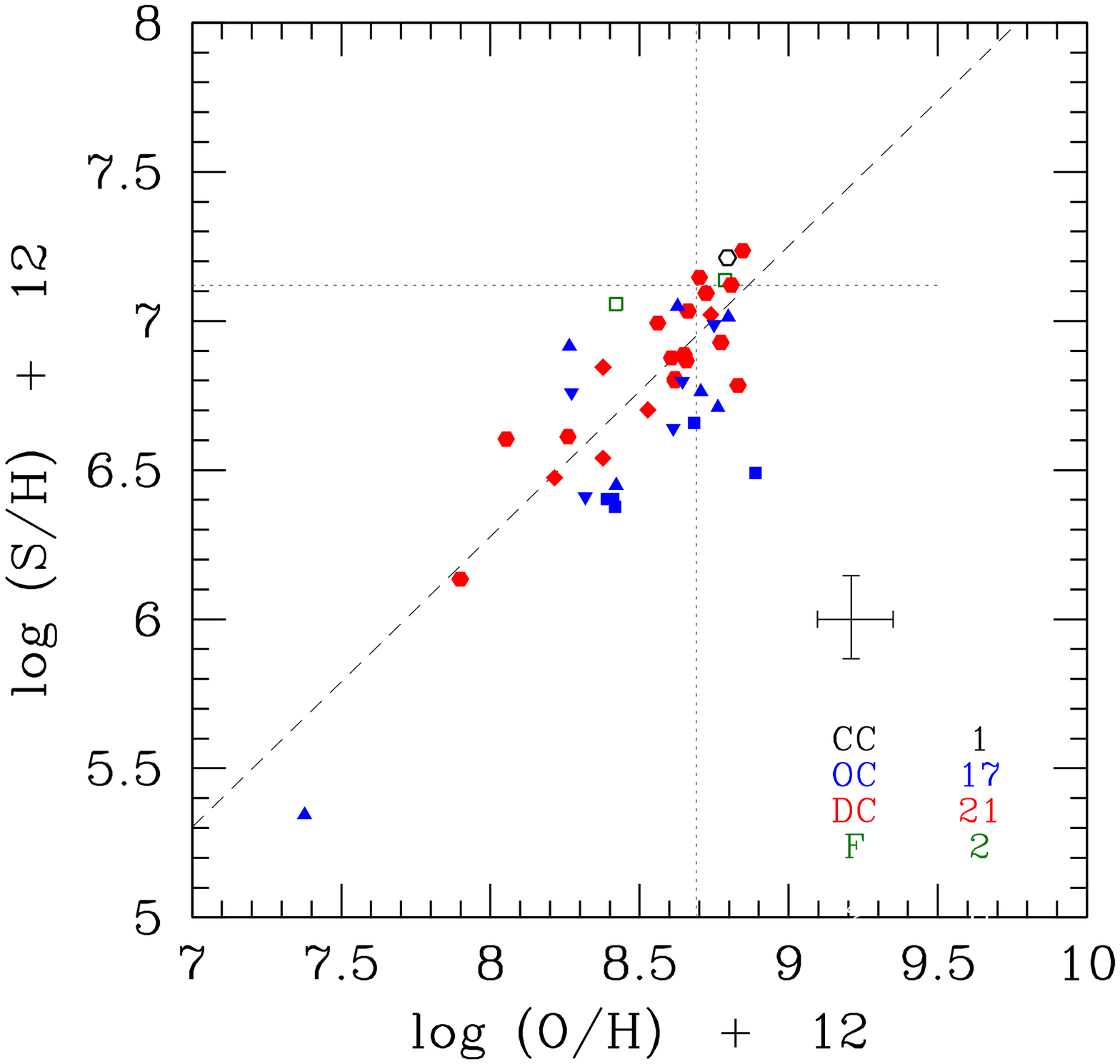}}
\caption[]{
 Diagrams of nebular abundance ratio O/H vs.\ S/H for Galactic disc PNe (left
 panel) and Galactic bulge PNe (right panel). The same meaning of symbols as in
 Figs. 5 and 6. The horizontal/vertical lines mark the solar abundance values,
 while the dashed line marks the relation derived by \citet{Izotov06}.
}
\label{abund_sh_db}
\end{figure*}

\begin{figure*}
\resizebox{0.47\hsize}{!}{\includegraphics{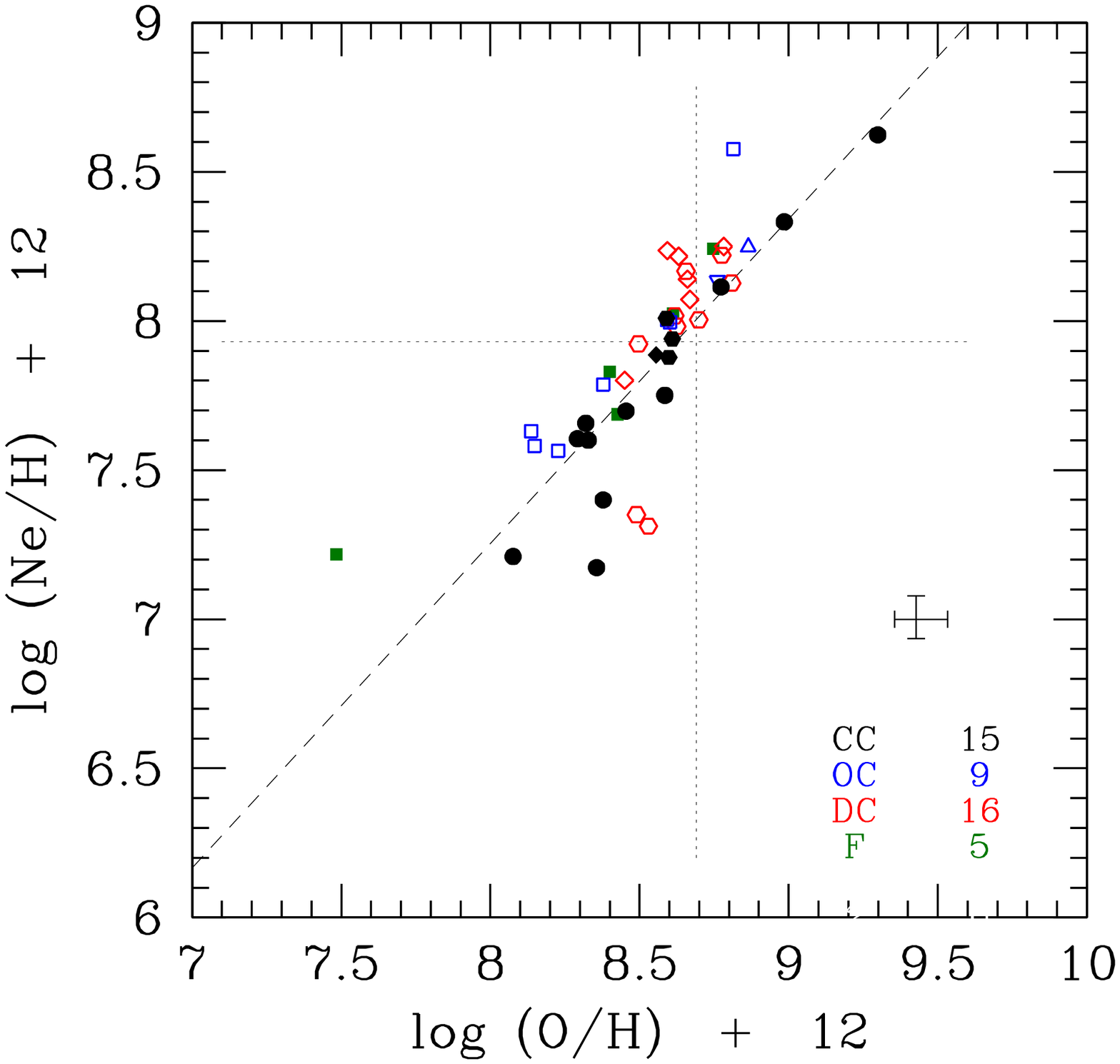}}
\resizebox{0.47\hsize}{!}{\includegraphics{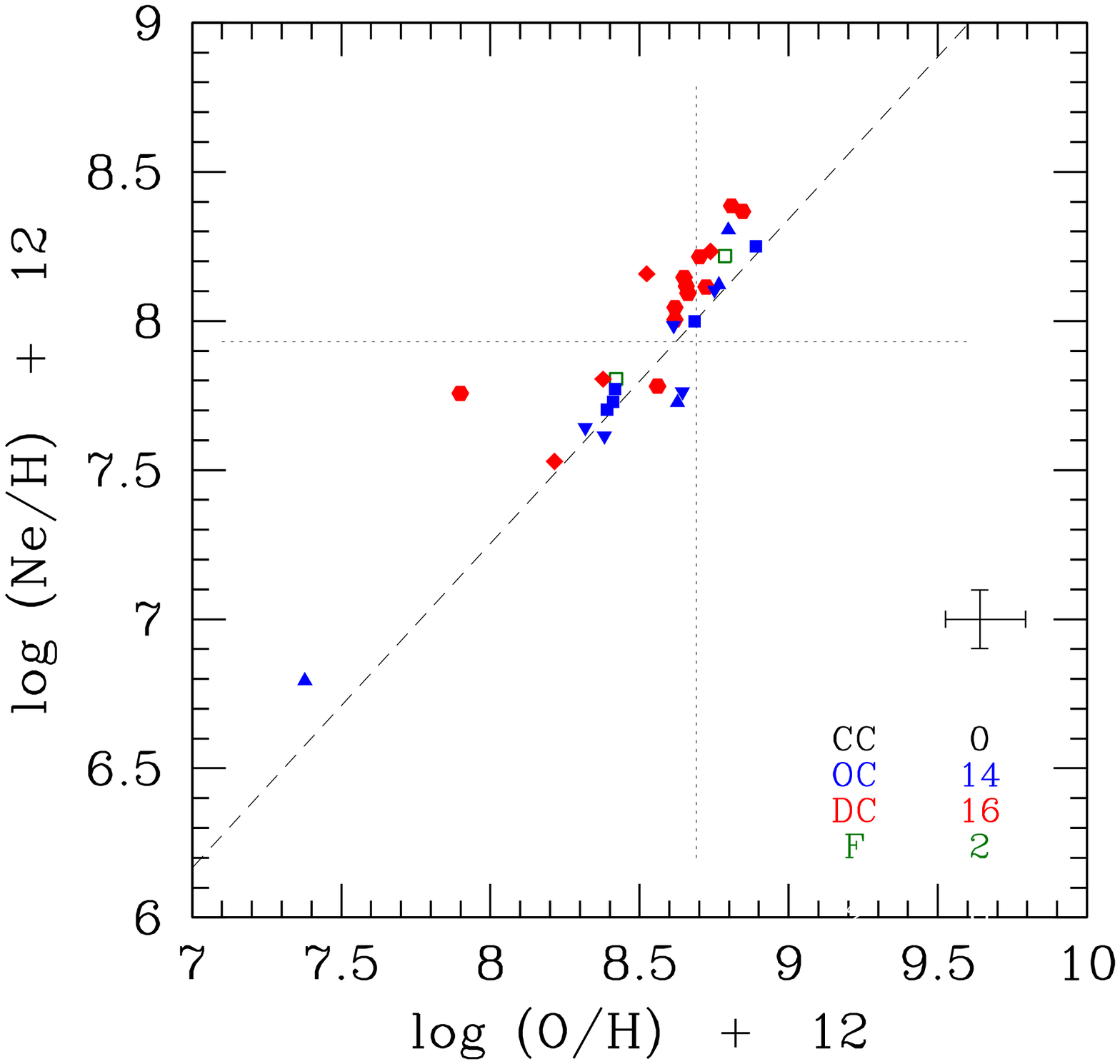}}
\caption[]{
 Diagrams of nebular abundance ratio O/H vs.\ Ne/H for Galactic disc PNe (left
 panel) and Galactic bulge PNe (right panel).  The same meaning of symbols as in
 Figs. 5 and 6. The horizontal/vertical lines mark the solar abundance values,
 while the dashed line marks the relation derived by \citet{Izotov06}.
}
\label{abund_neh_db}
\end{figure*}

\subsection{PNe nebular abundances vs.\ major {\it Spitzer} dust types}

In studying the derived nebular gas abundances vs.\ the major {\it Spitzer} dust
types, we found several abundance diagrams (e.g.\ N/H vs.\  He/H, N/O vs.\ O/H)
to be the most useful ones. Figures 5 and 6 display some of these useful
abundance diagrams for the compact Galactic disc PNe by \cite{Stanghellini2012}
and the bulge PNe by \cite{Stanghellini2012}, \cite{PereaCalderon2009}, and
\cite{Gutenkunst2008}. The several major {\it Spitzer} dust types CC, OC, DC,
and F (and their subtypes) are marked with different colours or symbols in these
plots. We also carried out Kolmogorov--Smirnov (K-S) and Wilcoxon (W)
statistical tests to confirm or discard possible differences in the chemical
abundances among the major {\it Spitzer} dust types. This is shown in Table 7
where we present the results of both statistical tests for bulge PNe (OC vs.\
DC) and disc PNe (OC vs.\ DC, OC vs.\ CC, and DC vs.\ CC). In addition, Table 8
lists the results of our K-S and W statistical tests for OC and DC PNe (disc
vs.\  bulge). It is to be noted here that we consider the differences in
abundances to be real when the probability that the two compared samples stem
from the same parent distribution is $\leq$1\% from both K-S and W statistical
tests. Thus, if both probabilities are low (0.01 or less), then the difference
is regarded as real, and they are marked with boldface characters in Tables 7
and 8. The median plasma parameters and abundance values for the Galactic disc
and bulge PNe are presented in Tables 9 and 10, respectively. The median values
(and in particular the percentiles) in Tables 9 and 10 make no sense when we
only have five or six objects. This is the case for the Galactic disc and bulge
F PNe, and their values in these tables should be taken with some caution. In
addition, as we see in Section 4.3, the median abundance patterns amongst the
several major {\it Spitzer} dust types are dominated by the specific dust
subtype, which are more numerous in each major dust group. Finally, we note that
in the following we adopt  the most recent \cite{Asplund2009} solar abundances
of He (10.93), N (7.83), O (8.69), Ne (7.93), S (7.12), Cl (5.50), and Ar
(6.40).\footnote{The solar abundances between brackets are given on the usual
scale $\log\epsilon$(X) = $\log$(X/H) $+$ 12.} 

Interestingly (although there are some clear outliers), Galactic disc DC PNe
are, on average, mainly located towards higher N and He abundances and N/O
abundance ratios, while just the opposite is seen for the OC objects (see Fig.\
5).  The DC disc PNe show median N and He abundances of $\log\epsilon$(N)=8.38
and $\log\epsilon$(He)=11.10, while OC disc PNe display lower median values of
$\log\epsilon$(N)=7.82 and $\log\epsilon$(He)=11.02; however, CC and F disc PNe
may populate the N/H vs.\  He/H and N/O vs.\  O/H diagrams more uniformly,
although most CC disc PNe display N and He abundances (with median values of
$\log\epsilon$(N)=7.76 and $\log\epsilon$(He)=11.03), similar to those of the OC
disc PNe. Similar behaviour is seen for the Galactic bulge PNe shown in Fig.~6,
with the exception that CC PNe are lacking towards this direction. The OC bulge
PNe display median values of $\log\epsilon$(N)=7.74 and
$\log\epsilon$(He)=11.04, while DC bulge PNe display higher median abundances of
$\log\epsilon$(N)=8.30 and $\log\epsilon$(He)=11.09. Our K-S and W statistical
tests shown in Table 7 confirm the differences in N/H, He/H, and N/O (also for
Ar/H, Ar/O, and S/O, see below) between the DC and OC PNe in both Galactic
environments. In addition, OC and DC PNe located in the Galactic disc and
compared to those pertaining to the Galactic bulge share the same abundance
pattern (see Tables 8--10). The only difference seems to be in the S/Ar
abundance ratio (see below).

Our preliminary interpretation is that Galactic disc DC PNe (with higher N and
He contents) probably have, on average, more massive and more metal-rich central
stars than the OC sources. At constant metallicity, high He/H and N/O ratios are
characteristic of proton-capture nucleosynthesis \citep{Stanghellini2006}; for
example, bipolar Type I PNe display high He/H and N/O ratios, indicating that
their progenitors are probably HBB AGB stars with initial masses $\sim$3$-$8
M$_{\odot}$ \citep{Peimbert1978, KingsburghBarlow1994, Stanghellini2006,
Karakas2009}. On the other hand, we also expect different metallicities through
our samples of Galactic PNe. Abundances of S, Cl, and Ar in PNe are usually
assumed to be unaltered by AGB nucleosynthesis, while the Ne content can be
slightly altered during the AGB \citep[][and references therein]{Karakas2009}. 
Under the assumption that S, Cl, and Ar are not altered by AGB evolution, we
find Ar to be most useful and reliable metallicity indicator.  This is because S
may be depleted into dust \citep[especially in CC PNe;][see also
below]{PottaschBernard-Salas2006} and because we have fewer objects with Cl
abundance determinations. The median Ar abundances are $\log\epsilon$(Ar)=6.56,
6.04, and 6.07 for the DC, OC, and CC disc PNe, respectively.  In Fig.\ 7 we
display Ar/H vs.\  O/H for the Galactic disc and bulge PNe depending on the
major {\it Spitzer} dust types.  Ar/H and O/H seem to be rather well correlated
(although with some scatter) in CC, OC, and F Galactic disc PNe (Fig.\  7, left
panel), while most of the DC Galactic disc PNe display the highest Ar abundances
(mostly supra-solar). The observed Ar-O correlation in CC and OC disc PNe may be
interpreted as due to the different metallicity (Ar it is expected to evolve in
lockstep with O) in disc PNe of rather low-mass central stars; O is expected to
be almost unaltered in low mass non-HBB stars \citep[e.g.][]{Karakas2009}.  The
F disc PNe with very little dust in their circumstellar shells are simply more
evolved objects, as suggested by their lower electron densities and higher
ionization (Table 9; see Section 5).

Thus, DC disc PNe (mostly with supra-solar metallicities) could be the
descendants of more massive HBB stars where N production takes place. Another
possibility could be that they evolve from supra-solar metallicity and less
massive stars (say $\sim$1.5$-$2.5 M$_{\odot}$) with some kind of extra mixing
where the central star may or may not be converted to C-rich depending on the
TDU episodes experienced. However, the detailed AGB evolution and
nucleosynthesis of such supra-solar metallicity stars remain to be theoretically
investigated. At present, we favour the high-mass interpretation because DC PNe
are mainly located towards the Galactic centre, as expected if they pertain to a
more massive and younger population. There are a few DC sources in Fig.\ 7 (left
panel) with supra-solar Ar abundances, together with slightly low O abundances. 
Strong HBB also may produce some O destruction.  Alternatively, some O depletion
could be possible if this element is being trapped in dust grains.  

Another interesting feature in Fig.\ 7 (left panel) is that most OC disc PNe
display subsolar Ar abundances (also an important fraction of CC disc PNe). This
implies that the \cite{Stanghellini2012} sample is dominated by relatively
low-mass and low-metallicity OC (and CC) PNe.  This is consistent with the high
rate ($>$60\%) of OC and CC PNe with unevolved/unprocessed circumstellar dust -
that is, OC$_{am}$ and CC$_{al}$ PNe (see Table 6) - found in the Galactic disc;
for instance, CC$_{al}$ PNe are very frequent in low-metallicity environments
dominated by low-mass stars, such as those of the Magellanic Clouds
\citep{Stanghellini2007}.  Higher central star mass OC PNe in the Galactic disc,
with their precursors being massive HBB AGB stars \citep[see e.g.][]{GH06,GH07},
evolve too fast to be detected in the optical
\citep[see][]{Garcia-Hernandez2007b}.  The only exception is the OC outlier (PNG
011.1$+$07.0) in the N/H vs.\  He/H diagram (Fig.\  5), which is the PN with the
highest N and He abundances in the disc sample.  PNG 011.1$+$07.0 is very
probably the unique very high-mass HBB PN in the \cite{Stanghellini2012}
sample\footnote{PNG 011.1$+$07.0 is the only OC$_{am}$ PN with the
amorphous silicate features in absorption, indicating an extremely thick
circumstellar envelope, and that the central star has just reappeared in the
optical (e.g. Garcia-Hernandez et al. 2007b).}.

The situation is less clear for the bulge PNe shown in Figure 7 (right panel). 
Again, OC bulge PNe seem to follow an Ar-O correlation (this time with an even
higher dispersion), and most DC bulge PNe are concentrated at solar and
supra-solar Ar abundances. DC PNe in the bulge show a very similar median Ar
abundance ($\log\epsilon$(Ar)=6.51) to the DC disc ones, while OC PNe show a
higher median Ar abundance ($\log\epsilon$(Ar)=6.29) than their Galactic disc
counterparts.\footnote{The higher median Ar content in OC bulge PNe is related
to the fact that OC$_{cr}$ PNe (with higher Ar abundances) are more numerous in
the bulge (see Section 4.3 for more details).}  This time, however, there are
more DC bulge sources (and a few OC$_{cr}$ ones) with somewhat low O content. 
Curiously, these DC bulge PNe with low O display even higher N abundances than
their disc counterparts; similar objects seem to be absent in the Galactic disc
sample (see Fig.  5). A higher N production at the expense of O could be an
indication of strong HBB (i.e., the activation of the O-N cycle) but this is
preferentially expected in lower metallicity environments, such as the
Magellanic Clouds (e.g. Chiappini et al.\  2009). The few (3--4) DC bulge PNe
with reliable Ar abundances, however, display high Ar abundances (nearly solar),
and the apparent N overproduction (and higher N/O ratios) could be due to
stronger HBB in higher mass stars.

Figures 8 and 9 display S/H vs.\ O/H and Ne/H vs.\ O/H in our Galactic PNe
samples, respectively. As for Fig.\ 7, we plot the Galactic disc and bulge PNe
in the left-hand and right-hand panels, respectively, and the major {\it
Spitzer} dust types (and their subtypes) are marked with different symbols and
colours. Figure 8 (left panel) shows that there seems to be some correlation
between S and O for disc PNe, as expected. The scatter for each major {\it
Spitzer} dust type is similar to what is found for Ar, but there seems to be a
different S-O relation among each dust type. In general, we find subsolar S
abundances for all major {\it Spitzer} dust types, with the CC disc PNe being
the most S-depleted objects. The median S abundances are $\log\epsilon$(S)=7.02,
6.66 and 6.47 for the DC, OC, and CC disc PNe, respectively. PNe are known to
display subsolar S abundances (e.g.\ Pottasch \& Bernard-Salas 2006) although
the importance of this finding is still unclear (see below). The most
interesting facts from Fig.\ 8 (left panel) are that i) DC disc PNe are again
clearly located towards the highest S abundances; and ii) CC disc PNe are, on
average, more depleted in S than the OC and DC disc PNe. As mentioned above, the
highest degree of S depletion in CC disc PNe may be understood if S is depleted
into dust in these objects (e.g.\ Pottasch \& Bernard-Salas 2006). We note,
however, that the S anomaly in PNe is still an open question. For example,
\citet{Henry2012} claim that a link with dust depletion of S can be ruled out
and suggest that the S anomaly in PNe is merely a problem with the ICFs. This S
depletion problem is significantly reduced when S abundances are determined
directly from IR measurements. Our S results would tend to support the depletion
hypothesis, but this should be evaluated more carefully in the future (e.g.\ by
using the S lines in the {\it Spitzer} spectra). As in the case of Ar, the S/H
vs.\ O/H diagram for bulge PNe is more complex, but again we find evidence of a
different S-O relation among each major {\it Spitzer} dust type (Fig.\ 8, right
panel). Both OC and DC bulge PNe populate this abundance diagram uniformly,
although subsolar S abundances are also generally found. The median S abundances
are $\log\epsilon$(S)=6.87 and 6.64 for the DC and OC bulge PNe, respectively. A
few (3--4) DC PNe in the bulge are located at low S and O abundances. It is
intriguing that these sources pertain to the group of DC bulge PNe with
relatively high Ar abundances and showing the apparent signature of the O-N
cycle. 

Taking the abundance errors (typically $\pm$0.1-0.15 dex) and our previous
results on Ar and S into account, we find a surprisingly tight correlation
between Ne and O for Galactic PNe in both the disc and bulge (Figure 9).  Ne
could be slightly modified by AGB nucleosynthesis when $^{22}$Ne is
overproduced.  An increase in $^{22}$Ne is theoretically predicted to occur in a
rather narrow mass range of $\sim$2.5--3.5 M$_{\odot}$ or in higher mass stars
as a consequence of some O destruction by strong HBB and slight Ne enrichments
via dredge-up; although the latter effect seems to be more important at low
metallicity \citep[][]{KarakasLattanzio2003,Karakas2009}.  In addition, in very
low mass AGB models, the partial mixed zone (PMZ) can produce $^{22}$Ne, further
increasing the Ne/O ratio \citep[e.g.][see also next section for more
details]{Karakas2009}. The tight Ne-O correlations displayed in Fig.\ 9 seem to
suggest that, in most cases, Ne is not strongly altered by AGB evolution and
that it closely follows the O content. Indeed, this would agree with
the remarkably constant Ne/O ratio observed in PNe with very different chemical
compositions \citep[see e.g.][]{Stasinska1998}. As we see in the next
section, however, the median Ne/O values observed in our Galactic PNe are
slightly higher than the AGB nucleosynthesis predictions with no PMZ, and they
seem to require including the PMZ in the theoretical models.

Also, some outliers may be identified in the Ne/H vs.\ O/H diagram for Galactic
disc PNe shown in Figure 9 (left panel). For example, a clear outlier in Fig.\ 9
(left panel) is again PNG 011.1$+$07.0 (probably a high-mass HBB PN), which
shows a huge Ne overabundance (or Ne/O ratio) and which clearly deviates from
the observed Ne-O trend. Also, some DC disc PNe display some Ne enhancements,
although more modest than that in PNG 011.1$+$07.0. The Ne abundances in the
latter DC disc PNe are higher than those in other objects at similar O/H and
than those observed in other types of PNe (e.g.\ most of the OC and CC disc
PNe). The median Ne abundances are $\log\epsilon$(Ne)=8.10, 8.00, and 7.75 for
the DC, OC, and CC disc PNe, respectively. The apparent Ne enhancements in some
DC disc PNe outliers could indicate that their central stars are more massive
than most OC and CC disc PNe but less massive than the high-mass HBB PN PNG
011.1$+$07.0. We note also that there are a few DC and CC disc PNe outliers that
show unusually low Ne contents. On the other hand, the Ne/H vs.\ O/H diagram for
bulge PNe (Fig.\ 9, right panel) shows an even tighter Ne-O correlation, where
less clear outliers are identified. A clear outlier is PNG 358.6$+$01.8, which
shows a strong Ne overabundance that deviates from the Ne-O relation.
Remarkably, PNG 358.6$+$01.8 is the lowest metallicity ($\log\epsilon$(Ar)=6.0)
object among the small group of DC bulge PNe showing the apparent signature of
the O-N cycle and with all abundances available. Its abundance pattern suggests
the activation of the O-N cycle by strong HBB in a low metallicity and
relatively massive progenitor. DC PNe in the bulge display similar median Ne
abundances ($\log\epsilon$(Ne)=8.12), while OC PNe show a median Ne abundance
($\log\epsilon$(Ne)=7.7) that is lower than their counterparts in the Galactic
disc. 

In summary, we conclude that, on average, DC disc PNe probably evolve from AGB
progenitors that are more massive and metal rich than most of the OC and CC disc
PNe, as already suggested by their higher N and He abundances. The same
conclusion seems to be applicable to the sample of PNe in the bulge, with the
important peculiarity of the complete lack of CC PNe in this Galactic
environment (see Section 6).

\subsection{Major {\it Spitzer} dust types vs.\ AGB nucleosynthesis predictions}

For comparison with the chemical abundances observed in our sample of PNe
with {\it Spitzer} spectra, in Table 11 we present the nucleosynthesis
theoretical predictions for He, N/O, Ne/O, and S/O in AGB stars of different
progenitor masses (from 1 to 6 M$_{\odot}$) and metallicities (z=0.02,
0.008, and 0.004) by \cite{Karakas2010}.\footnote{\cite{Karakas2010} does not
give predictions for Ar and Cl but these elements are expected to be
unaltered by AGB evolution (e.g.\ Karakas et al.\  2009, and references
therein).}  These models assume the \cite{VassiliadisWood1993} mass loss on
the AGB and do not include either convective overshooting or a partially mixed
zone (PMZ), which is required for including a $^{13}$C pocket that
produces neutron capture elements \citep[see e.g.][]{Lugaro2012,
Kamath2012}.  The only exception are the 3 M$_{\odot}$ models, for which
Karakas (2010) also give the nucleosynthesis predictions with PMZ.  The
predictions of the latter 3 M$_{\odot}$ models with PMZ are also given in
Table 11.  We note that the Karakas (2010) AGB nucleosynthesis predictions
are calculated with \cite{AndersGrevesse1989} solar abundances, while here
we assume the most recent solar abundances by Asplund et al.\   (2009) (see
below).  The He abundance and the abundance ratios given in Table 11
(Karakas 2010) are consistent with similar models calculated with Asplund et
al.\  (2009) solar abundances (A. Karakas 2013, private communication).  The
comparison of the median chemical abundances observed in our sample of
Galactic disc PNe (Table 9) with the theoretical predictions for AGB stars
listed in Table 11 seems to support our preliminary interpretation above.

The DC disc PNe display median O (8.66), Ar (6.56), Ne (8.10), and S (7.02)
abundances that compare reasonably well (within the errors) with the
corresponding O, Ar, Ne, and S solar abundances of 8.69, 6.40, 7.93, and 7.12,
respectively (Asplund et al.\  2009).  The median He abundance (11.10) and N/O
(-0.27) abundance ratios observed in DC disc PNe agree very well (especially the
N/O ratio, which is much more sensitive to AGB nucleosynthesis) with the
theoretical predictions for a solar metallicity (z$\sim$0.02) $\sim$5
M$_{\odot}$ AGB star (see Table 11). According to Karakas (2010) models, such a
star is with in the limit for HBB occurrence. A relatively weak HBB activation would
be consistent with the nearly solar O abundances observed in DC disc PNe (there
is no important O destruction by too strong an HBB). The S/O and Ne/O ratios
are less sensitive to AGB evolution (see Table 11), as expected.  The median S/O
(-1.62) in DC disc PNe is consistent with the AGB predictions (Table 11),
suggesting a low degree of S depletion in these objects. The median Ne/O
(-0.57), however, is somewhat higher than AGB predictions in Table 11.  

This apparent Ne overproduction in DC disc PNe seems to be real and the median
Ne/O in DC disc PNe is further increased by the few sources with some Ne
enhancements mentioned above (see Figure 9). A higher mass limit for the DC
disc PNe is put by the chemical abundances observed in PNG 011.1$+$07.0, the
only very high-mass HBB PN in the disc sample. The high He (11.25) and N/O
(0.09) in this source are consistent with a progenitor mass of 6 M$_{\odot}$
(see Table 11). The S/O (-1.57) ratio also agrees with the 6 M$_{\odot}$ model
predictions, but the observed Ne/O (-0.24) is about 0.5 dex higher than these
predictions (Table 11). Higher mass (7$-$8 M$_{\odot}$) nucleosynthesis models
predict slightly higher He and N/O (and S/O) (A.\,Karakas 2013; private
communication) but still consistent (within the errors) with the abundances
observed in PNG 011.1$+$07.0. The extremely high Ne/O in PNG 011.1$+$07.0 is
intriguing. At z=0.02, Ne/O increases only slightly with increasing stellar
mass. Interestingly, the Ne/O ratio can increase significantly (e.g.\ up to
0.2--0.3 dex in a 5 M$_{\odot}$ star) in massive AGB models with delayed
superwinds \citep[which also includes a small PMZ of 1 $\times$ 10$^{-4}$
M$_{\odot}$;][]{Karakas2012} when compared to the Karakas (2010) models. The
study of the Ne/O ratio in massive HBB AGB stars with delayed superwinds should
be extended to higher masses, PMZ sizes, and supra-solar metallicities in the
future.  Apart from the unusually high Ne/O ratio, we conclude that PNG
011.1$+$07.0 should have evolved from an AGB star of at least 6 M$_{\odot}$.

The OC disc PNe show median O (8.57), Ar (6.04), Ne (8.00), and S (6.66)
abundances that correspond to a subsolar metallicity of [M/H]$\sim$-0.4 (or
z$\sim$0.008). The median O (8.57) abundance seems to only be a little depleted
only, but the median Ne (8.00) content is nearly solar. The median N/O (-0.81)
(and He within the errors) in OC disc PNe is consistent with the theoretical
predictions for a z=0.008 $\sim$1 M$_{\odot}$ AGB star (see Table 11). However,
the median Ne/O (-0.60) is similar to that in DC disc PNe, being 0.24 dex higher
than the 1 M$_{\odot}$ model predictions in Table 11. In very low-mass AGB
models, the PMZ can produce $^{22}$Ne, further increasing the Ne/O ratio (e.g.\
Karakas et al.\ 2009). The PMZ effect on the Ne/O ratio can be seen in Table 11,
where the 3 M$_{\odot}$ model predictions with and without PMZ are
compared.\footnote{Unfortunately, Karakas (2010) does not give the
nucleosynthesis predictions for AGB models with PMZ for masses different that 3
M$_{\odot}$.} It can be seen that Ne/O can be further enhanced by $\sim$0.2 dex
(with He, N/O, and S/O less altered) in the low-metallicity models with PMZ. The
Ne/O ratio increases with increasing size of the PMZ; e.g.\ metal-poor (z=0.01)
1.25 M$_{\odot}$ models with PMZ show that Ne/O can vary from $\sim$-0.7 to -0.4
dex for PMZ sizes between 1 $\times$ 10$^{-3}$ and 8 $\times$ 10$^{-3}$
M$_{\odot}$ (A. I. Karakas 2013, private communication). Thus, the relatively
high Ne/O abundances in OC disc PNe may be reflecting the effect of PMZ during
the previous AGB phase. Indeed, our chemical abundances in OC disc PNe could
perhaps be used to put observational constraints on the size of the PMZ in the
very low-mass AGB models.   

Furthermore, CC disc PNe show median O (8.52) and Ar (6.07) abundances similar
to those in OC disc PNe. The median Ar abundance also indicates a subsolar
metallicity ([M/H]$\sim$-0.3$-$-0.4 or z$\sim$0.008) in CC disc PNe. However,
both median abundances of Ne (7.75) and S (6.47) in CC disc PNe are lower than
those in the OC ones by about 0.2$-$0.3 dex. The median N/O (-0.70) is higher
than the corresponding value in OC sources, which could suggest higher
progenitor masses in CC disc PNe (the predicted N/O ratios increase with
increasing stellar mass; see Table 11), but this possible difference is not
confirmed by our statistical tests (Table 7). At z=0.008, only models with
progenitor masses 1.9 $\leq$ M $<$ 3 M$_{\odot}$ become C-rich. This simple fact
already suggests that CC disc PNe should, on average, have higher central star
masses than the OC disc PNe. The best fit to the observed median N/O and Ne/O
ratios in CC PNe is given by the 1.9 M$_{\odot}$ model. 

Comparison of AGB theoretical predictions with the median chemical abundances
observed in our sample of Galactic bulge PNe (where CC PNe are absent) turns out
to be very simple. The DC PNe in the bulge and disc share the same median
abundance pattern, which is consistent with the theoretical predictions for a
solar-metallicity (z$\sim$0.02) $\sim$5 M$_{\odot}$ AGB star (see above). The OC
PNe in the bulge display median chemical abundances that are almost identical to
their Galactic disc counterparts. The only difference (although not confirmed by
our statistical tests) is that OC bulge PNe seem to show a higher median Ar
abundance. We will see in the following sections that this apparently higher Ar
abundance is due to the fact that OC PNe with crystalline silicate dust features
(OC$_{cr}$), being more numerous in the bulge, are more metal rich than the
other types of OC PNe with amorphous silicates (OC$_{am}$ and OC$_{am+cr}$). We
anticipate that the median abundances of OC bulge PNe with amorphous silicates
are identical to those of the bulk of OC PNe in the Galactic
disc,\footnote{OC$_{cr}$ PNe are less frequent in the Galactic disc and have a
negligible impact on the median abundances derived in OC disc PNe.} in agreement
with the model predictions for a z=0.008 $\sim$1 M$_{\odot}$ AGB star. The
median abundances of the OC$_{cr}$ subtype are more consistent with higher
metallicity (nearly solar) and slightly more massive AGB progenitors (see
Section 4.4).

Finally, it is to be noted here that the lower mass limits to produce C-rich
(C/O $>$ 1) AGB stars is a longstanding issue. This problem is related with with
how convection is theoretically treated in stellar interiors. Including
convective overshooting \citep[see e.g.][]{Herwig2000} is a way to increase the
efficiency of the third dredge-up (TDU) and to reduce the lower mass limit for
C-star formation. The mass limit for the HBB ocurrence is also another parameter
that strongly depends on the theoretical details \citep[convection
treatment, mass loss prescription, etc.; see e.g.][and references
therein]{Garcia-Hernandez2013}. Thus, we would like to point out that the exact
progenitor masses of Galactic PNe with {\it Spitzer} spectra that we infer from
our comparison with the Karakas (2010) AGB nucleosynthesis models may be rather
uncertain.  However, our main finding of different average progenitor masses and
metallicities amongst the several major {\it Spitzer} dust types is assured.

\subsection{PNe nebular abundances versus {\it Spitzer} dust subtypes}

Analysing the derived nebular gas abundances vs.\ the {\it Spitzer} dust
subtypes is more difficult and uncertain due to the low number of objects in
each dust subtype. Taking the results discussed in Section 4.1 into account,
where we found that OC and DC PNe in the Galactic disc and bulge display very
similar abundance patterns (Table 8), we decided to merge both galactic PNe
samples with the intention of obtaining a higher statistical significance of any
potential result. The results of our K-S and W statistical tests for the {\it
Spitzer} dust subtypes DC (DC$_{cr}$ vs.\ DC$_{am+cr}$), CC (CC$_{ar}$ vs.\
CC$_{al}$), and OC (OC$_{cr}$ vs.\ OC$_{am}$, OC$_{cr}$ vs. OC$_{am+cr}$, and 
OC$_{am}$ vs.\ OC$_{am+cr}$) in disc and bulge PNe are shown in Table 12.  We
still have somewhat low numbers ($\sim$10) of OC$_{cr}$ and OC$_{am+cr}$ PNe and
an even lower number ($\sim$6) of CC$_{ar}$ PNe, and the results for these
{\it Spitzer} dust subtypes (especially for CC$_{ar}$ objects) should be taken
with caution. A higher number of sources ($>$10), however, is present among the
two DC subtypes and the OC$_{am}$ and CC$_{al}$ PNe. Tables 13 and 14 list the
median plasma parameters and abundances for all Galactic PNe (disc and bulge)
for the OC subtypes and DC (and CC) subtypes, respectively.

Remarkably, O-rich PNe with crystalline silicate dust features (OC$_{cr}$)
display median N and He abundances of $\log\epsilon$(N)=8.01 and
$\log\epsilon$(He)=11.06, which are slightly higher than those in the OC$_{am}$
and OC$_{am+cr}$ subtypes with amorphous silicate dust features. OC$_{am}$ and
OC$_{am+cr}$ PNe show very similar median N and He abundances;
$\log\epsilon$(N)=7.78 and $\log\epsilon$(He)=11.02 are observed in OC$_{am}$
PNe while these values are 7.75 and 11.03 in the OC$_{am+cr}$ sources. The
median N/O ratio in OC$_{cr}$ PNe (-0.65) is also slightly higher than those in
the OC$_{am}$ (-0.80) and OC$_{am+cr}$ (-0.88) objects. Unfortunately, our K-S
and W statistical tests do not confirm the possible differences in the He/H, N/H
and N/O between the different OC subtypes (Table 12). In addition, OC$_{cr}$ PNe
display a high (nearly solar) median Ar abundance ($\log\epsilon$(Ar)=6.44),
which contrasts with the subsolar Ar abundances of $\log\epsilon$(Ar)=6.03 and
6.13 obtained in OC$_{am}$ and OC$_{am+cr}$ PNe, respectively. 

This difference in the Ar/H abundance seems to be real and it is confirmed by
the K-S and W statistical tests in Table 12. We note that two OC$_{cr}$ PNe (PNG
007$+$03.2 and PNG 357.6$+$01.7, both in the bulge) display an abundance pattern
(with high He, N, Ar, and N/O; see Table~3) almost identical to the one shown by
the DC PNe. Their classification as OC$_{cr}$ PNe is based on the lack of
PAH-like dust features in the $\sim$5$-$10 $\mu$m region. The S/N in the
$\sim$5$-$10 $\mu$m region of their {\it Spitzer} spectra is very low and both
sources only display a very weak PAH-like emission feature at 11.3 $\mu$m. Thus,
PNG 007$+$03.2 and PNG 357.6$+$01.7 should be truly DC PNe, where the weak
PAH-like features have probably escaped detection from {\it Spitzer}. The rest
of the OC$_{cr}$ PNe dominate the median abundances mentioned above (see also
Table 13). Thus, our chemical abundances confirm that OC$_{cr}$ PNe are more
metal-rich than the other OC subtypes with amorphous silicates (see also the
next section). Furthermore, the O-rich PNe with amorphous (OC$_{am}$ and
OC$_{am+cr}$) and crystalline (OC$_{cr}$) silicate dust features are more
numerous in the disc and bulge, respectively, dominating the observed
distributions of chemical abundances (median values) in the major OC-type PNe
and studied in the previous sections (see also Section 6).   

With only crystalline silicate dust features (DC$_{cr}$), the DC PNe display a
median abundance pattern (He, N, O, Ne, Ar, and S; see Table 14) almost
identical (differences $\leq$0.15 dex) to those PNe also showing amorphous
silicates (DC$_{am+cr}$). The K-S and W statistical tests (Table 12) indicate
that there are no statistically significant differences between both DC
subtypes. There is a tentative hint that DC$_{cr}$ PNe could be slightly more
metal-rich (i.e.\ higher Ar abundances) and more N-rich than their DC$_{am+cr}$
counterparts, as suggested by their slightly higher Ar and N median abundances
($\log\epsilon$(N)=8.39 and $\log\epsilon$(Ar)=6.56). Indeed, we find relatively
low statistical probabilities for these two elements (Table 12) but not low
enough to consider these possible differences to be real. Thus, we conclude
that, on average, both DC subtypes are chemically indistinguishable and that
they probably evolve from the same AGB progenitors (see also Section 4.2). 

Regarding the CC PNe, we note that there are few objects amongst the different
CC dust subtypes (CC$_{ar}$, CC$_{al}$, and CC$_{ar+al}$). In particular, we
only have data for eight CC$_{ar}$ PNe and three CC$_{ar+al}$ PNe (with the
simultaneous presence of aromatic and aliphatic dust features).  Curiously, two
(PNG 006.1$+$08.3 and PNG 345.2-08.8) of these CC$_{ar+al}$ sources are PNe
where the complex fullerene (e.g.\ C$_{60}$ and C$_{70}$) molecules have
recently been found \citep[e.g.][]{Garcia-Hernandez2010, Garcia-Hernandez2012b}.
The third object (PNG 041.8$+$04.4) is an infrared spectroscopic twin of PNG
006.1$+$08.3 (PN M 1-20), but the fullerene features are not detected in its
{\it Spitzer} spectrum. Thus, we do not list the three CC$_{ar+al}$ objects
mentioned above\footnote{The chemical composition of the three CC$_{ar+al}$ in
the sample is very similar to the bulk of CC$_{al}$ PNe (Table 14).} in Table
14, and they have not been considered in our K-S and W statistical test
displayed in Table 12. 

More interesting is that C-rich PNe with aliphatic dust features (CC$_{al}$)
display median N ($\log\epsilon$(N)=7.66) and He ($\log\epsilon$(He)=11.03)
abundances lower than those in the CC$_{ar}$ subtype with aromatic (PAH-like)
dust features; $\log\epsilon$(N)=8.20 and $\log\epsilon$(He)=11.12 are observed
in CC$_{ar}$ PNe. The median N/O ratio of -0.71 in CC$_{al}$ PNe is also
considerably lower than in the CC$_{ar}$ (-0.38) sources.  Also, CC$_{al}$ PNe
show a median subsolar Ar abundance ($\log\epsilon$(Ar)=5.97), while CC$_{ar}$
PNe show a nearly solar Ar content ($\log\epsilon$(Ar)=6.44). The median S
abundances are $\log\epsilon$(S)=6.71 and 6.35 for the CC$_{ar}$ and CC$_{al}$
PNe, respectively. However, taking the metal-poor character of CC$_{al}$ PNe
into account, the S depletion (of $\sim$0.4 dex) is similar in both types of CC
PNe. 

Our K-S and W tests give relatively low statistical probabilities ($\leq$8\%)
for all elements (He, N, O, Ar, and S) available in the CC$_{ar}$ sources (Table
12). However, the number of CC$_{ar}$ PNe is still too low, preventing us from
reaching a firm statistical confirmation of these differences in the median
chemical abundances for both subtypes of CC PNe. We believe that these
differences (at least to some extent) may be real. Indeed, at least half (four)
of the CC$_{ar}$ objects (i.e.\ PNG 010.6+03.2, PNG 309.5-02.9, PNG 336.3-05.6,
and PNG 355.7-03.0) show an abundance pattern (with high He, N, Ar, and N/O; see
Table~3) identical to the one shown by the DC PNe. This is also evident in
Figures 5 and 6, where it can be seen that several CC$_{ar}$ PNe are located in
the same regions as occupied by the DC PNe. Their classification as CC$_{ar}$
relies on the apparent lack of clear crystalline silicate features in the
$\sim$20$-$38 $\mu$m spectral region.  As pointed out by
\cite{Stanghellini2012}, some dust classifications may be uncertain. Indeed, the
shape of the dust continuum emission seen in their {\it Spitzer} spectra is not
very different to the one in DC PNe; e.g., PNG 010.6+03.2 even seems to show
some tentative crystalline silicate features in its {\it Spitzer} spectrum
\citep[see Fig.~4 in][]{Stanghellini2012}. Thus, the last four sources should be
true DC PNe where the weak crystalline silicate features were possibly not
detected by {\it Spitzer}.  By inspecting the {\it Spitzer} spectra of the
remaining (four) CC$_{ar}$ PNe, we also find some doubtful classifications. In
two sources (PNG 297.4+03.7 and PNG 344.8+03.4), the aromatic (PAH-like) dust
features are very weak, while the other two sources (PNG 014.3-05.5 and PNG
285.4+01.5) show somewhat unusual infrared spectra with the presence of a strong
bump emission around $\sim$24 $\mu$m \citep[see fig.~4 in][]{Stanghellini2012}. 
In short, our chemical abundances confirm that CC$_{al}$ PNe are metal-poor
objects and that they dominate the median chemical abundances of the CC-type PNe
in our PNe samples (Sections 4.1 and 4.2).

\subsection{{\it Spitzer} dust subtypes vs.\ AGB nucleosynthesis predictions}

Comparison of the chemical abundances among the different {\it Spitzer} dust
subtypes with the theoretical nucleosynthesis predictions (see also Section 4.1)
turns out to be straightforward. This is because the median abundances in the
major {\it Spitzer} CC and OC dust types are dominated by specific dust subtypes
(the more numerous ones) while these values in DC PNe are representative of the
two DC subtypes. Thus, the comparison done in Section 4.1 for the major {\it
Spitzer} dust types holds here for the particular dust subtypes dominating the
abundance distributions among the major dust classes. 

In particular, DC PNe with both crystalline and amorphous silicate dust features
(DC$_{cr}$ and DC$_{am+cr}$) are high-metallicity related objects, and their
median He abundances and N/O abundance ratios agree very well with the
theoretical predictions for the solar metallicity (z$\sim$0.02) $\sim$5
M$_{\odot}$ AGB star shown in Table 11. This contrasts with Gorny et al.\  
(2010), who found unusual chemical compositions of the nebular gas in the few
DC$_{am+cr}$ PNe in the Galactic bulge they analysed and linked this fact to
their dust subtype. We have shown here that when using larger and homogeneous
PNe samples \citep[i.e.\ the][sample]{Stanghellini2012}, any possible difference
in the chemical composition of both DC subtypes disappear. Still, a difference
between bulge and disc DC PNe remains and there are a few DC bulge PNe (and a
few OC$_{cr}$ ones) with somewhat low O content and the highest N/O ratios that
are not present in the disc sample. This O-N anticorrelation could indicate even
higher progenitor masses and a stronger HBB (i.e.\ the activation of the O-N
cycle) in these peculiar DC PNe in the bulge. As indicated by the derived
nebular gas abundances, a few OC$_{cr}$ PNe and some CC$_{ar}$ PNe in our sample
may be truly DC PNe where the PAH-like and the crystalline silicate features,
respectively, may have escaped detection by {\it Spitzer} (see Section 4.3 for
more details).

On the other hand, OC PNe with amorphous silicate dust features (OC$_{am}$ and
OC$_{am+cr}$) seem to evolve from the same low metallicity progenitors.  Their
median He abundances and N/O abundance ratios are consistent with predictions
for a low metallicity (z$\sim$0.008) $\sim$1 M$_{\odot}$ AGB star (Table 11). 
More interesting is that the bulk of OC$_{cr}$ PNe are of higher metallicity
(nearly solar) than the other OC subtypes with amorphous silicates. Their
median He and N/O of $\log\epsilon$(He)=11.06 and -0.65 suggest solar
metallicity (z$\sim$0.02) $\sim$1-1.5 M$_{\odot}$ AGB stars (see Table 11) as
likely progenitors. The observed median Ne/O ratio of -0.63 is also higher than
the model predictions. As discussed in Section 4.1, this disagreement
may be solved when using AGB nucleosynthesis models with PMZ. Furthermore, the
unusual chemical compositions (e.g.\ very low abundances of N) found by Gorny et
al.\  (2010) in a few OC$_{am+cr}$ PNe in the Galactic bulge do not seem so
pronounced and usual in the more homogeneous sample of Galactic disc PNe (see
Fig.\  6). The suggestion that they are rather peculiar and rare objects that
may originate in binary systems (see Gorny et al.\  2010 for more details) has
not been confirmed by photometric observations (Hajduk et al.\  2014, in press;
Hajduk, private communication).

Similar to the OC PNe, we find that the CC PNe with aliphatic dust features 
(CC$_{al}$ and CC$_{ar+al}$, which are mostly located in the Galactic disc)
probably evolve from low-metallicity and relatively low-mass progenitors. They
completely dominate the chemical pattern observed in the major CC type, and the
conclusions reached in Section 4.1 for the CC PNe hold here. Although the
observed median He abundances and N/O abundance ratios in CC PNe with aliphatic
dust are not completely reproduced by the theoretical predictions, they should
evolve from low-metallicity (z$\sim$0.008) and relatively low-mass (1.9 $\leq$ M
$<$ 3 M$_{\odot}$) AGB stars. Finally, as mentioned above, a large number of the
rather rare (and unusual) CC PNe with aromatic dust (CC$_{ar}$) show He
abundances and N/O abundance ratios consistent with predictions for a solar
metallicity (z$\sim$0.02) $\sim$5 M$_{\odot}$ AGB star. 

\section{Nebular properties and evolutionary status}

\begin{figure*}
\resizebox{0.47\hsize}{!}{\includegraphics{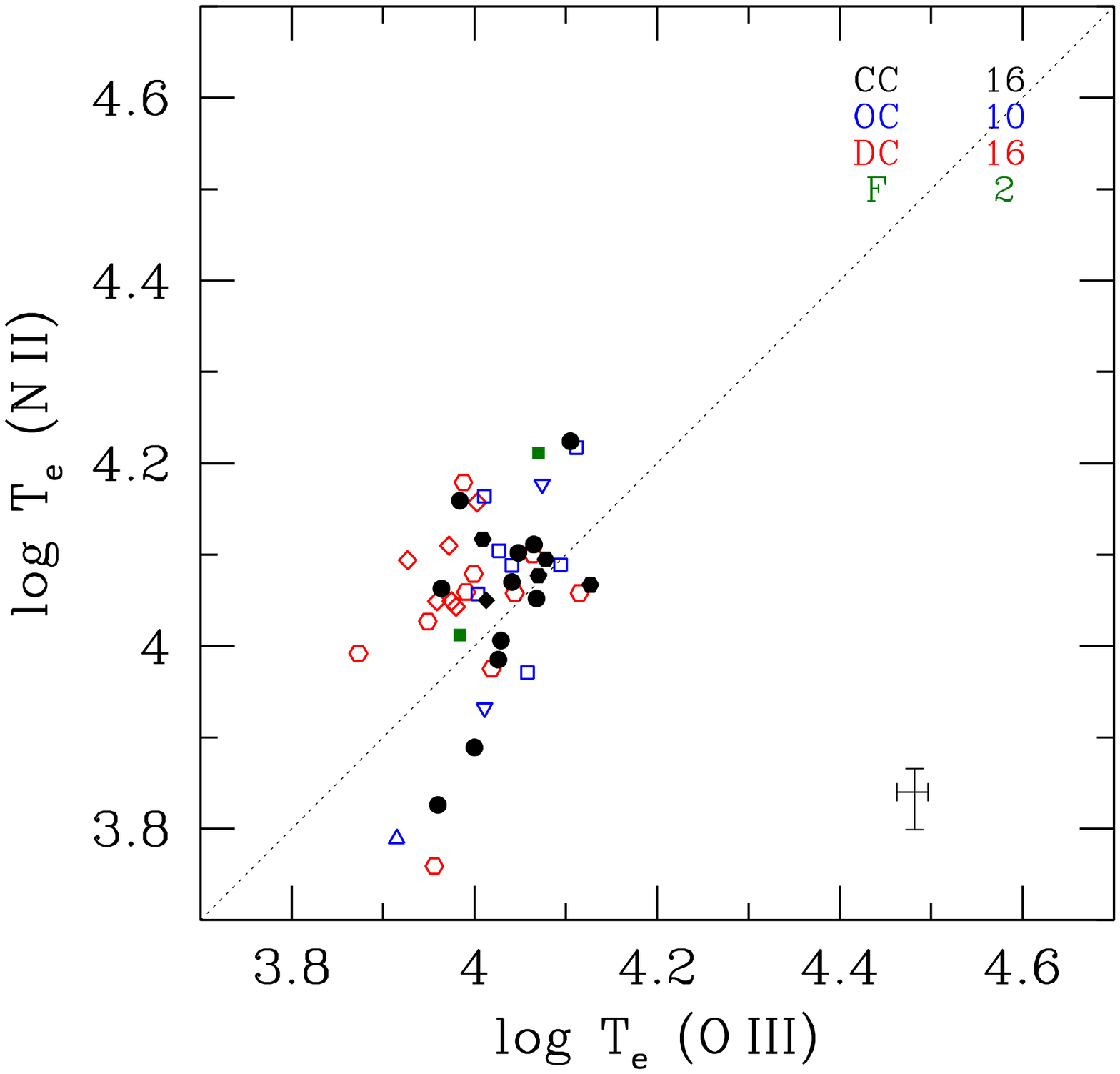}}
\resizebox{0.47\hsize}{!}{\includegraphics{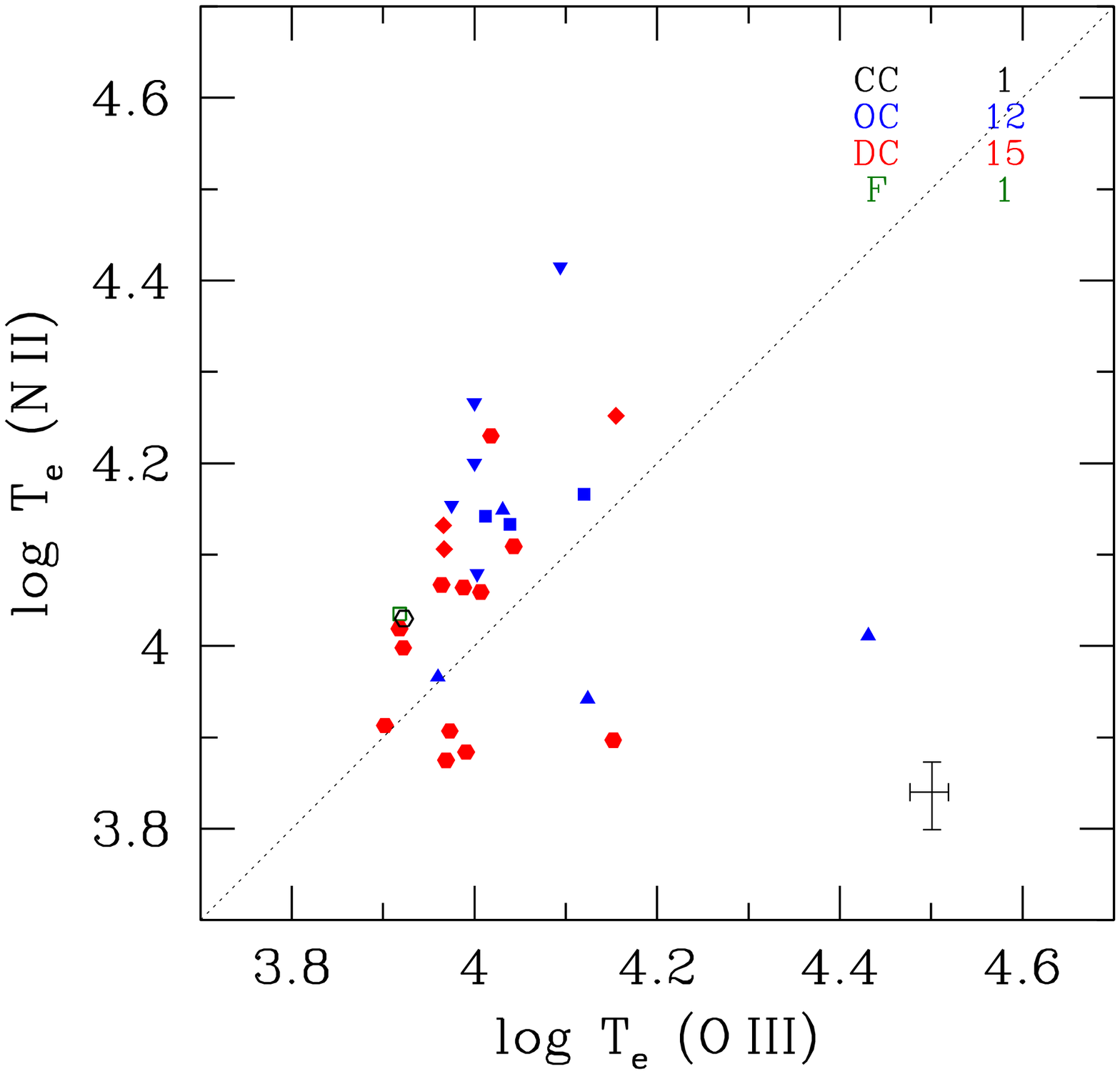}}
\caption[]{
 Diagrams of nebular electron temperature T$_e$ from the  
 [O~{\sc iii}] $\lambda$4363/5007  vs.\  T$_e$ from the 
 [N~{\sc ii}] $\lambda$5755/6584 ratios for Galactic disc
 (left panel) and Galactic bulge PNe (right panel).  
 The same meaning of symbols as in Figs. 5 and 6.
}
\label{to_tn}
\end{figure*}

\begin{figure*}
\resizebox{0.49\hsize}{!}{\includegraphics{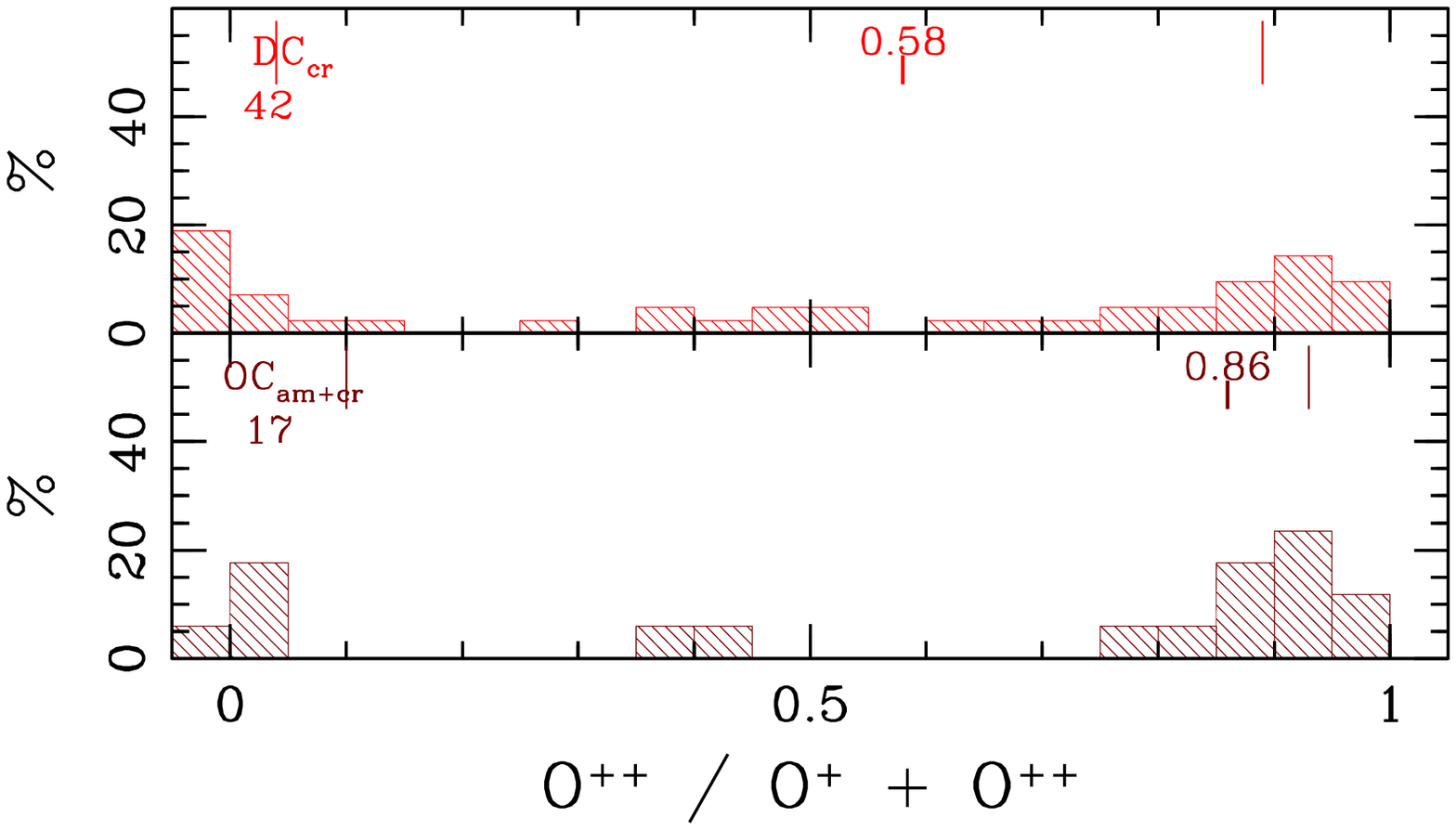}}
\resizebox{0.49\hsize}{!}{\includegraphics{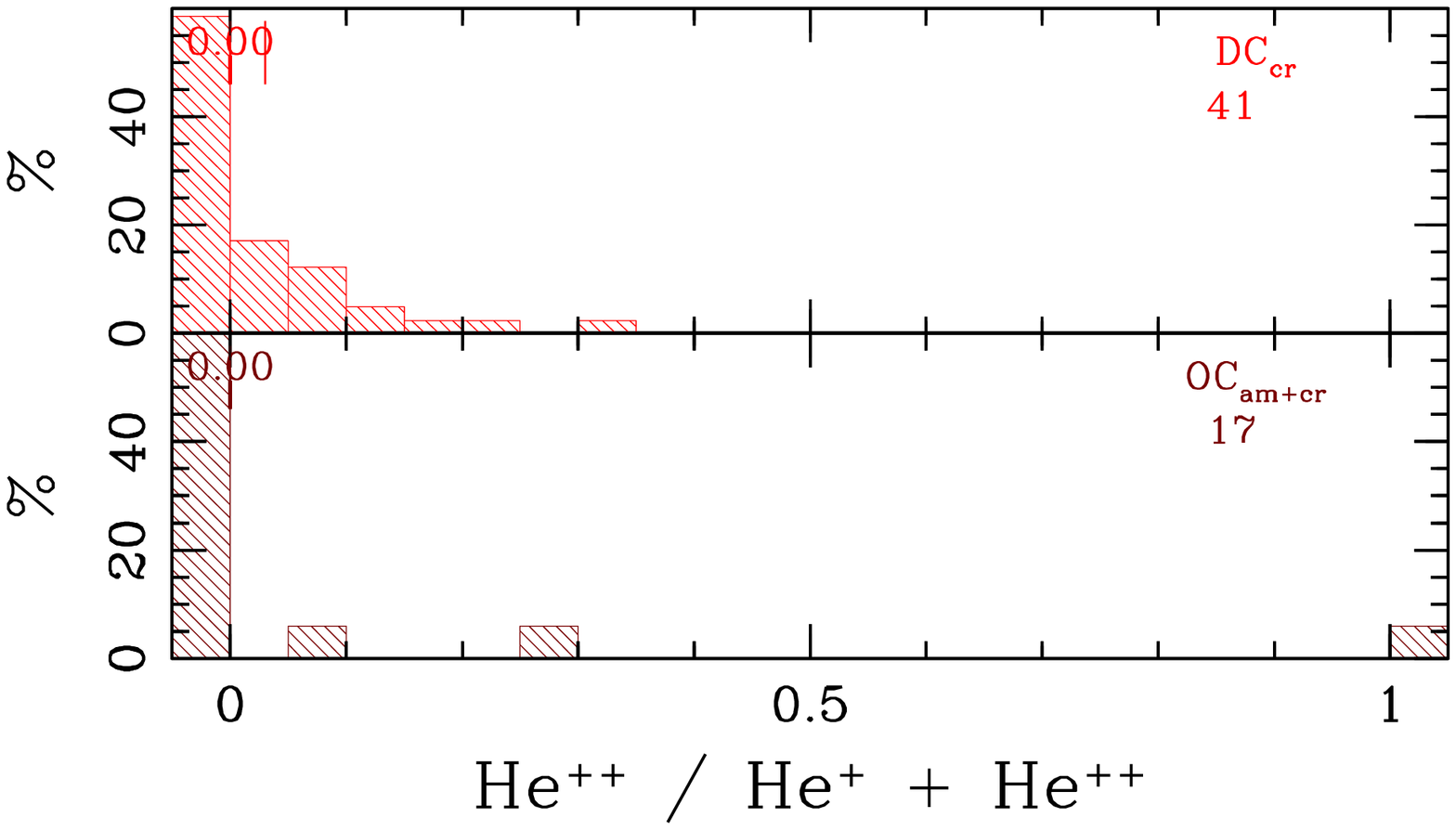}}

\resizebox{0.49\hsize}{!}{\includegraphics{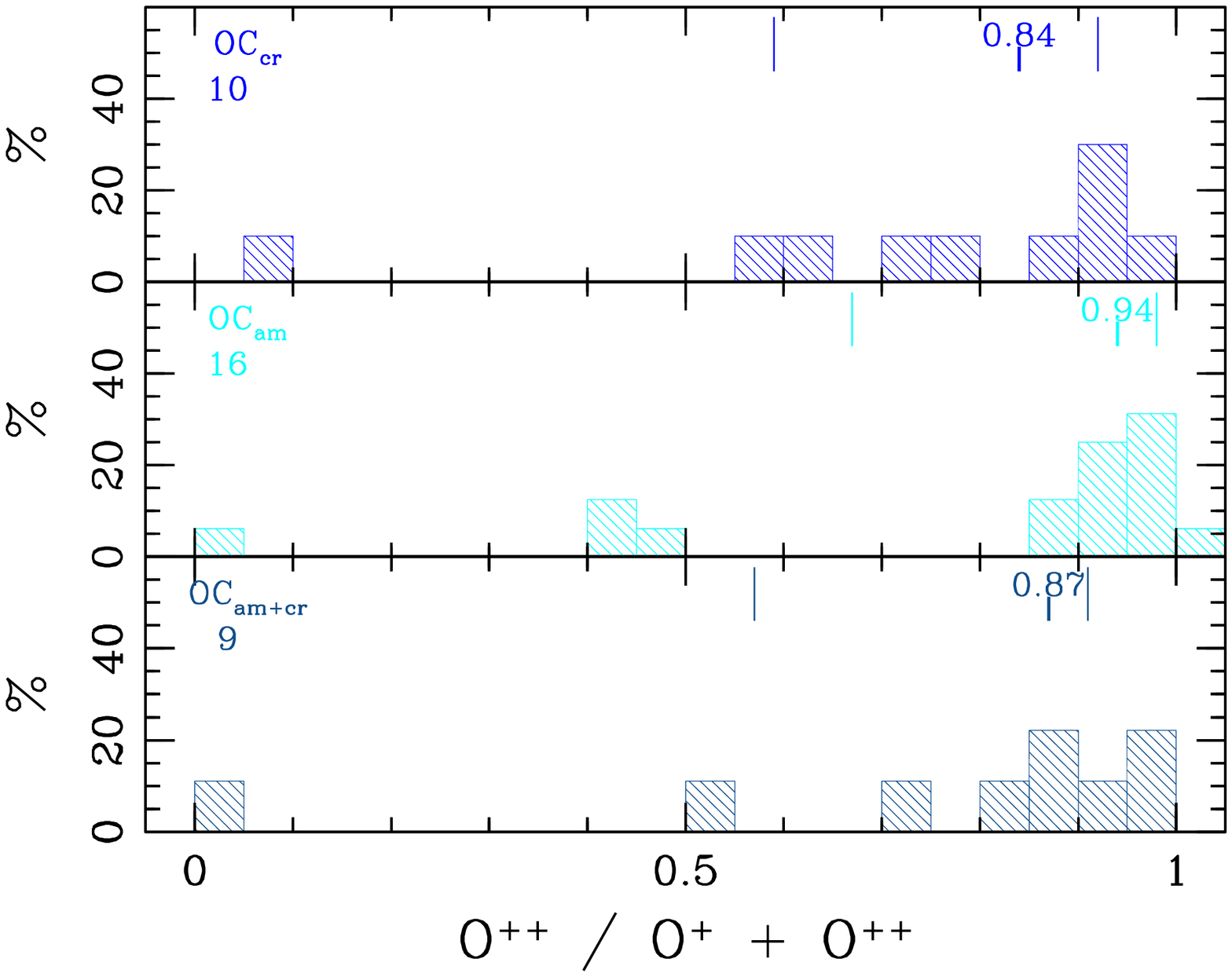}}
\resizebox{0.49\hsize}{!}{\includegraphics{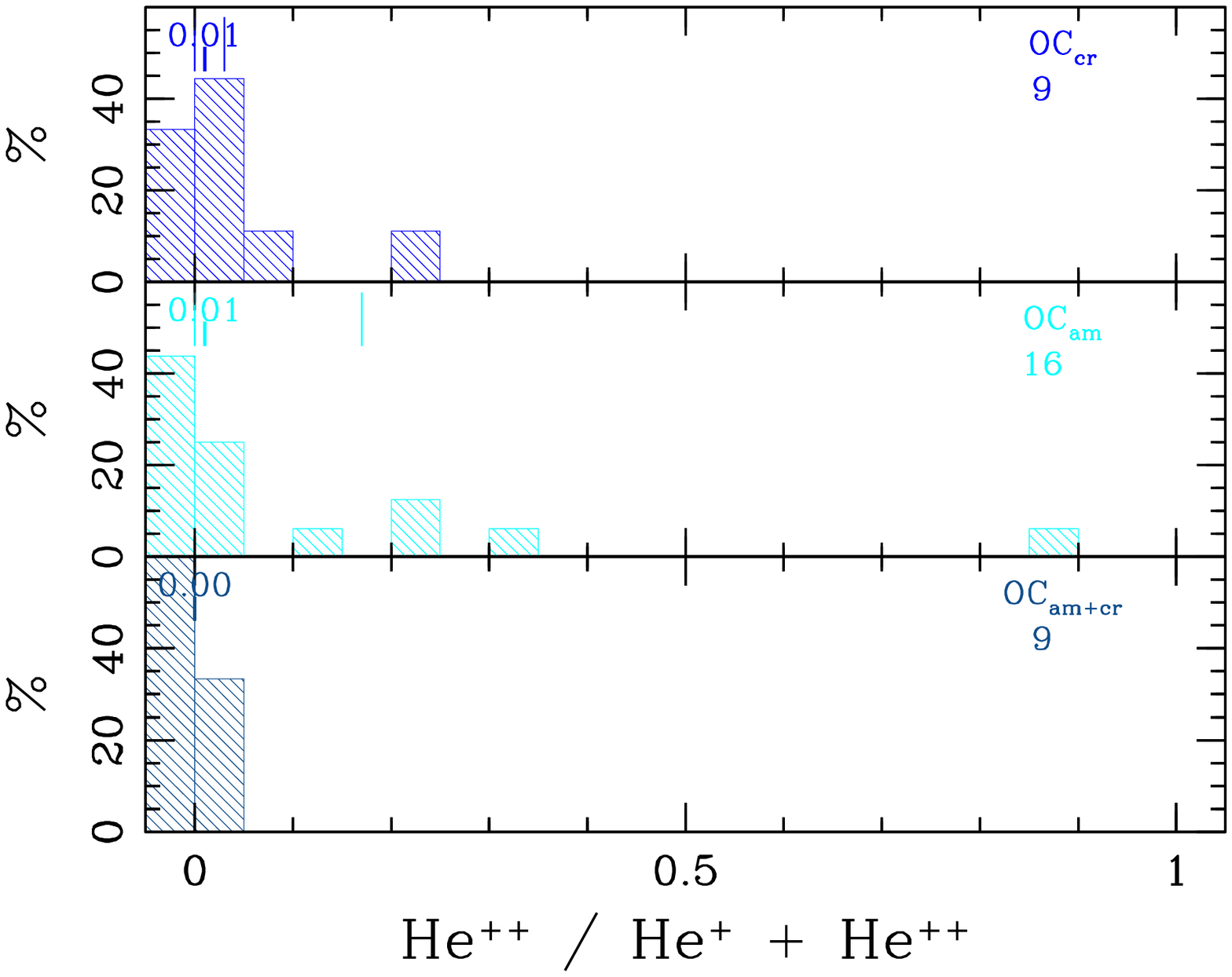}}

\resizebox{0.49\hsize}{!}{\includegraphics{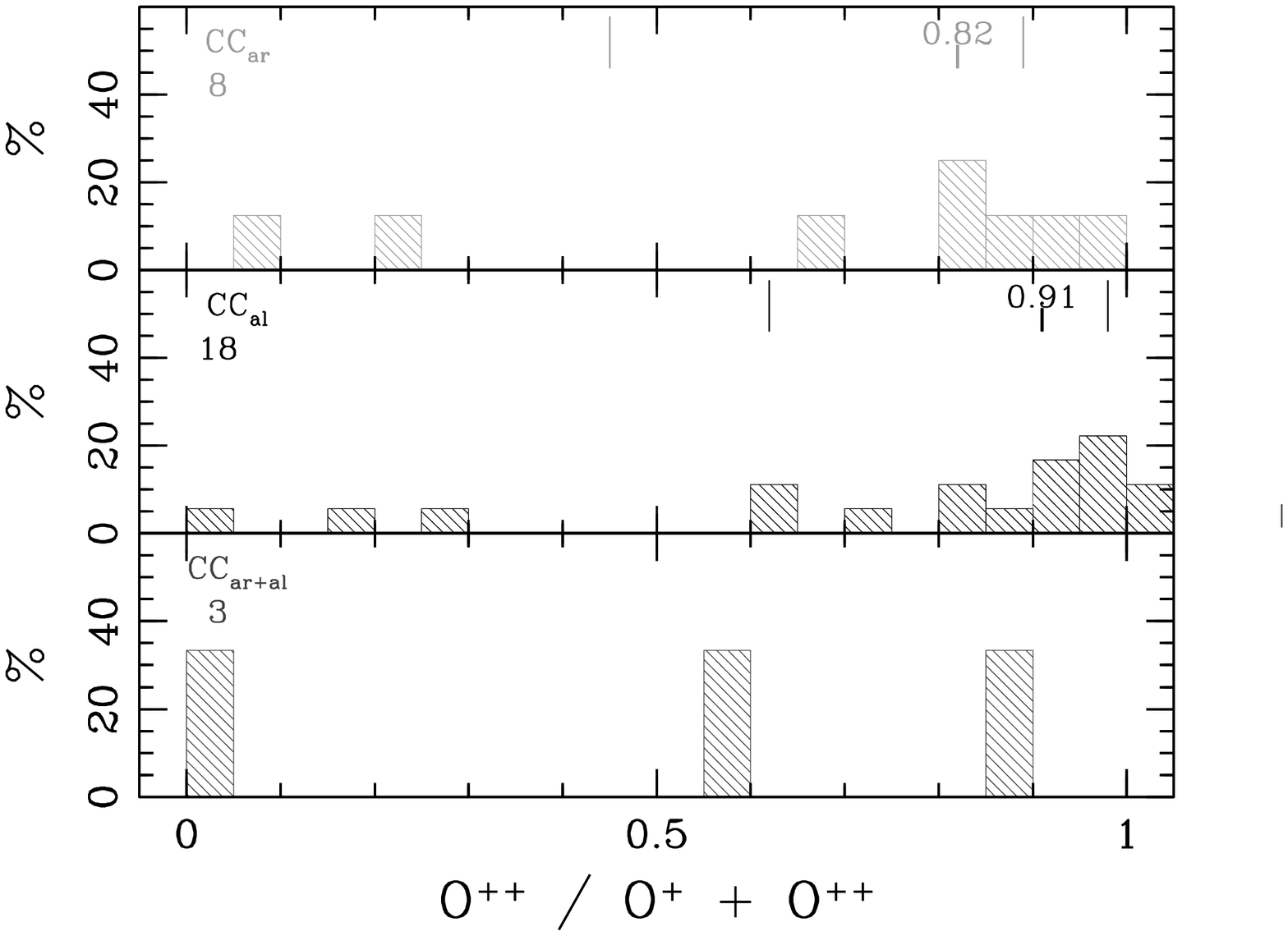}}
\resizebox{0.49\hsize}{!}{\includegraphics{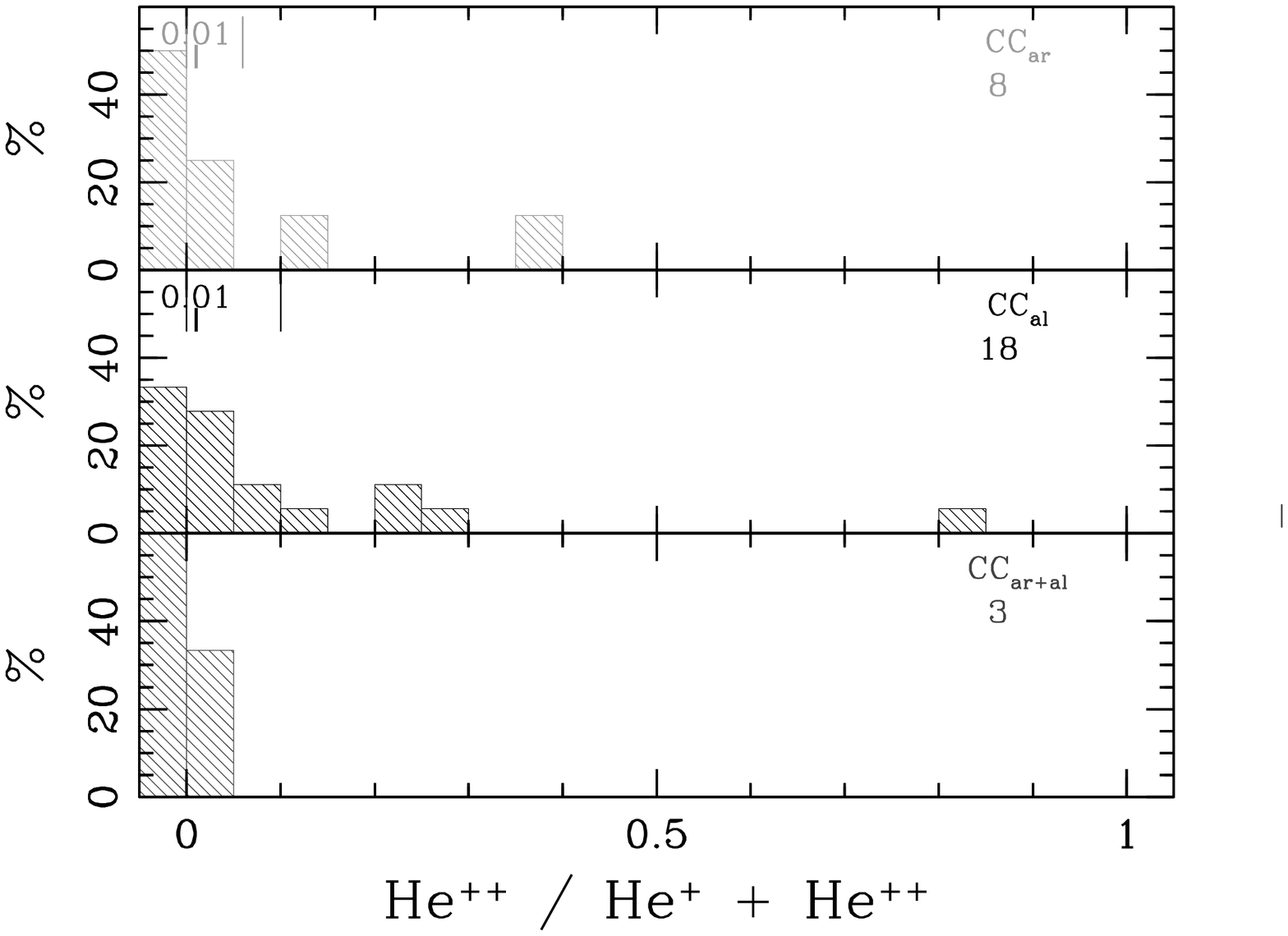}}

\caption[]{
 The distribution of \O++/O\ (left panel) and \He++/He\ (right panel) for {\it Spitzer}
 dust subtypes in combined Galactic disc and bulge samples.  From top to
 bottom: distributions for dust subtypes of DC, OC and CC PNe
 respectively.
}
\label{hist_ion}
\end{figure*}

\begin{figure*}
\resizebox{0.47\hsize}{!}{\includegraphics{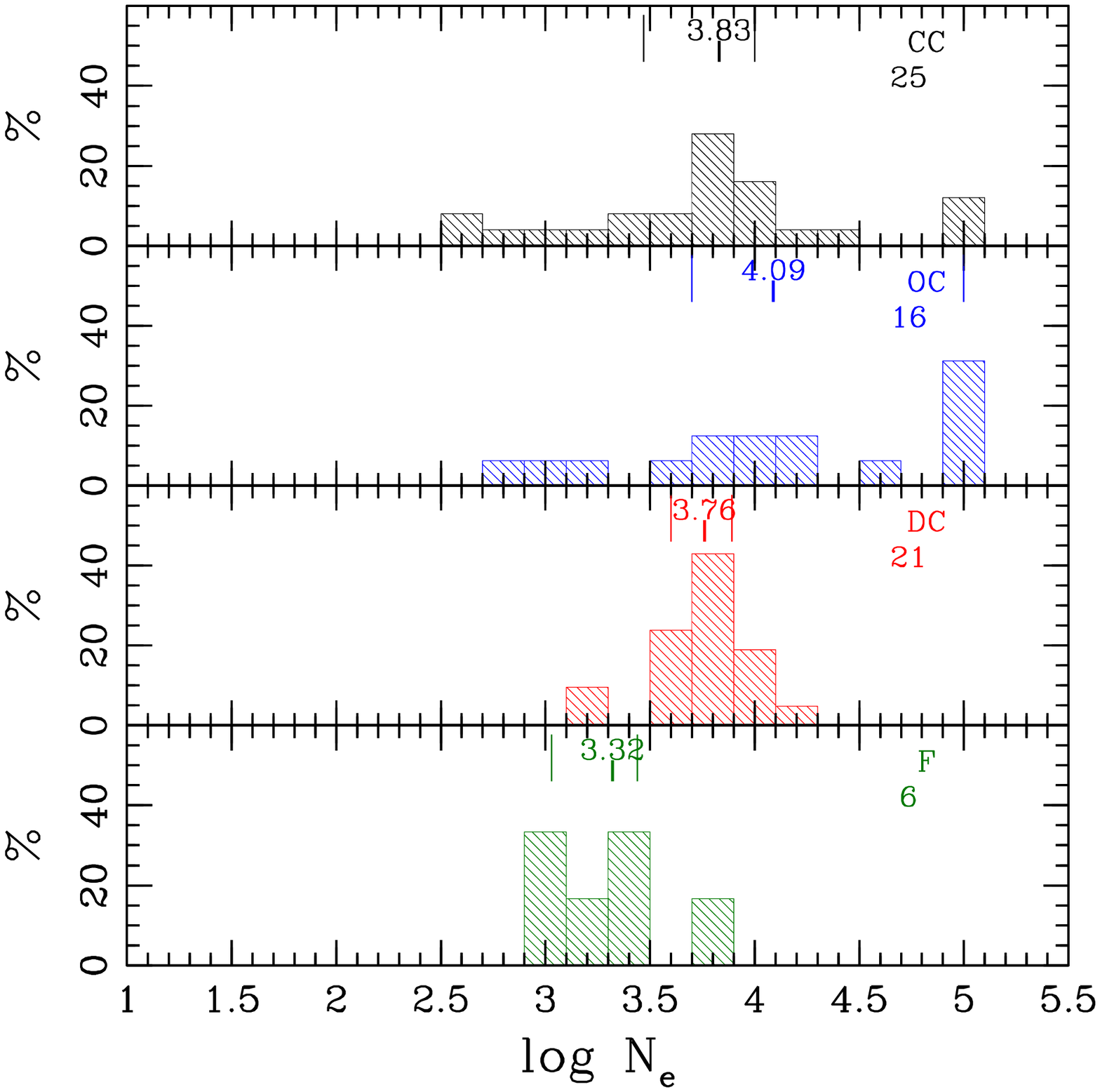}}
\resizebox{0.47\hsize}{!}{\includegraphics{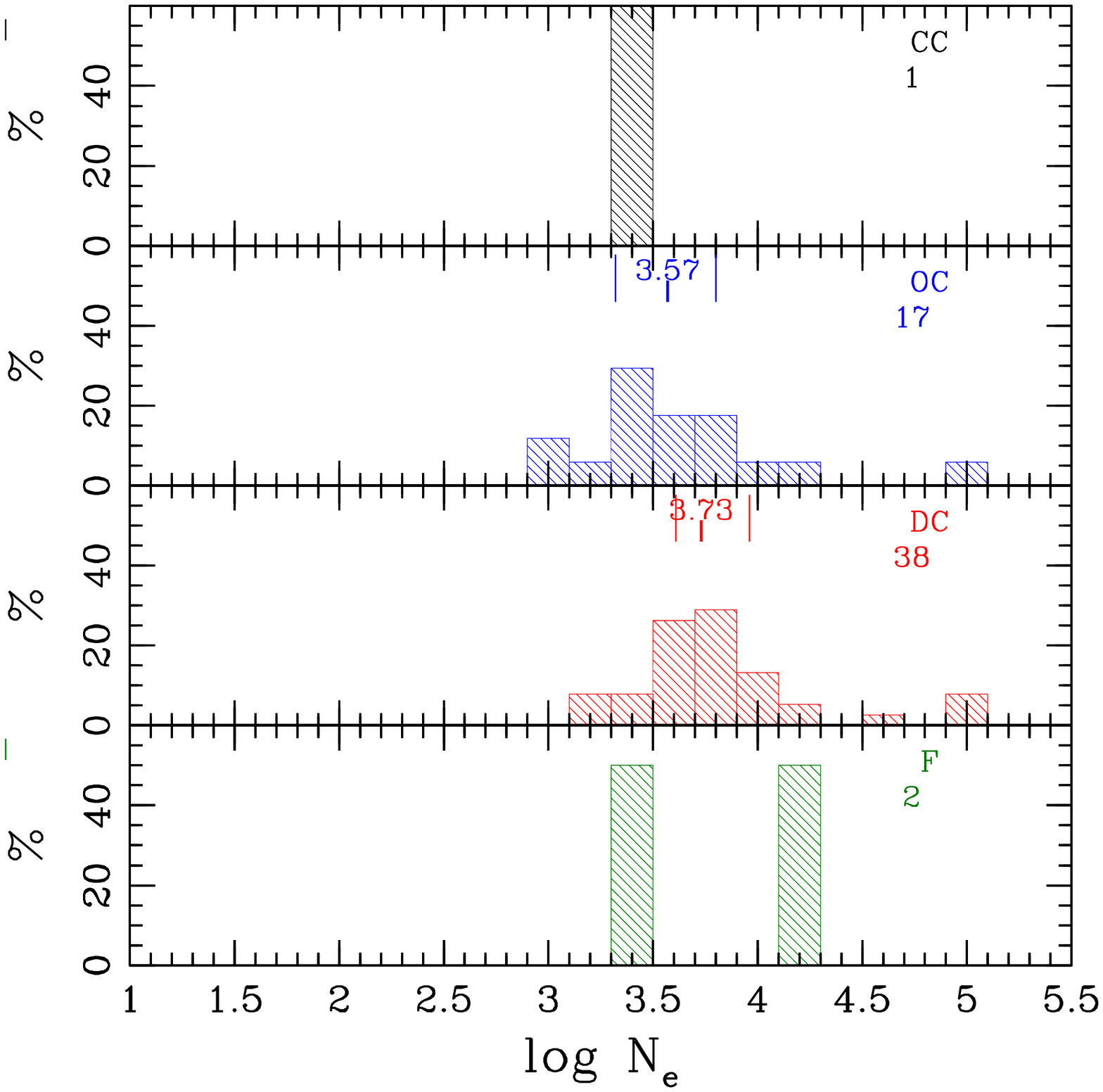}}
\caption[]{
 Histograms of nebular electron density N$_e$ derived from \rSii\ ratio 
 for Galactic disc PNe (left panel) and Galactic bulge PNe (right panel).
}
\label{hist_Ne}
\end{figure*}

\begin{figure*}
\resizebox{0.33\hsize}{!}{\includegraphics{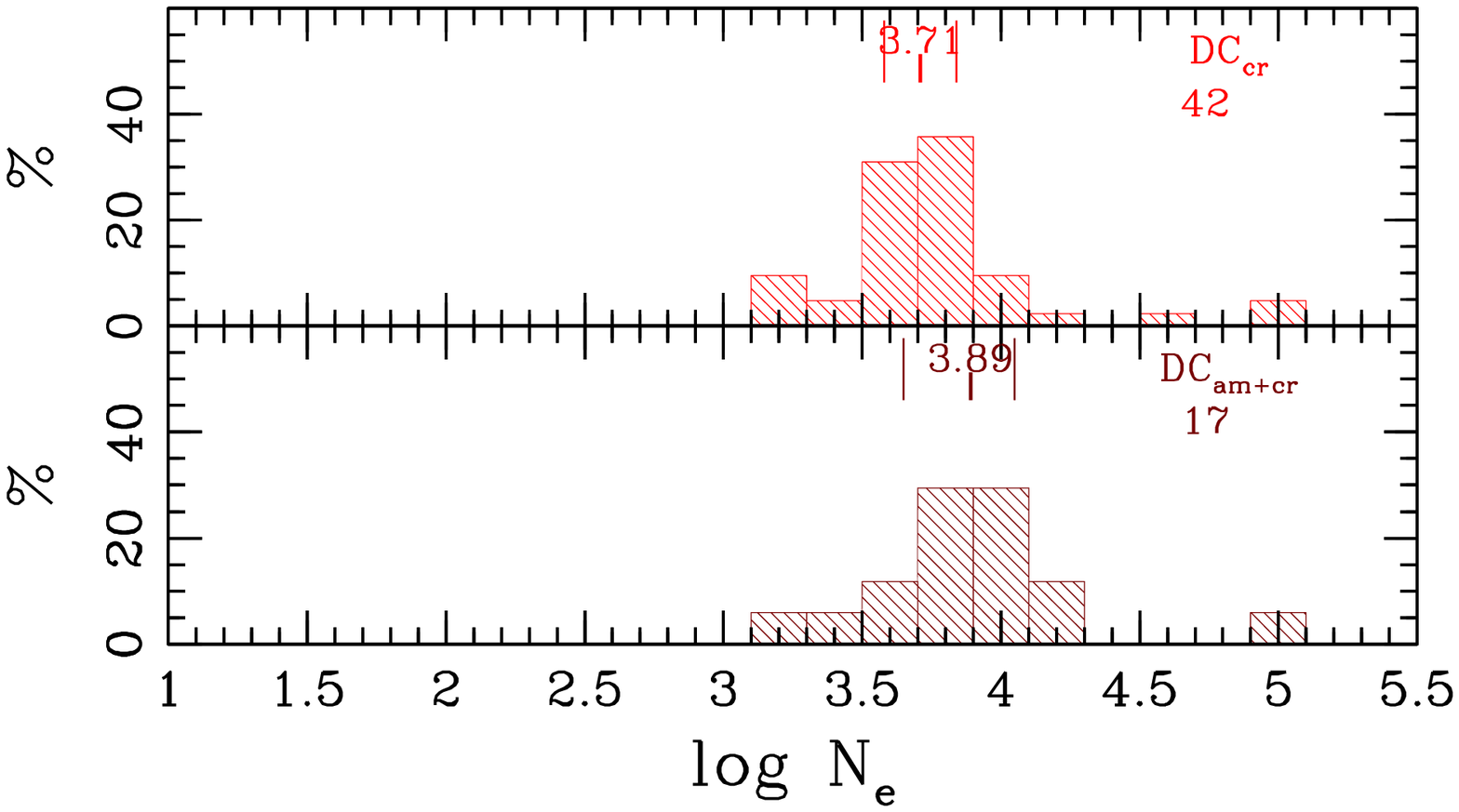}}
\resizebox{0.33\hsize}{!}{\includegraphics{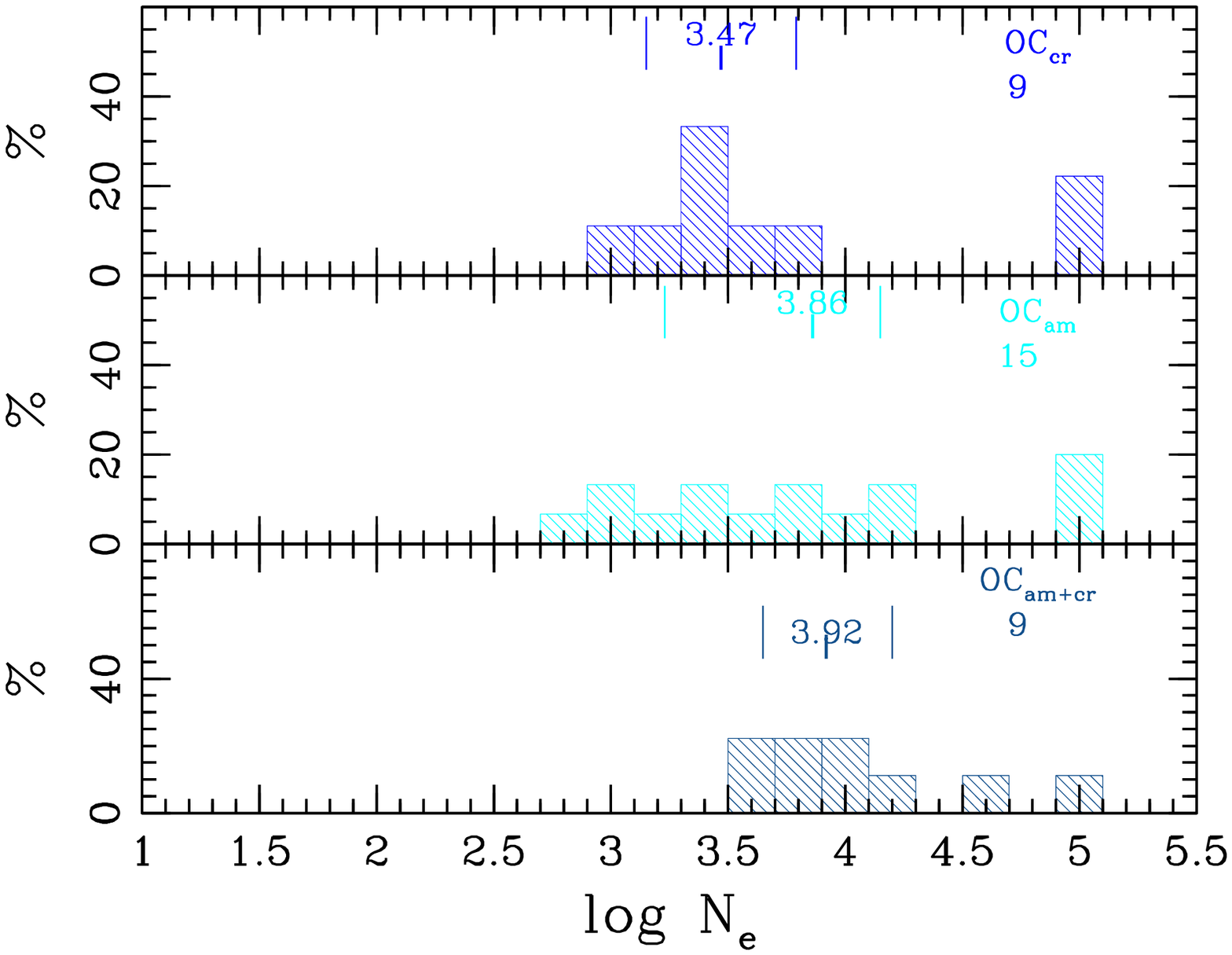}}
\resizebox{0.33\hsize}{!}{\includegraphics{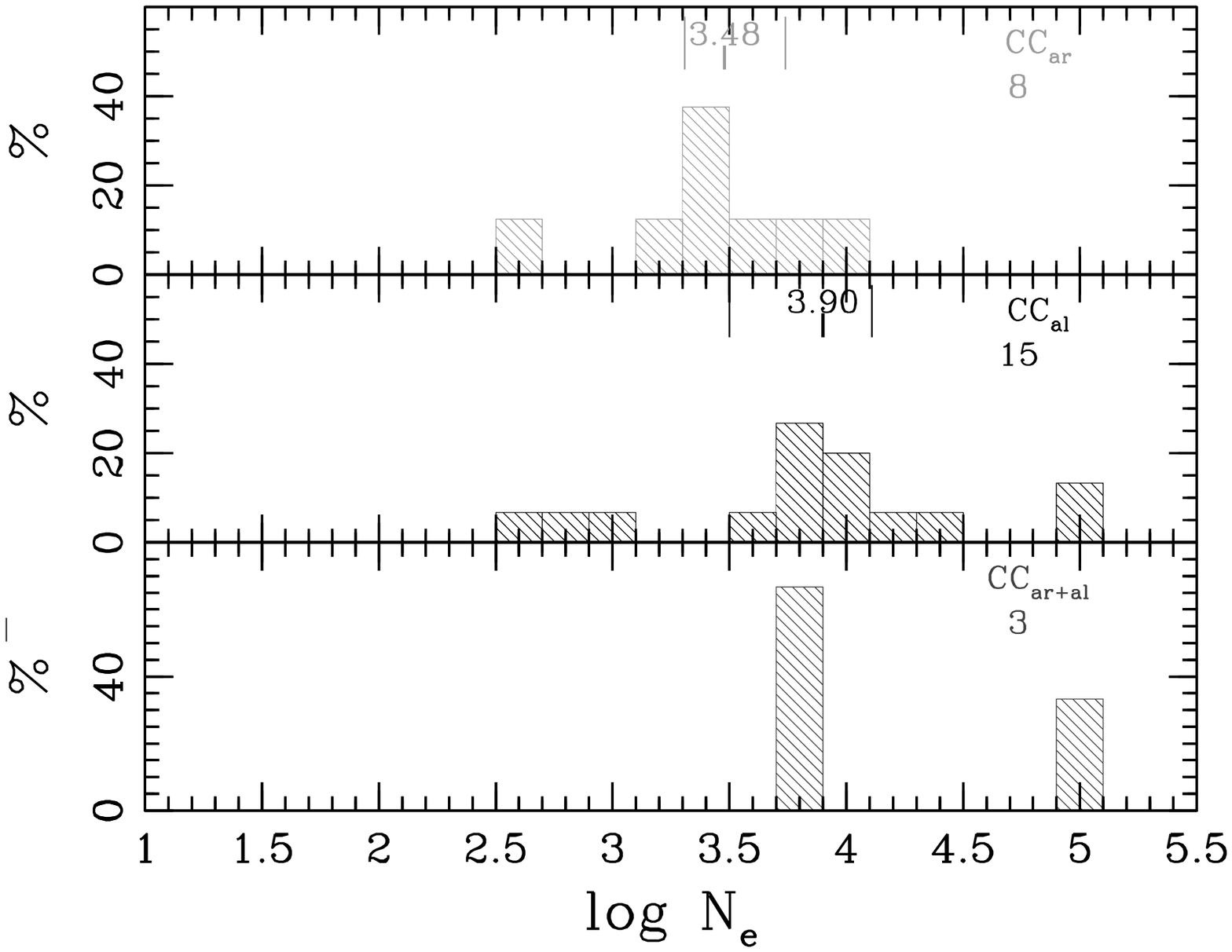}}
\caption[]{
 Histograms of nebular electron density N$_e$ derived from
 \rSii\ ratio for combined samples of Galactic disc and bulge PNe.
 The left to right panels present distributions for dust subtypes of 
 DC, OC, and CC PNe, respectively.
}
\label{hist_Ne_sub}
\end{figure*}

\begin{figure*}
\resizebox{0.47\hsize}{!}{\includegraphics{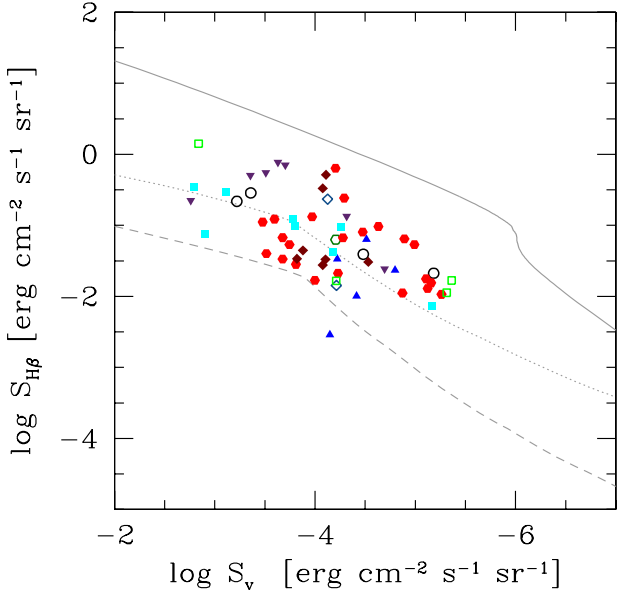}}
\caption[]{
  Surface brightness S$_{H\beta}$ versus S$_V$ for the different {\it Spitzer}
dust subtypes of  Galactic disc and bulge PNe combined. Three theoretical
tracks, representing models with 0.56 (dashed line), 0.60 (dotted line), and
0.64 (solid line)\,\msun\ central stars are also shown for comparison (see text
for more details). The same meaning of symbols as in \figref{fig_lb} is used: 
\CCar - open hexagons; \CCal - open circles; \CCaral - open diamonds; \OCcr -
triangles; \OCam - filled squares; \OCamcr - reversed triangles; \DCcr - filled
hexagons; \DCamcr - filled diamonds; F - open squares. 
}
\label{ss}
\end{figure*}

\begin{figure*}
\resizebox{0.33\hsize}{!}{\includegraphics{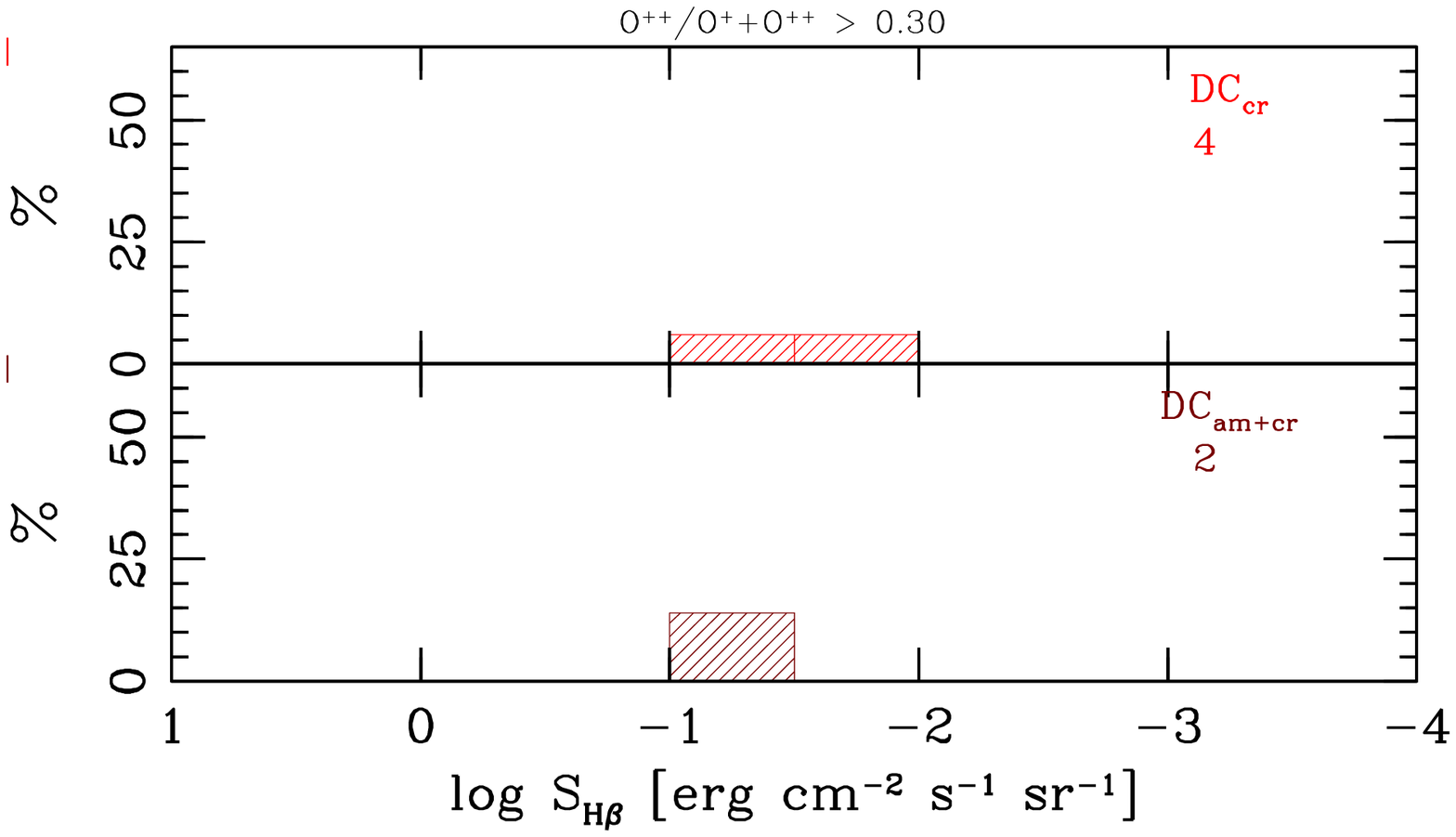}}
\resizebox{0.33\hsize}{!}{\includegraphics{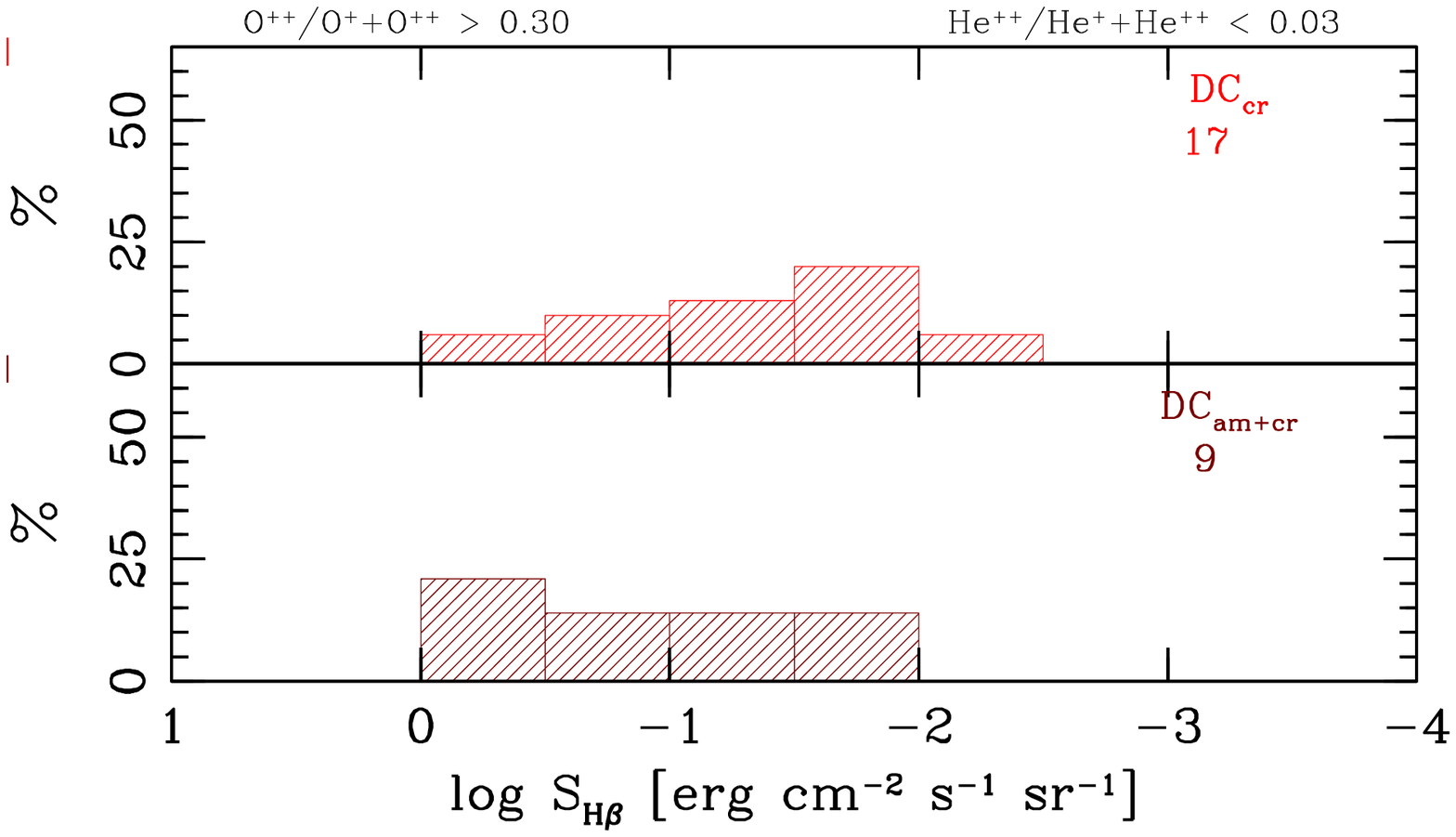}}
\resizebox{0.33\hsize}{!}{\includegraphics{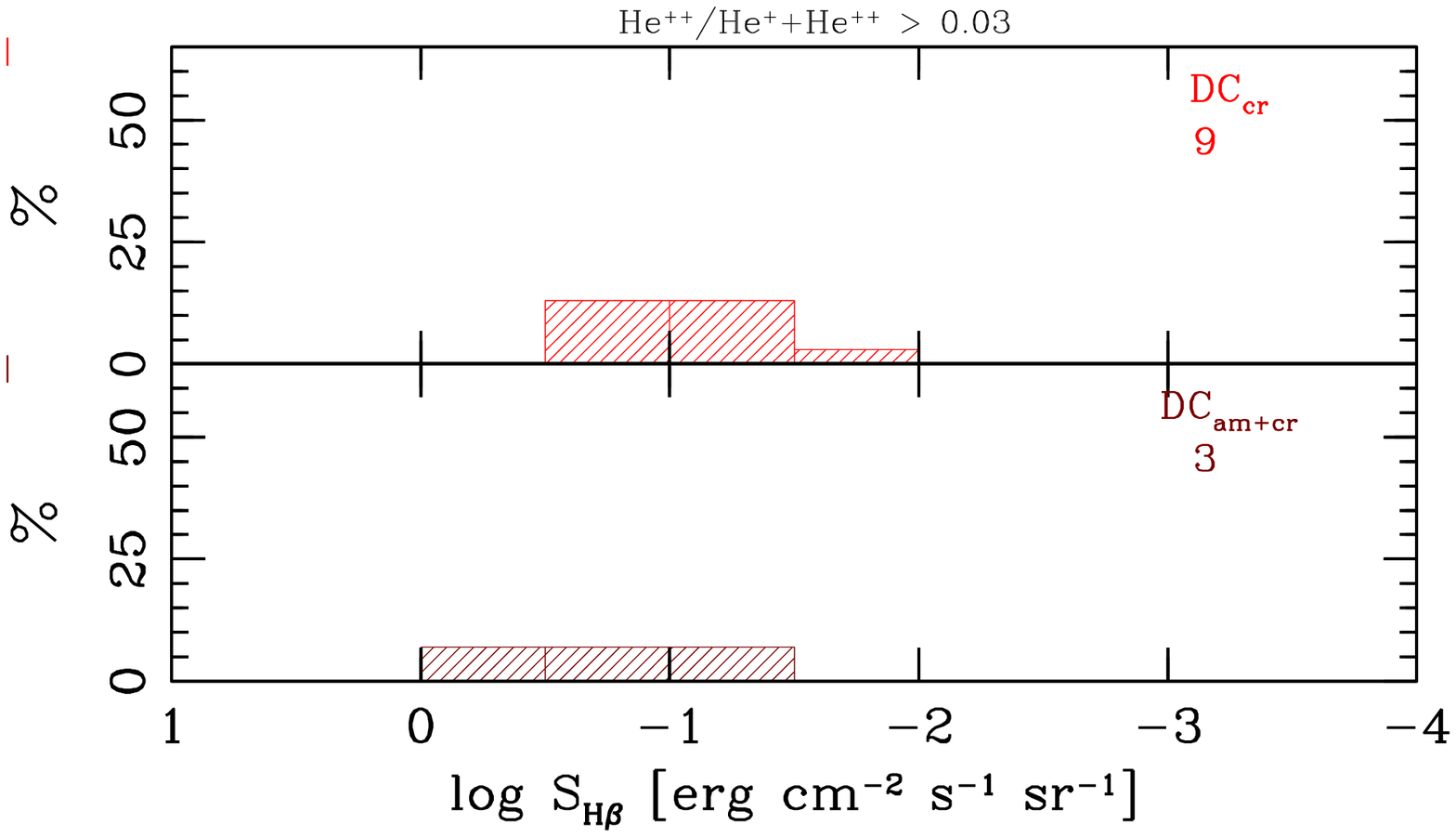}}

\resizebox{0.33\hsize}{!}{\includegraphics{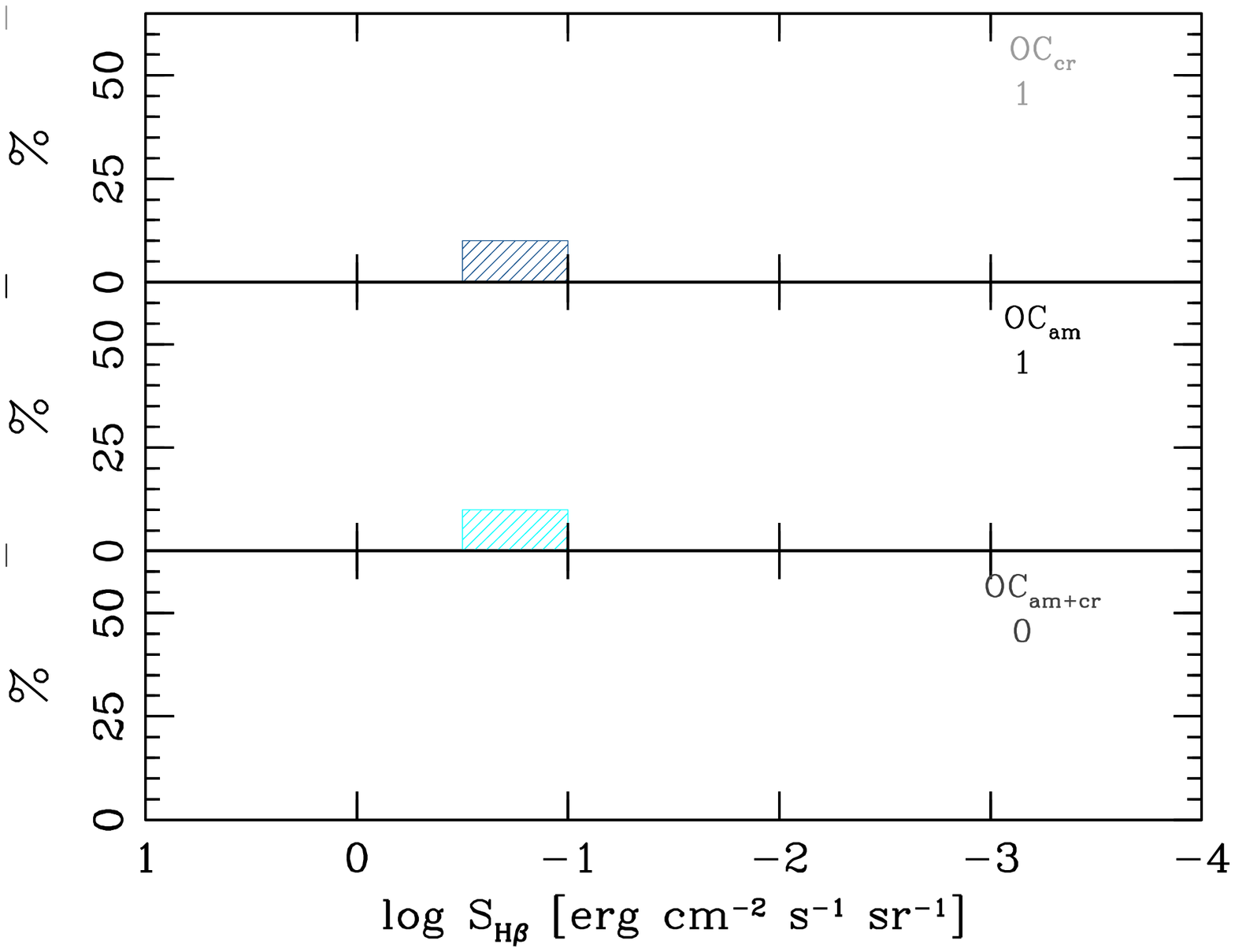}}
\resizebox{0.33\hsize}{!}{\includegraphics{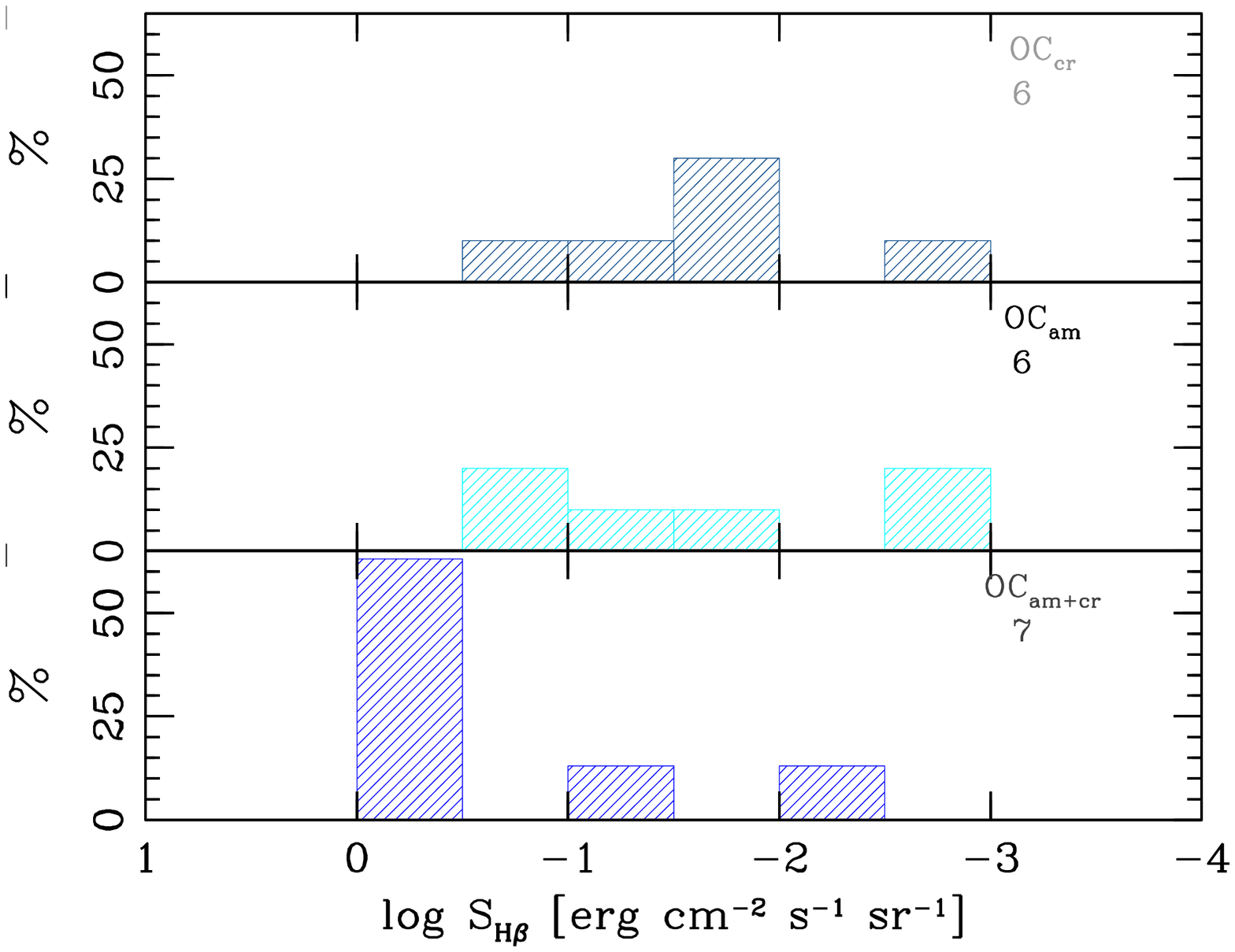}}
\resizebox{0.33\hsize}{!}{\includegraphics{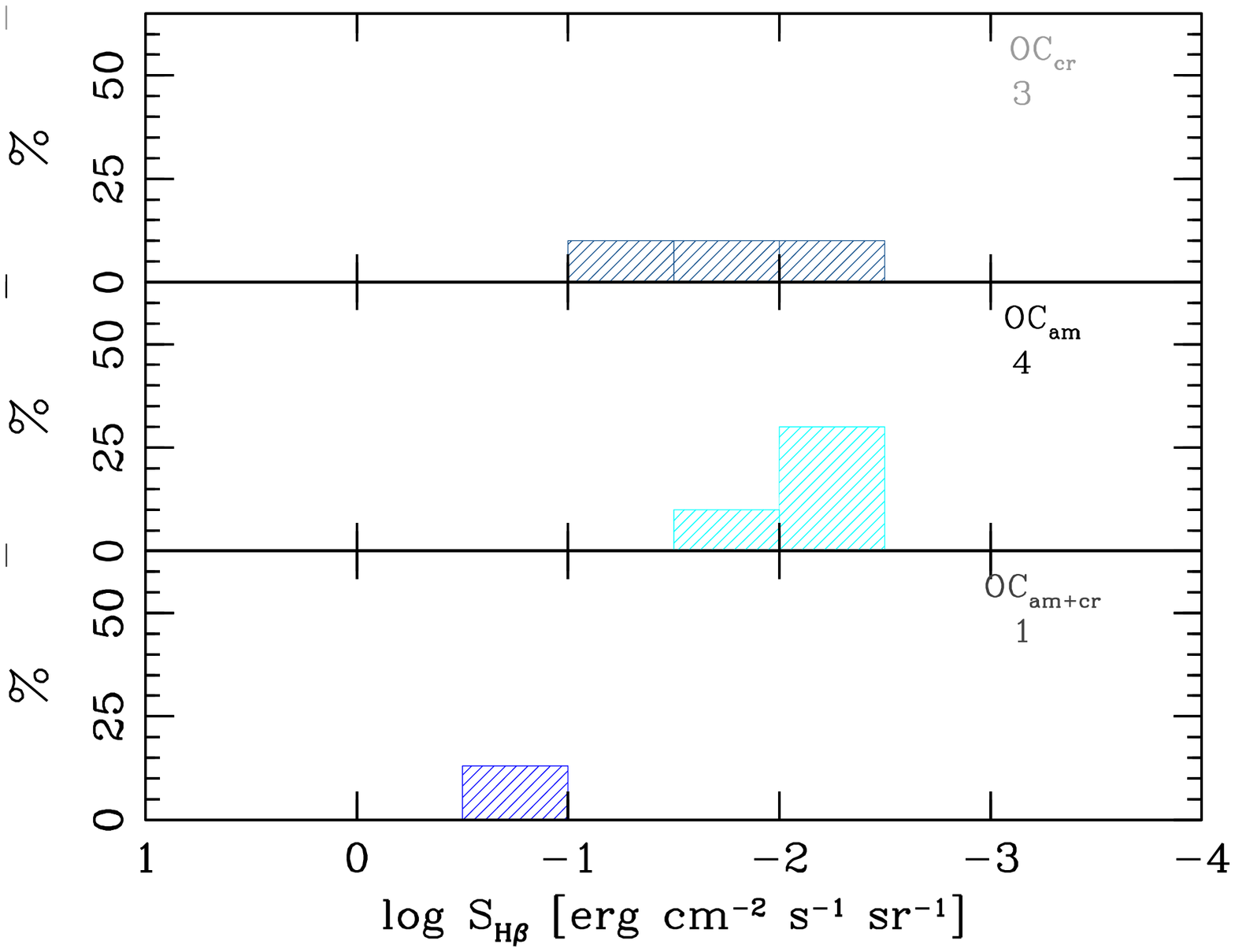}}

\resizebox{0.33\hsize}{!}{\includegraphics{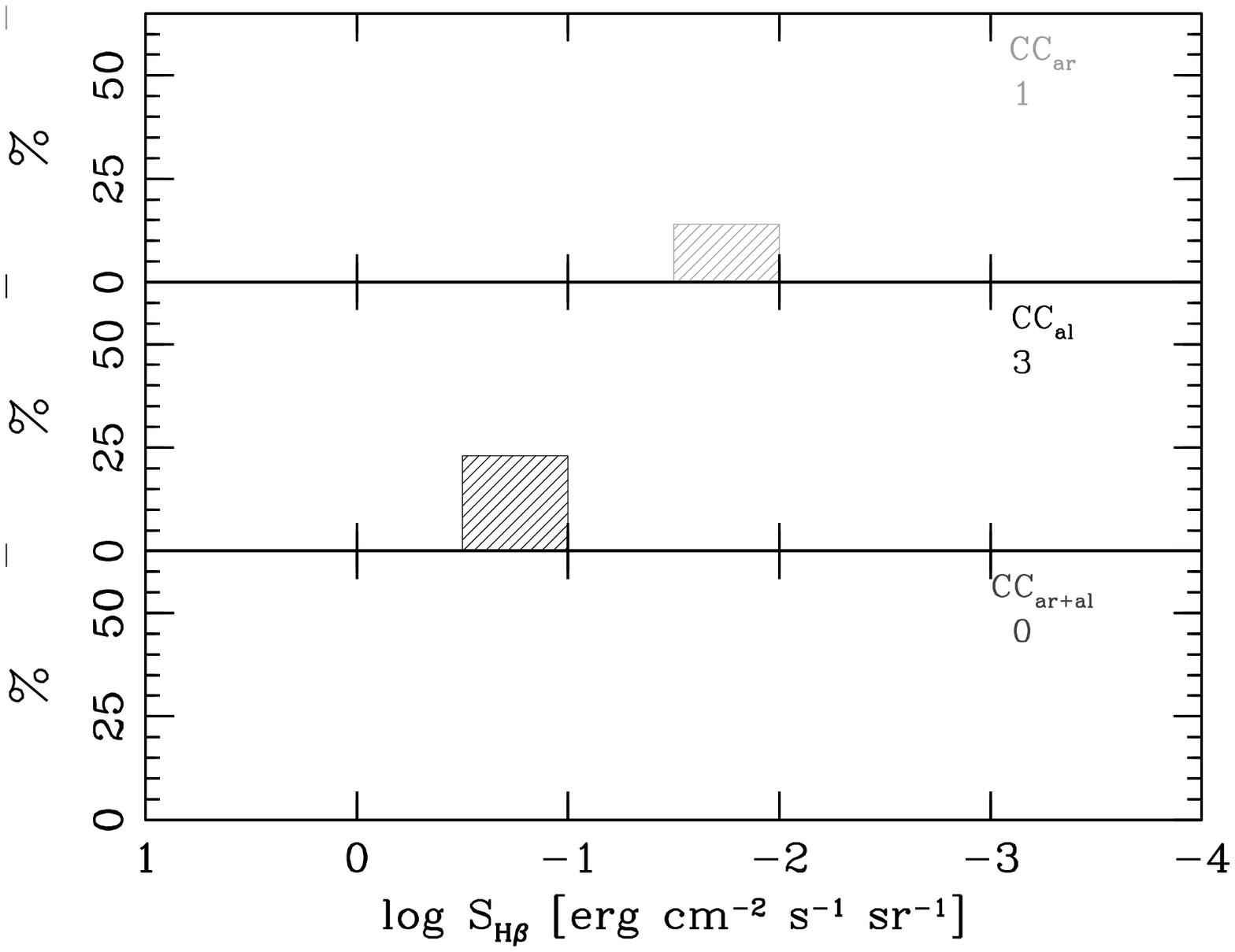}}
\resizebox{0.33\hsize}{!}{\includegraphics{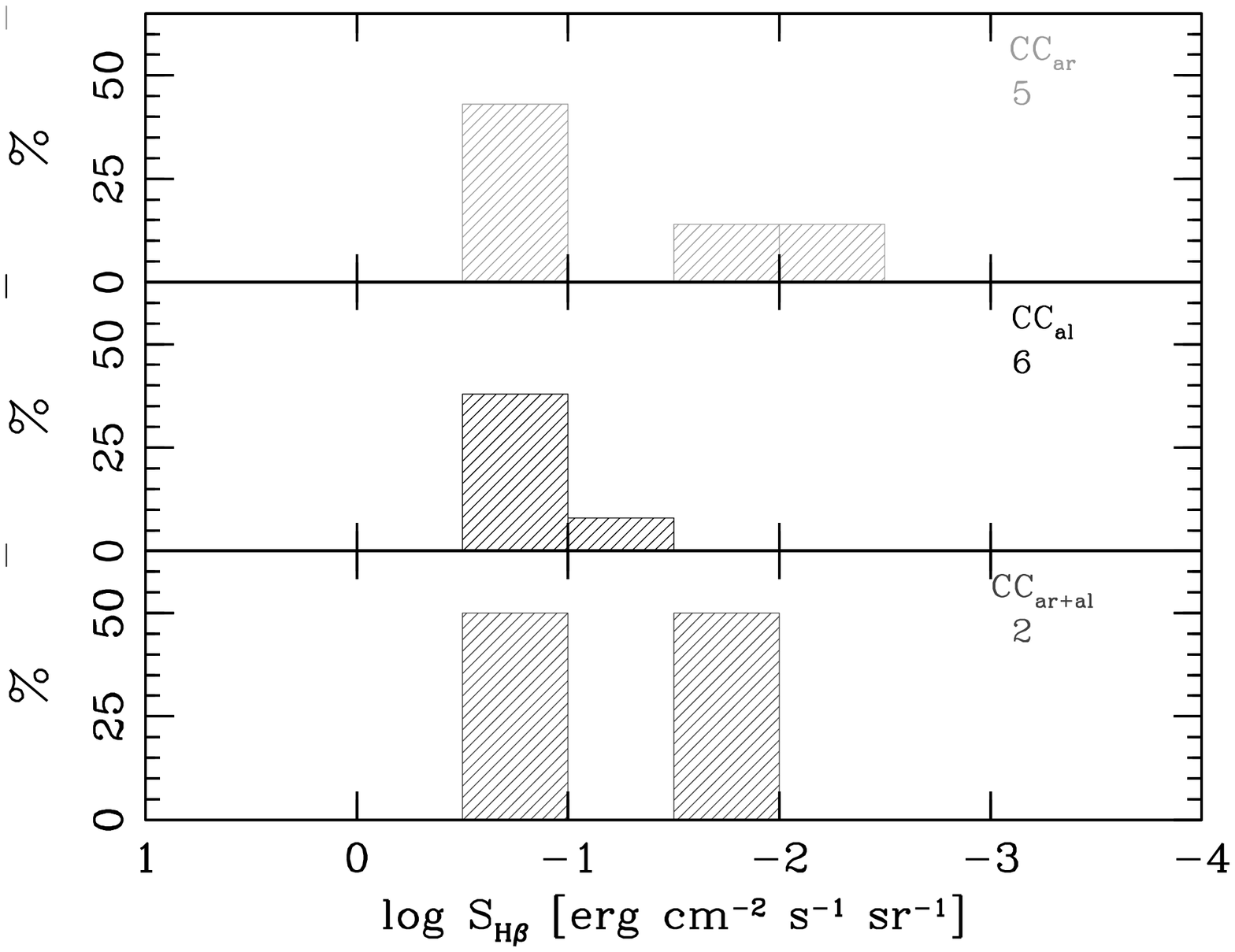}}
\resizebox{0.33\hsize}{!}{\includegraphics{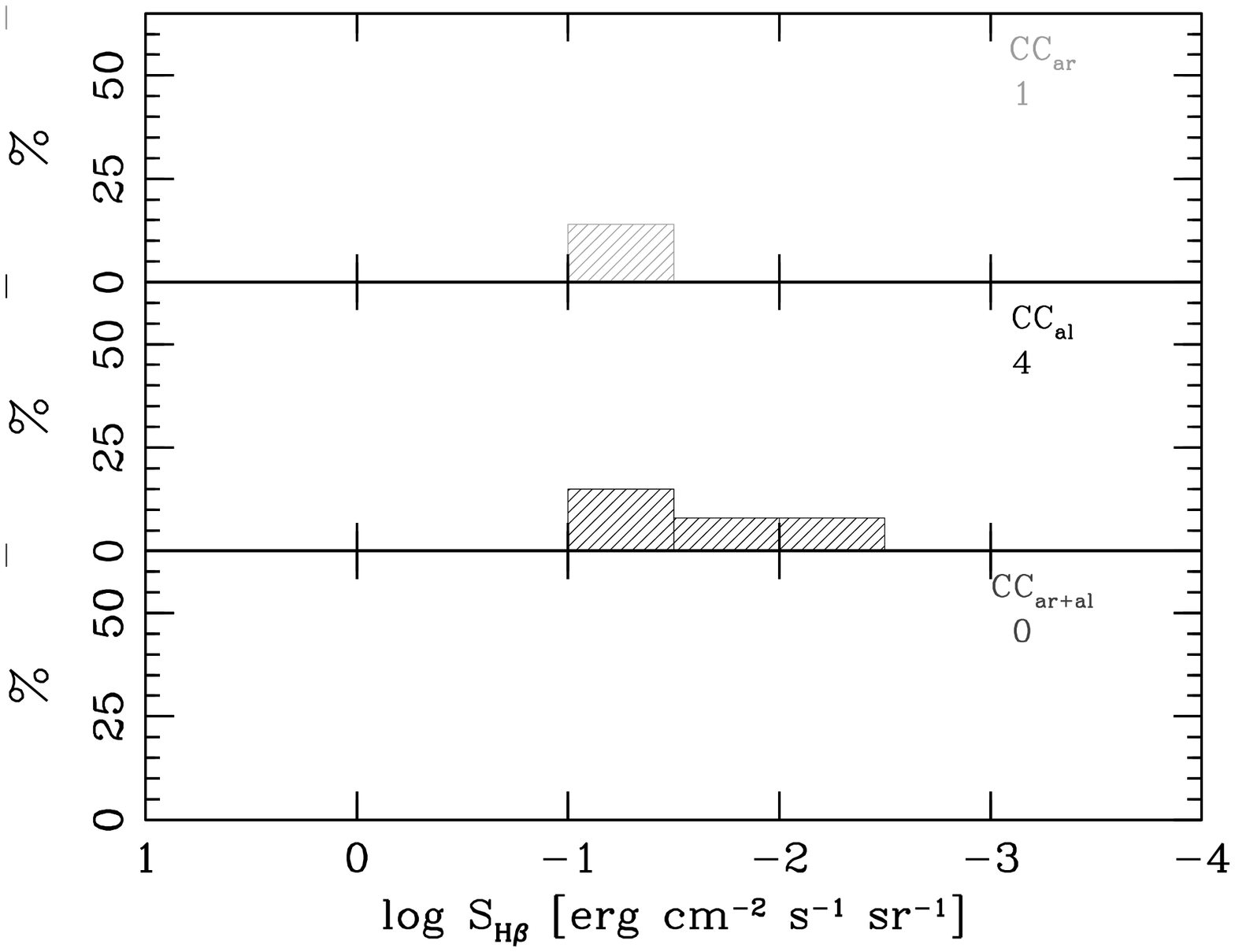}}

\caption[]{
 The distribution of S$_{H\beta}$ surface brightness for {\it Spitzer} dust
 subtypes and different ionization classes. From top to bottom:
 distributions for dust subtypes of DC, OC, and CC PNe respectively. 
 From left ro right: distributions for low ionization (\O++/O\ < 0.3);
 intermediate ionization (\O++/O\ > 0.3, and \He++/He\ < 0.03)), and high
 ionization (\He++/He\ > 0.03) are presented.
}
\label{hist_shb}
\end{figure*}

In this section the physical properties of the PNe with different {\it Spitzer}
dust types/subtypes and their evolutionary status are presented. We start with
the electron temperatures (T$_e$) derived from the [O~{\sc iii}]
$\lambda$4363/5007 and [N~{\sc ii}] $\lambda$5755/6584 ratios. Their median
values for the different {\it Spitzer} dust types in the Galactic disc and bulge
are presented in Tables~9 and 10, respectively. The results of our K-S and W
statistical tests are presented in Table~7, and in \figref{to_tn} we present the
relations T$_e$(O\,III) vs.\  T$_e$(N\,II) separately for PNe in the Galactic
disc and bulge regions. It can be seen that, even though the two ratios can in
principle provide an electron temperature of slightly different regions of the
nebula (zones of ions of lower and higher ionization potentials), both are
relatively well correlated in \figref{to_tn}. Only in the case of OC PNe located
in the Galactic bulge is the median T$_e$(N\,II) s 0.13 dex higher than
T$_e$(O\,III), but this sample is obviously not uniform with a few clear
outliers as seen in \figref{to_tn}. Concerning the differences between the major
dust types, DC PNe in the Galactic disc differ from OC and CC PNe by having 
lower T$_e$(O\,III) (the median values are 3.98 vs. identical 4.05 in OC and CC
PNe). The DC bulge PNe, however, differ from OC PNe by having lower T$_e$(N\,II)
temperatures (the median values are 3.94 and 4.14, respectively). The physical
explanation of such behaviour could be that DCs in the disc are more oxygen
abundant so that these atoms can cool the nebulae more efficiently.

The ionization level of the nebular gas can be evaluated with the  \O++/O\ and
\He++/He ionic ratios. In Tables 9 and 10 the median values of these parameters
are given for PNe of different dust types in the Galactic disc and bulge
separately. In \figref{hist_ion} their histograms for dust subtypes are
presented.  It can be seen that in both regions of the Milky Way most of the
groups analysed here (disregarding only the featureless F-type PNe that are
expected to represent more advanced evolutionary stages) show a similar
intermediate level of ionization. A median \He++/He=0 is found for all of them,
and the median \O++/O\ ranges from 0.83 in CC disc PNe to 0.92 for OC disc PNe. 
The exceptions are DC PNe in the Galactic bulge with a median \O++/O=0.55.  The
results of our statistical tests\footnote{The Wilcoxon test is not
reliable in the case of \He++/He\ because the derived ratios are mostly 0 and
the distributions are flat cut on one side.} presented in Table~7 confirm that
in the Galactic bulge there is indeed a statistically significant difference
between DC and OC PNe.  On the other hand, the difference between DC PNe in the
Galactic disc vs.\ DC PNe in the bulge is not confirmed (Table~8), although the
median \O++/O\ values (0.84 and 0.55, respectively) seem to suggest that it
could be present.

The distinction between \O++/O\ in Galactic disc and bulge DC PNe is in fact a
consequence of the difference between the nebular properties of the \DCcr\ and
\DCamcr\ subtypes. The former dominate the Galactic bulge DC sample, and at the
same time, are found more frequently in a lower ionization state (see Tables 6
and 14). As can be seen in the top left-hand panel of \figref{hist_ion}, in
reality the difference is, however, not that profound. Indeed, the median
\O++/O\ value is lower for \DCcr\ than for \DCamcr, but in both groups their
members can be found in a whole range of ionization stages. What is intriguing
is that proportionally more \DCcr\ show the presence of doubly ionized He$^{++}$
ions than in the case of \DCamcr\; (\figref{hist_ion} top right). This issue
could indicate that \DCamcr\ evolve towards the \DCcr\ stage, but it 
contradicts the \O++/O\ distribution. For the other dust subtypes, there are no
clear differences in the  ionization level either for OC or for CC PNe, as can
be inferred from \figref{hist_ion} and Table~14. This is confirmed by the
statistical tests in Table~12.

The electron density (N$_e$) is a parameter that can provide information on the
evolutionary status of the individual nebula. However, the distributions
analysed for some PNe subsamples will also depend on the properties of the
central stars. Older nebulae show lower electron densities, but if the nebular
gas expands more rapidly, then the drop in the electron density is also rapid.
Furthermore, the more massive central stars have a higher probability of
detecting the surrounding nebulae when they are still relatively dense. With our
own observations and literature spectra, we derived electron densities N$_e$
from the \rSii\ ratio. The median values of N$_e$ for all {\it Spitzer} main
dust types are given in Tables~9 and 10 for the Galactic disc and bulge PNe,
respectively. The histograms of the corresponding distributions are presented in
\figref{hist_Ne}. Only DC and OC PNe are numerous enough in both Galactic
regions to allow for statistically meaningful comparisons. The median N$_e$
values and their distributions seem to be very similar for the DC PNe in the
Galactic disc and bulge. In contrast, the N$_e$ distribution for the OC PNe in
the Galactic disc is flatter but has a  higher median value than in the bulge,
where it also has a more peaked shape. These differences are not strictly
confirmed by the statistical tests, though (see Tables~7 and 8). In any case, 
we have to remember that the histograms in \figref{hist_Ne} are the
superposition of the distributions of the various {\it Spitzer} dust subtypes
within each main dust group (see below). We do not have enough data for the CC
and F {\it Spitzer} dust types to analyse them in the Galactic bulge. In the
Galactic disc, the major CC and F {\it Spitzer} dust types display the lowest
median N$_e$ in accordance with the assumption that they represent later stages
of nebular evolution. In the case of Galactic disc CC PNe, the median electron
density is very similar to the DC PNe, but the N$_e$ distribution clearly has a
low-density tail, indicating that more evolutionarily advanced nebulae with
C-rich dust features are still present in this group.

In the left-hand and centre panels of \figref{hist_Ne_sub}, we present the
distributions of the derived N$_e$ separately for \DCcr\ and \DCamcr\ and for
\OCcr, \OCam, and \OCamcr, respectively, by using merged samples from the
Galactic disc and bulge. It can be seen that the log N$_e$ distributions for the
\DCcr\ and \DCamcr\ PNe clearly peak around their median values of 3.71 and
3.89, and the statistical significance of this difference is close to being
confirmed by our statistical tests (probability 0.05 and 0.03 for K-S and W
tests in Table~12). In the case of the OC dust subsamples, there are probably
not enough objects to reach firm conclusions. However, the N$_e$ distribution
of \OCam\ PNe seems to be fairly flat. Concerning \OCamcr\ PNe, they clearly
seem to be denser (or younger) than the \OCcr\ and \OCam\ subsamples. Since
\OCam\ dominate in the Galactic disc sample (Table~6), the general N$_e$
distribution for disc OC PNe (\figref{hist_Ne}) was flatter and shifted to
higher values than for bulge OC PNe where \OCcr\ PNe compose half of that
sample. The differences in spatial distribution (i.e.\  galactic latitude) and
chemical abundances among the several OC subtypes mentioned in the previous
sections rule out the possibility that \OCam\ and \OCamcr\ sources could evolve
into \OCcr\ ones as a result of silicates crystallization. On the other hand,
evolution of \OCam\ objects into \OCamcr\ ones or \DCamcr\ into \DCcr\ would
still be possible and consistent with the idea that the amorphous silicates are
transformed/processed to crystalline silicates from the AGB to the PNe phase. In
the right-hand panel of \figref{hist_Ne_sub}, the log N$_e$ distributions for the
several subtypes of PNe with carbon-rich dust features are also presented.
Although the distributions seems to be rather broad, they indicate that the
\CCar\ sources generally are less dense (or older) than the \CCal\ ones.
However, the much lower median metallicity displayed by the \CCal\ objects with
respect to the \CCar\ ones (with a nearly solar metal content) seem to discard
any possible evolutionary link between both types of PNe. 

We note that any difference in the N$_e$ distributions and their corresponding
median values can also be interpreted as due to individual subsamples
originating in parent stellar populations of different masses. Another way to
analyse possible evolutionary links between PNe with different {\it Spitzer}
dust subtypes is to compare objects in diagrams that follow the evolution of the
planetary nebula itself and the central star. Such diagram is plotted in
\figref{ss} where we present the nebular H$\beta$ surface brightness
S$_{H\beta}$ versus S$_V$ - a similarly defined parameter based on the stellar
flux in the visual V band and nebular parameter $\theta$;
S$_V$=F$_V$/($\pi$$\theta^{2}$) \citep{GornyTylenda2000}. Both parameters are
independent of distance so we plot them in \figref{ss} for the Galactic disc and
bulge PNe of all {\it Spitzer} dust subtypes. Unfortunately, many objects from
our samples could not be placed in \figref{ss} since the stellar flux in the
optical is needed to compute one of the parameters, and it is generally 
available for central stars of PNe alone. The difficulty of measuring the
central star V magnitude seems natural for F type PNe that are presumably older
and their very hot central stars can therefore be faint in the
optical.\footnote{All analysed PNe have small apparent diameters. This usually
makes measurements of stellar flux difficult, but it should influence all
subsamples to a comparable extent.} For the PNe with O-rich and mixed chemistry
dust, the relative percentages of central stars with measured stellar optical
flux are 36\% and 39\%, respectively. Surprisingly, only seven (18\%) PNe with
C-rich dust have this data available. Most of them are found in the Galactic
disc, and their smaller average distance (as compared to other groups consisting
also of bulge objects) should make the measurements easier. 

There are three theoretical tracks in \figref{ss} shown for comparison 
representing models with 0.56, 0.60, and 0.64\,\msun\ central stars or
progenitor masses of 1, 3, and 4 \msun, respectively. They have been computed
combining the evolution of a central star adopted from models by Bl\"ocker
(1995) with a simple model of spherical nebula with a total nebular gas mass of
0.20 \msun, filling factor of 0.75 and an expansion velocity of 20 kms$^{-1}$.
These tracks have been overplotted only for reference purposes, and their exact
locations will change not only with variations in the nebular model parameters
but also with the details of stellar evolution as the evolutionary timescale.
The evolutionary timescale depends, for example, on the exact moment or thermal
pulse phase when the central star left the AGB, but also on the mass loss rate
from the central star during the AGB, post-AGB, and PNe phases (see
\cite{Gorny1994} for more details and \cite{GornyTylenda2000} for some examples
on how much track locations can be modified). Thus, the three simple theoretical
tracks displayed in \figref{ss} are probably not representative of the several
types of PNe with {\it Spitzer} spectra. For example, the high-metallicity
(solar/supra-solar) DC PNe are known to be non-spherical sources
\citep{GuzmanRamirez2011}, where a late thermal pulse could produce the observed
mixed chemistry \citep{PereaCalderon2009}. Also, several CC and OC dust subtypes
(e.g.\ \CCal, \OCam, and \OCamcr) display very low median metallicities that
could significantly affect their evolutionary timescales (i.e.\ their mass loss
rates). An exhaustive study of the space parameters affecting these theoretical
tracks is beyond the scope of this paper, but one needs to keep these
limitations in mind when comparing different PNe populations in \figref{ss}. In
any case, we have found useful the exploration of \figref{ss}, and some
interesting features have emerged. 

Analysing the locations of DC PNe in \figref{ss} shows that they seem to occupy
the largest space between the plotted 0.56 and 0.64\,\msun\ tracks with roughly
half of them above and below the 0.60\,\msun\ model.\footnote{It can be added
here that the six \DCcr\ PNe located closest to the plotted 0.64\msun\ track in
\figref{ss} all come from the \cite{PereaCalderon2009} sample (M1-31, H1-50,
M1-25, M1-40, M1-33, and Hb\,4). This sample is biased towards PNe with various
types of emission-line central stars, but among the six objects mentioned there
is diversity with two [WC], two wels, one unknown and one confirmed non-emission
line central star PNe.} This seems to be at variance with our conclusions from
Section~4 that the bulk of DC PNe probably represent the objects with the
highest stellar masses among the PNe analysed in this work, as deduced from
their enlarged N and He abundances. However, as underlined above, the
theoretical tracks shown in \figref{ss} may not be appropiate for these
non-spherical sources. Also, better agreement with higher mass central stars
could be reached if the evolution is faster than in the Bl\"ocker (1995) models
or if the masses of the nebulae are higher than the assumed value of 0.2 \msun.
If the apparent dichotomy (with roughly half of the sources above and below the
0.60\,\msun\ track) of DC PNe in \figref{ss} is real, then one may speculate
that DC PNe could evolve through two different high-metallicity channels, such
as from high-mass (M$\geq$3 \msun) HBB stars and from less massive (1$<$ M $<$ 3
\msun) stars with some extra mixing. We should add that among the \DCcr\ objects
closest to the low-mass 0.56\,\msun\ track in \figref{ss}, there are three
low-metallicity Galactic bulge PNe that are suspected of having undergone the
O-N cycle out of four with available S$_V$ surface brightness\footnote{These
objects are H\,1-43, M\,3-40, and M\,4-6. In fact, the S$_V$ for the last two
are upper limits, as is also the case for the fourth, H\,2-17, which is located
close to 0.60\msun\ track in \figref{ss}.}  (see also comments in Sect.~4.1).
Again, a detailed theoretical study of AGB evolution and nucleosynthesis at
supra-solar metallicity would be desirable to get some insight into possible
evolutionary channels for DC PNe. It is noteworthy that the \DCcr\ and \DCamcr\
PNe occupy practically the same locations in \figref{ss}, and it is hard to
reconcile it with the idea of an evolutionary link or transition from \DCamcr\
into \DCcr.

The situation of PNe with various forms of oxygen-rich dust in \figref{ss} is
totally different. This group is composed of three dust subtypes (\OCcr, \OCam,
and \OCamcr), and all of them clearly occupy distinct locations in the plot.
\OCamcr\ are found among objects with largest S$_{H\beta}$ and S$_V$, whereas
the opposite is true for \OCcr.  Taking the different metallicity for both OC
subtypes (see above) into account, this cannot be interpreted as an evolutionary
sequence but rather suggests that both groups originate from different
progenitor populations. However, the locations in \figref{ss} and dust
properties do not exclude a possible evolutionary link from \OCam\ to \OCamcr\
(see also below). Interestingly, the \OCam\ PNe span the whole range of S$_V$
surface brightness because they are placed very close to the 0.6 \msun\ track.
This suggests that \OCam\ PNe  probably originates in progenitor stars in a
narrow mass range. This is also the case for \CCal\ PNe (although the number of
objects with available data is  even smaller). For \CCar\ and the scarcer
\CCaral\ PNe, no statistically significant conclusions can be derived using
\figref{ss}. The situation of F type PNe with featureless {\it Spitzer} spectra
is not much better, although the three low surface-brightness objects plotted in
\figref{ss} confirm the natural assumption that F-type sources represent the
later stages of nebular life. There could be, however, some special objects, as
is probably the case for the high surface-brightness F-type outlier NGC\,6833.

To analyse the evolutionary status of the different {\it Spitzer} dust subtypes
by using larger samples, one can limit the discussion to the S$_{H\beta}$
distributions. This is because S$_{H\beta}$ is available for almost all PNe
(except the smallest star-like objects with unmeasured diameters). To
simultaneously preserve some information on the central stars, we decided to
divide PNe into those of low ionization (defined approximately as having \O++/O\
< 0.3), intermediate ionization (if \O++/O\ > 0.3 but \He++/He\ < 0.03), and
high ionization PNe (\He++/He\ > 0.03). This should qualitatively divide PNe
into groups with central stars below 35kK, between 35kK and 70kK, and above
70kK, respectively. The S$_{H\beta}$ distributions for the different {\it
Spitzer} dust subtypes and ionization states are presented in
\figref{hist_shb}.  The theoretical predictions of S$_{H\beta}$ distributions
for central stars of 0.56, 0.60, and 0.64\msun~(surrounded by the same nebula as
are used to construct tracks presented in \figref{ss}) can be found in Fig.\,7
of \cite{Gorny2010}. For a qualitative comparison it is enough to realize that,
owing to huge differences in the theoretical evolutionary timescales, it is
reasonable to expect only PNe with 0.56 and 0.60 \msun\ central stars in the low
and intermediate ionization groups and the latter will generally have higher
S$_{H\beta}$ values. Only in the high ionization group should any PNe with 0.64
\msun\ central stars  be expected. In this last high-ionization group, the PNe
with 0.60 \msun\ stars should have their peak-of-discovery probability, whereas
nebulae with 0.56 \msun\ stars should either be very faint or already too
dispersed to be noticed (very low S$_{H\beta}$).

In the left-hand panel of \figref{hist_shb} we present the S$_{H\beta}$
distributions for the DC, OC, and CC dust subtypes with the lowest ionization
(or central star temperatures). Even though we are investigating small PNe with
clear signs of the presence of dust that should therefore be expected to be at
earlier evolutionary phases than the {\it general} Galactic PN population, we
find very few low-ionization PNe in our samples. This can mean either that there
are not many PNe with low-mass central stars among them or that the theoretical
evolutionary timescales predicted by models of \cite{Bloecker1995} are too slow
compared to reality. The lack of OC PNe in the plots is particularly intriguing
since, based on their chemical abundances (e.g.\  low N/O ratios, see
Section~4), we expect them to come from the lowest mass AGB stars.

The middle panel of \figref{hist_shb} shows the distributions of S$_{H\beta}$
for intermediate ionization PNe divided into the several {\it Spitzer} dust
subtypes. Most of the objects we analysed fall into this category irrespective
of their dust type. \DCcr\ and \DCamcr\ both span a similar range in
S$_{H\beta}$, suggesting also similar central star masses. The S$_{H\beta}$ of
\OCcr\ and \OCam\ PNe also display a similar range among intermediate ionization
PNe, but they are shifted (roughly by 0.5 dex) towards lower S$_{H\beta}$
values. This S$_{H\beta}$ shift would be consistent with the \OCcr\ and \OCam\
PNe being less massive than the DC PNe, as expected from our chemical abundances
analysis (Section~4). This is not observed, however, for the \OCamcr\ PNe, which
display an apparent peak in their S$_{H\beta}$ distribution towards high values,
indicating higher central stars masses (as could have already been deduced from
their locations in \figref{ss}). Concerning the objects with carbon-based dust
among intermediate ionization PNe (bottom middle panel of \figref{hist_shb}),
there are no clear signs of any significant difference in their central star
masses.

The right-hand panel of \figref{hist_shb} presents S$_{H\beta}$ for high
ionization PNe in our samples. In general, there are two to three times fewer 
PNe in this ionization class than in the intermediate ionization group. This is
the natural consequence of our PNe samples being dominated by compact (and
relatively young) objects. By comparing the S$_{H\beta}$ distributions for high
ionization PNe depending on the different dust types and subtypes, we reach
similar conclusions to the intermediate ionization case. In particular, the OC
PNe (at least the \OCcr\ and \OCam\ ones) are shifted towards lower S$_{H\beta}$
values with respect to the DC PNe. This again suggests that DC PNe are more
massive than the \OCcr\ and \OCam\ PNe.\footnote{Our comparisons with
theoretical tracks have to remain qualitative at this stage. For example, for
every group of {\it Spitzer} dust type, about 1/4 of the analysed population
have high-ionization PNe.  Assuming the model predictions from
\cite{Bloecker1995} tracks and the model calculations presented in fig.\ 7 of
\cite{Gorny2010}, which agree with the  predicted probability for the 0.56\msun\
model. At the same time it is  discordanr with very low S$_{H\beta}$ predictions
from these calculations.  Model fine-tuning is, however, beyond the scope of the
present paper before new data on stellar V magnitudes are collected.}

\section{Discussion: A link between {\it Spitzer} dust properties,
chemical abundances, and evolutionary status}

This paper shows that the chemical abundances in our samples of Galactic PNe in
the disc and bulge are intimately related to their {\it Spitzer} dust
properties; i.e.\ the several {\it Spitzer} dust types/subtypes correspond to
different populations of PNe (with different average progenitor masses and
metallicities) in both Galactic environments. 

The Galactic disc population of compact and presumably young PNe is dominated by
low-metallicity OC and CC PNe, which are the descendants of very low-mass
($\sim$1$-$1.5 M$_{\odot}$) and low-mass (1.9 $\leq$ M $<$ 3 M$_{\odot}$) AGB
stars, respectively.  These populations of OC and CC PNe in the disc are
dominated by PNe with amorphous (OC$_{am}$) and aliphatic (CC$_{al}$) dust
features, respectively; i.e.\ those PNe with unevolved/unprocessed dust (see
below). DC and F PNe represent a significant fraction ($\sim$20\%) of the
Galactic disc PNe population. The DC PNe (both with amorphous and crystalline
silicates) represent a younger population of high-metallicity sources evolving
from relatively massive ($\sim$3$-$5 M$_{\odot}$, depending on the theoretical
modelling details) HBB AGB stars. High-metallicity and more massive (e.g.\ 
$\geq$ 6 M$_{\odot}$) HBB PNe are (almost) completely lacking in our Galactic
disc sample, probably because these sources evolve too fast (and are obscured by
dust) to be detected in the optical.  The disc F PNe are the most evolved
objects and seem to be a mix of low- and high-metallicity PNe. The
low-metallicity F PNe may be very low-mass and old PNe (with very low surface
brightness) similar to those found in the Galactic halo.

Interestingly, the Galactic bulge population of PNe (most of them also compact
and young) is dominated by a young population of high-metallicity DC PNe, with
their progenitors being relatively massive ($\sim$3$-$5 M$_{\odot}$) HBB AGB
stars. The OC sources also represent a significant fraction ($\sim$30\%) of the
Galactic bulge PNe population, but CC (and F) PNe are very uncommon. The
population of OC PNe in the bulge is dominated by those PNe with crystalline
silicate dust features (OC$_{cr}$), which also have higher metallicity than PNe
with amorphous silicate dust features (OC$_{am}$ and OC$_{am+cr}$). Although the
number of sources in each OC subtype is small, it seems that some OC$_{cr}$ PNe
are also massive, while the rest could be the result of the evolution of
high-metallicity and very-low mass ($\sim$1$-$1.5 M$_{\odot}$) AGB stars. The
low-metallicity and very low-mass ($\sim$1$-$1.5 M$_{\odot}$) OC PNe with
amorphous silicates - which dominate in the Galactic disc - are more scarce
towards the bulge.  The lack of CC bulge PNe is not surprising if we consider
that C-rich AGB stars are very rare in this Galactic environment  \citep[see
e.g.][and references therein]{Uttenthaler2007}.  The high Ar abundances that we
obtain for most bulge PNe indicate that high metallicity is an intrinsic
property towards the bulge.  Theoretical models predict that a high metallicity
can prevent the formation of C-rich AGB stars because a larger amount of C needs
to be dredged up and the TDU is less efficient \citep[e.g.][]{Karakas2002,
Marigo2013}. Our results in bulge PNe are consistent with the recent findings in
the bulge of M31 by \cite{Boyer2013}, which also show a complete lack of C-rich
AGB stars in this high-metallicity environment and suggests that there is a
metallicity threshold above which O-rich AGB stars cannot be converted into
C-rich ones.

Remarkably, our work shows that the double-dust chemistry (DC) phenomenon in PNe
mainly takes place in high-metallicity and relatively high-mass ($\sim$3$-$5
M$_{\odot}$) young PNe.  It is worth mentioning here that
\cite{Garcia-Rojas2013} also confirm the relatively massive nature (e.g.\  high
He and N) of DC PNe by obtaining precise chemical abundances in a small sample
of PNe with [WC]-type central stars (although their sample is biased towards
early [WC] types, see also below) from high resolution and high quality spectra.
The dominance of such a young population of PNe in the Galactic bulge naturally
explains the high detection rate of dual-dust chemistry found in this
environment (Perea-Calder\'on et al.  2009). Although dual-dust chemistry is
highly frequent in PNe with [WC]-type central stars, this phenomenon is also
seen in weak emission-line stars (wels), as well as in other PNe with normal
central stars (being neither [WC] nor wels) (Perea-Calder\'on et al.\  2009). 
Unfortunately, for a majority of DC PNe there is no information about their
central stars (e.g.\ [WC], wels, normal) but we find similar results to those
of  Perea-Calder\'on et al.\ (2009). The Perea-Calder\'on et al.\   sample of
bulge PNe is heavily biased towards emission-line objects; twelve [WC] and VL
PNe (all DC), seven wels (six DC and one OC), and 16 normal (nine DC, five OC,
and one CC).\footnote{The smaller Gutenkunst et al.\  (2008) sample of bulge PNe
also includes one [WC] PN and one wels (both with DC).}  In contrast, the
\cite{Stanghellini2012} sample of compact (presumably young) PNe seems to be
underpopulated in [WC] PNe: two [WC] PNe (both with DC), four wels (all of them
with OC), and 44 normal (24 DC, nine OC, four CC, and seven F) are found in the
bulge, while five [WC] and VL PNe (two DC and three CC), 14 wels (eight CC, four
OC, one DC, and one F), and 79 normal (21 CC, 27 OC, 14 DC, and 17 F) are found
in the disc. Thus, the last numbers suggest that the wels phenomenon is common
to very different types of PNe (DC, CC, and OC). In addition, PNe in the
\cite{Stanghellini2012} sample are very faint in the optical (i.e.\ fainter than
average PNe), which could explain why there are so few [WC] objects in their
sample; [WC] central stars can be associated primarily with brighter PNe
\citep[see][]{Gorny2009}.

DC PNe with O-rich amorphous dust features seem to be more common in the
Galactic disc than in the bulge, but it is not clear if this is just a selection
effect. One expects the evolution of the O-rich dust features to proceed
from amorphous silicates (in the AGB/post-AGB stage) to crystalline
silicates (in the PN phase) (Garc\'{\i}a-Lario \& Perea-Calder\'on 2003; see
also Garc\'{\i}a-Hern\'andez 2012 and references therein). Several models
for the silicates crystallization have been proposed: i) low-temperature
crystallization in circumbinary discs \citep[e.g.][]{Molster1999}, and ii)
high-temperature crystallization at the end of the AGB due to the strong
mass loss \citep[e.g.][]{Waters1996}. In the single-star evolutionary
framework, the high-metallicity and high-progenitor masses of DC PNe would
naturally explain an efficient crystallization of the silicates in their
circumstellar envelopes as a consequence of the higher mass loss rates (and
thus a higher amount of dust) experienced by these sources at the end of the
previous AGB phase. In addition, DC PNe with O-rich amorphous and
crystalline silicates dust features (DC$_{am+cr}$ and DC$_{cr}$) are, on
average, chemically indistinguishable (i.e.\ they probably evolve from the
same AGB progenitors). The higher median electron densities shown by the
DC$_{am+cr}$ PNe tentatively suggest that they may be the precursors of the
DC$_{cr}$ ones, in agreement with the expected AGB-PN evolution of the dust
features in O-rich circumstellar shells. However, this is not confirmed by
the S$_{H\beta}$ versus S$_V$ diagram presented in \figref{ss}, which
otherwise seems to suggest that DC PNe could evolve through an additional
high-metallicity channel, i.e. from less massive (1$<$ M $<$ 3 \msun) AGB
stars with some extra mixing. This should encourage theoretical modellers to
explore AGB evolution and nucleosynthesis at supra-solar metallicity for a
complete range of progenitors masses.  Finally, at present we cannot
completely discard the binary star scenario and the circumbinary discs as
the responsible mechanism for the crystallization of silicates around DC
PNe. This, however, seems to be less probable \citep[see e.g.][]{PereaCalderon2009, Miszalski2009}.

The simultaenous presence of PAHs and silicates in DC PNe is still unclear. 
Perea-Calder\'on et al.\  (2009) discuss several mechanisms to explain the
double-dust chemistry phenomenon in the Galactic bulge, and their most plausible
scenario is a final thermal pulse on the AGB (or just after), which may turn an
O-rich outflow into a C-rich one \citep[][]{Waters1998}. The crystallization of
amorphous silicates may be due to the enhanced mass loss, while the PAHs may
form the newly released carbon-rich material. On the other hand,
Guzm\'an-Ram\'{\i}rez et al.\  (2011) propose a chemical model where hydrocarbon
chains can form within O-rich gas through gas-phase chemical reactions, and they
conclude that the formation of PAHs in DC bulge PNe is best explained through
hydrocarbon chemistry in an ultraviolet (UV)-irradiated, dense torus. Both
proposals are based on the assumption that DC PNe in the bulge are low-mass
objects. However, we have shown here that the chemical abundances observed in DC
PNe (both in the bulge and disc) are consistent with their descendants being
high-metallicity and relatively massive ($\sim$3$-$5 M$_{\odot}$) AGB stars
experiencing HBB (see also Garc\'{\i}a-Rojas et al.  2013). For
intermediate-mass HBB stars, the strong mass loss may deactivate the HBB 
\citep[see e.g.][and references therein]{Garcia-Hernandez2006b} at the end of
the AGB phase.  The high temperature annealing may lead to the silicates
crystallization, while carbon, again allowed to be dredged up to the stellar
surface, may be used for the PAH formation. Indeed, recent exploratory synthetic
evolution calculations of solar-metallicity HBB AGB stars with delayed
superwinds (i.e.\ allowing for TPs near the tip of the AGB once HBB has ceased;
see Karakas et al.\  2012 for more details) predict that these stars could be
converted to C-rich ones.  We note that the chemical model presented by
Guzm\'an-Ram\'{\i}rez et al.\ (2011) to explain PAH formation in an O-rich
environment may be still applicable to intermediate-mass HBB stars.

Garc\'{\i}a-Rojas et al.\ (2013) have very recently reported precise C/O nebular
ratios in eight DC PNe with early [WC]-type central stars. Five DC PNe in their
sample (four DC$_{cr}$ and one DC$_{am+cr}$) have O-rich (C/O $<$ 1) nebulae,
but three of them (Cn 1-5, M 1-32, and NGC 2867, all DC$_{cr}$) have C-rich
nebulae. Interestingly, the DC PNe with O-rich nebulae are those with the
typical DC {\it Spitzer} spectrum, showing very weak PAH bands and
crystalline/amorphous silicates, while the C-rich ones display very unusual {\it
Spitzer} spectra with strong PAH bands (and a strong bump emission around
$\sim$24 $\mu$m) and very weak crystalline silicate features. Cn 1-5 is the most
extreme case, displaying the lowest C/O nebular ratio of 1.29 and the highest He
(11.21) and N enrichments (N/O=0.45), which suggest a progenitor star that is
even more massive than 5 M$_{\odot}$.  Further precise determinations of the C/O
ratios (e.g. based on optical recombination lines as in Garc\'{\i}a-Rojas et al.
have recently been done from high-resolution and high-quality spectra) in a
complete sample of DC PNe would be very useful for learning about the dominant
mechanism for PAH formation (HBB deactivation and/or hydrocarbon chemistry
within O-rich shells), as well as for a fine-tuning (e.g. the specific
conditions for the C-star conversion) of the synthetic evolution calculations of
high-metallicity HBB AGB stars with delayed superwinds.

The OC PNe with O-rich amorphous dust features are also more frequent in the
Galactic disc than in the bulge. In particular, \cite{Stanghellini2012} have
already pointed out that the OC$_{cr}$ disc PNe show lower Galactic latitudes
(b) than the OC$_{am}$ disc PNe (see below; Fig.\ 4). This would be consistent
with the OC$_{cr}$ PNe being more massive (and/or more metal rich) than the
OC$_{am}$ ones. Here we have shown that OC$_{cr}$ PNe (in the disc and bulge)
are more metal rich and slightly more massive than OC PNe with amorphous dust
features (OC$_{am}$ and OC$_{am+cr}$). In the single star scenario, the
metal-rich and probably more massive character of OC$_{cr}$ PNe would again be
consistent with a more efficient silicates crystallization as a consequence of
the higher mass loss rates (and dust production) at the tip of the AGB. The
metal-poor (and probably less massive) OC PNe with amorphous dust features would
experience much lower mass loss rates (and thus smaller amount of dust) in the
AGB phase. It is nevertheless not clear (i.e. not confirmed by our statistical
tests) that OC$_{am}$ and OC$_{am+cr}$ could be evolutionarily linked (e.g.\ the
OC$_{am}$ PNe being the precursors of the OC$_{am+cr}$ ones) or if they reflect
slightly different progenitor masses and metallicities. 

The CC PNe (being absent in the Galactic bulge, see above) with aliphatic dust
features (CC$_{al}$) dominate the population of compact (and young) C-rich PNe
in the Galactic disc. From the AGB to the PN stage, the evolution of C-rich dust
proceeds from the amorphous state to aliphatic, and then to PAHs (e.g.\
Garc\'{\i}a-Lario \& Perea-Calder\'on 2003; Garc\'{\i}a-Hern\'andez 2012); the
dust grains evolve from small clusters of molecules (e.g.\ PAH clusters) to
ionized PAHs. Thus, the apparent lack of CC$_{ar}$ disc PNe (i.e., with aromatic
PAH-like features) is explained by the dominant evolutionary stage of the dust
grains around the disc population of C-rich PNe. Our chemical analysis indicates
that most CC$_{al}$ PNe are the descendants of rather low-metallicity C-rich AGB
stars. Low-metallicity environments favour the detection of unprocessed dust
grains in PNe {\it Spitzer} spectra \citep[see][]{Stanghellini2007}.  

The transformation of aliphatic groups to aromatic ones \citep[e.g.\ as a
consequence of the UV radiation field of the central star][]{Kwok2001} is not
very efficient (and more slowly because of the low metallicity), and we detect
the early aliphatic stages of the circumtellar carbonaceous grains or the
precursors of the PAH-like molecules that are usually seen in more evolved PNe. 
The few higher metallicity CC$_{al}$ PNe in our disc sample should be very young
objects where there is not enough time for an efficient aliphatic to aromatic
dust conversion. Furthermore, the CC$_{ar}$ disc PNe are more evolved objects
(i.e. with higher median electron density), although they are not evolutionarily
linked to the CC$_{al}$ ones. CC$_{ar}$ PNe are higher metallicity (nearly
solar) and rather peculiar PNe. At least half of them seem to be truly DC PNe
where the weak crystalline silicates have escaped detection from {\it Spitzer}.
Interestingly, the two CC$_{ar+al}$ PNe with fullerenes also have low
metallicity and electron densities lower than the bulk of CC$_{al}$ objects,
suggesting that they are evolutionary linked (e.g.\ CC$_{ar+al}$ PNe are
slightly more evolved), as expected from the C-rich dust evolution during the
transition phase from AGB stars to PNe.

\section{Summary and conclusions}

We have combined new low-resolution (R$\sim$800) optical spectra with recent
data in order to construct representative samples of Galactic disc and bulge PNe
(mostly compact and presumably young) with {\it Spitzer} spectra. Depending on
the nature of the dust features - C-rich, O-rich, and both C- and O-rich dust
features (or double chemistry) -  seen in their {\it Spitzer} spectra, the
Galactic disc and bulge PNe are classified into four major dust types (oxygen
chemistry or OC, carbon chemistry or CC, double chemistry or DC, featureless or
F) and subtypes (amorphous and crystalline, and aliphatic and aromatic) and
their Galactic distributions are presented. Nebular gas abundances of He, N, O,
Ne, S, Cl, and Ar, as well as plasma parameters, were derived in a homogeneous
way by using the classical empirical method. We studied the median chemical
abundances in the Galactic disc and bulge PNe depending on their {\it Spitzer}
dust types and subtypes and compared them with the homogeneous dataset of AGB
nucleosynthesis predictions by Karakas (2010). Also, we analysed the nebular
properties among the several types and subtypes of PNe and discussed a possible
link between the {\it Spitzer} dust properties, chemical abundances, and
evolutionary status. 

The main results of our work can be summarized as follows.

\begin{enumerate}

\item{The several {\it Spitzer} dust types and subtypes of PNe (with the
exception of F PNe) are distributed differently between the Galactic disc
and bulge regions. In particular, DC PNe are less common in the Galactic
disc than OC and CC PNe but clearly dominate the Galactic bulge region,
which otherwise show an almost complete lack of CC PNe. Both OC and DC PNe with
amorphous silicates (OC$_{am}$ and DC$_{am+cr}$) are more uncommon in the
Galactic bulge than in the disc, while the opposite is seen for OC and DC PNe
with crystalline silicates (OC$_{cr}$ and DC$_{cr}$). CC PNe with aliphatic
features (CC$_{al}$) completely dominate in the Galactic disc. DC PNe in the
Galactic disc are mainly located towards the Galactic centre with no significant
difference between the DC$_{am+cr}$ and DC$_{cr}$ subtypes. OC PNe with
amorphous silicates (OC$_{am}$) and featureless (F) PNe are located further away
from the Galactic plane than the other types/subtypes of PNe.}

\item{The median abundance pattern (mainly N/H, He/H, Ar/H, and N/O) of DC PNe
is statistically different from that of OC PNe in both Galactic environments
with DC PNe and OC PNe in the Galactic disc and bulge sharing almost the same
abundance pattern. CC disc PNe (lacking in the bulge region) display an
abundance pattern that is quite similar to that of the OC disc PNe, with the
exception of S, which is found to be significantly more depleted in the CC
objects. Our dissection of the {\it Spitzer} dust subtypes vs.\ the nebular gas
abundances shows that i) the DC subtypes (DC$_{cr}$ and DC$_{am+cr}$) are
indistinguishable, and they show an abundance pattern identical to the major DC
type; ii) the OC subtypes with amorphous silicates (OC$_{am}$ and OC$_{am+cr}$)
are chemically very similar, dominating the abundance pattern observed in the
major OC type, while the less numerous OC$_{cr}$ PNe are more metal rich; iii)
the CC subtypes with aliphatic dust features (CC$_{al}$ and CC$_{ar+al}$)
display similar abundances, dominating the chemical pattern observed in the
major CC type, while CC$_{ar}$ PNe, being more metal-rich, are also scarcer; iv)
a few OC$_{cr}$ and about half ($\sim$50\%) of CC$_{ar}$ PNe display nebular gas
abundances that are almost identical to those seen in DC PNe, suggesting that
they are related objects.}

\item{The several {\it Spitzer} dust types and subtypes correspond to different
populations of PNe (with different average progenitor masses and metallicities)
in both Galactic environments. We note that although the progenitor masses that
we infer from comparisons with the AGB nucleosynthesis predictions may be rather
uncertain, our finding of different average progenitor masses and metallicities
among the several {\it Spitzer} dust types/subtypes is assured. In summary:

{\bf i)} DC PNe both with amorphous and crystalline silicates (DC$_{am+cr}$ and
DC$_{cr}$) are high metallicity (solar/supra-solar) sources evolving from
relatively massive $\sim$3--5 M$_{\odot}$ AGB stars experiencing HBB. They show
the highest median He abundances and N/O abundance ratios, which are consistent
with solar metallicity (z$\sim$0.02) $\sim$5 M$_{\odot}$ HBB AGB stars. However,
other possible evolutionary channels cannot be discarded now. In addition,  it
is still not clear that \DCamcr\ PNe could evolve towards the \DCcr\ stage.
Curiously, there are a few DC PNe in the bulge with some indications of the O-N
cycle activation, which are not present in the disc.  

{\bf ii)} OC PNe with amorphous silicates (OC$_{am}$) are low-metallicity
sources with median He abundances and N/O abundance ratios that agree well
with predictions of low-metallicity (z$\sim$0.008) and very low-mass ($\sim$1
M$_{\odot}$) AGB stars. OC PNe with both amorphous and crystalline silicates
(OC$_{am+cr}$) are chemically similar to the OC$_{am}$ ones, but we have found
tentative hints (i.e.\ in the S$_{H\beta}$ vs.\ S$_V$ diagram and S$_{H\beta}$
distributions), suggesting that OC$_{am+cr}$ PNe could be more massive than
OC$_{am}$ PNe. The bulk of OC PNe with only crystalline silicates (OC$_{cr}$)
are higher metallicity (nearly solar) objects with their median He
abundances and N/O abundance ratios consistent with solar metallicity
(z$\sim$0.02) $\sim$1--1.5 M$_{\odot}$ AGB stars.

{\bf iii)} CC PNe with aliphatic features (CC$_{al}$ and CC$_{ar+al}$) (mostly
located in the Galactic disc) are low-metallicity (z$\sim$0.008) objects that
probably evolve from relatively low-mass (1.9 $\leq$ M $<$ 3 M$_{\odot}$) AGB
stars. The best fit to the observed abundances is given by the models for the
low-metallicity (z$\sim$0.008) $\sim$1.9 M$_{\odot}$ AGB star.}

\item{In both Galactic regions (bulge and disc), most of the PNe dust classes
analysed here show a similar intermediate level of ionization, probably reflecting
that our PNe samples are dominated by relatively young PNe. The only exceptions
are the featureless F-type PNe that represent more advanced evolutionary stages.
There is an apparent dichotomy of DC PNe in the S$_{H\beta}$ vs.\ S$_V$ diagram
that could suggest an additional high-metallicity evolutionary channel for these
objects; e.g.\ less massive (1 $<$  M $<$ 3 M$_{\odot}$) with some kind of
extra mixing. Also, there is tentative evidence that OC$_{am+cr}$ PNe could be
slightly more massive than the OC$_{am}$ ones. This evidence, however,
comes from comparisons with simple theoretical tracks that may not be
representative of the several types of PNe with {\it Spitzer} spectra. More
interesting is that OC and CC with unevolved dust (OC$_{am}$ and
CC$_{al}$ PNe, respectively) seem to originate in progenitor masses in a
narrow mass range, in agreement with the nucleosynthesis predictions.} 

\end{enumerate} 

In short, our most remarkable result is that DC PNe, both with amorphous and
crystalline silicates, display high-metallicity (solar/supra-solar) and the
highest He abundances and N/O ratios, indicating that they likely evolve from
relatively massive ($\sim$3--5 M$_{\odot}$) HBB AGB stars; however, PNe with
O-rich and C-rich unevolved dust (amorphous and aliphatic) evolve from subsolar
metallicity (z$\sim$0.008) and lower mass ($<$3 M$_{\odot}$) AGB stars. Although
we have obtained very interesting results by studying more balanced samples of
the Galactic disc and bulge PNe, more optical spectroscopic observations would
be desirable. In particular, the available data for some {\it Spitzer} OC
(OC$_{cr}$ and OC$_{am+cr}$) and CC (CC$_{ar}$ and CC$_{ar+al}$) dust subtypes
are still very scarce, so more chemical abundances would be useful to discard or
confirm any possible evolutionary link among the several {\it Spitzer} dust
subtypes. More theoretical efforts (e.g.\ a detailed study of AGB evolution and
nucleosynthesis at supra-solar metallicity) are also encouraged.

\begin{acknowledgements} This work is based on observations made with the
Telescopio Nazionale Galileo operated on the island of La Palma by the Centro
Galileo Galilei of INAF (Istituto Nazionale di Astrofisica) at the Spanish
Observatorio del Roque de Los Muchachos of the Instituto de Astrof\'{\i}sica de
Canarias. We acknowledge A. I. Karakas for providing us with additional
unpublished model predictions. We also acknowledge A. I. Karakas, M. Lugaro,
and J. Garc\'{\i}a-Rojas for very useful discussions as well as L. Stanghellini
for sharing her Spitzer data on compact Galactic PNe with us. S.K.G.
acknowledges support from grant N203 511838 of the Science and High Education
Ministry of Poland. D.A.G.H. acknowledges support for this work provided by the
Spanish Ministry of Economy and Competitiveness under grants AYA$-$2011$-$27754
and AYA$-$2011$-$29060.  
\end{acknowledgements}

\bibliographystyle{aa}
\bibliography{tng_paper}


\clearpage

\begin{appendix}

\section{New emission-line central stars}

\begin{figure}
\resizebox{\hsize}{!}{\includegraphics{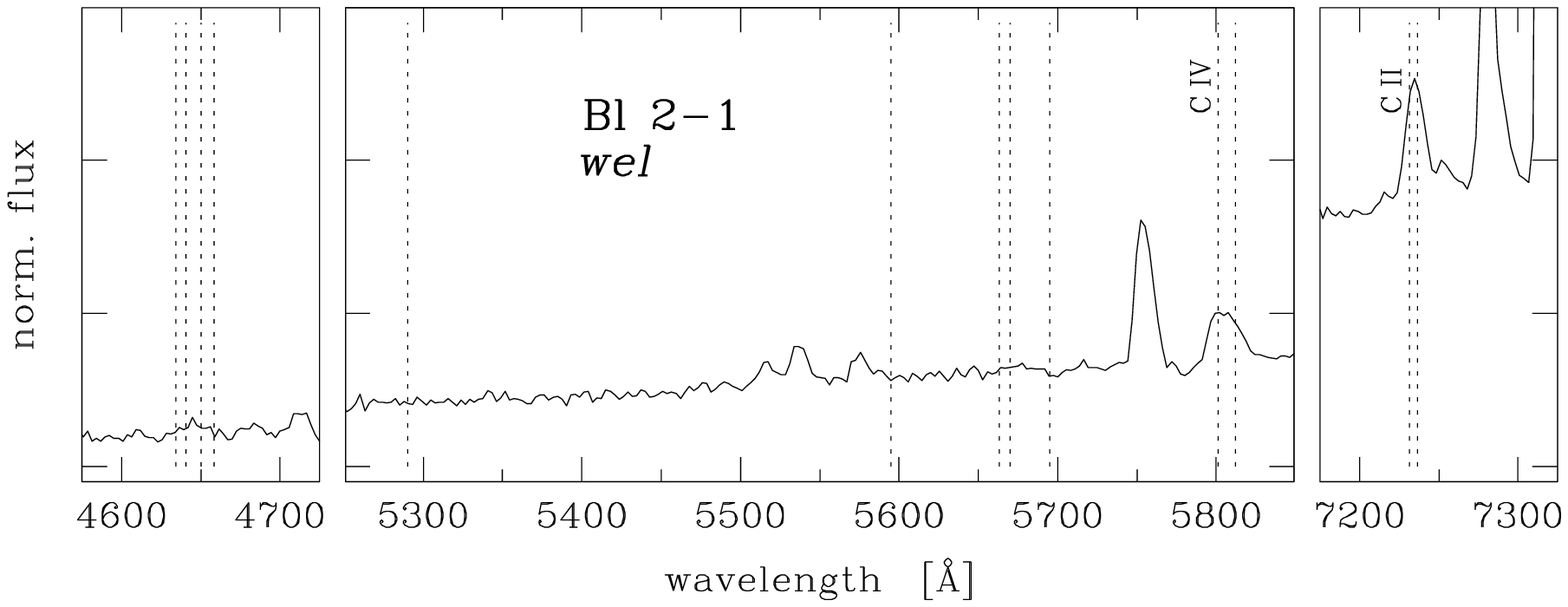}}
\resizebox{\hsize}{!}{\includegraphics{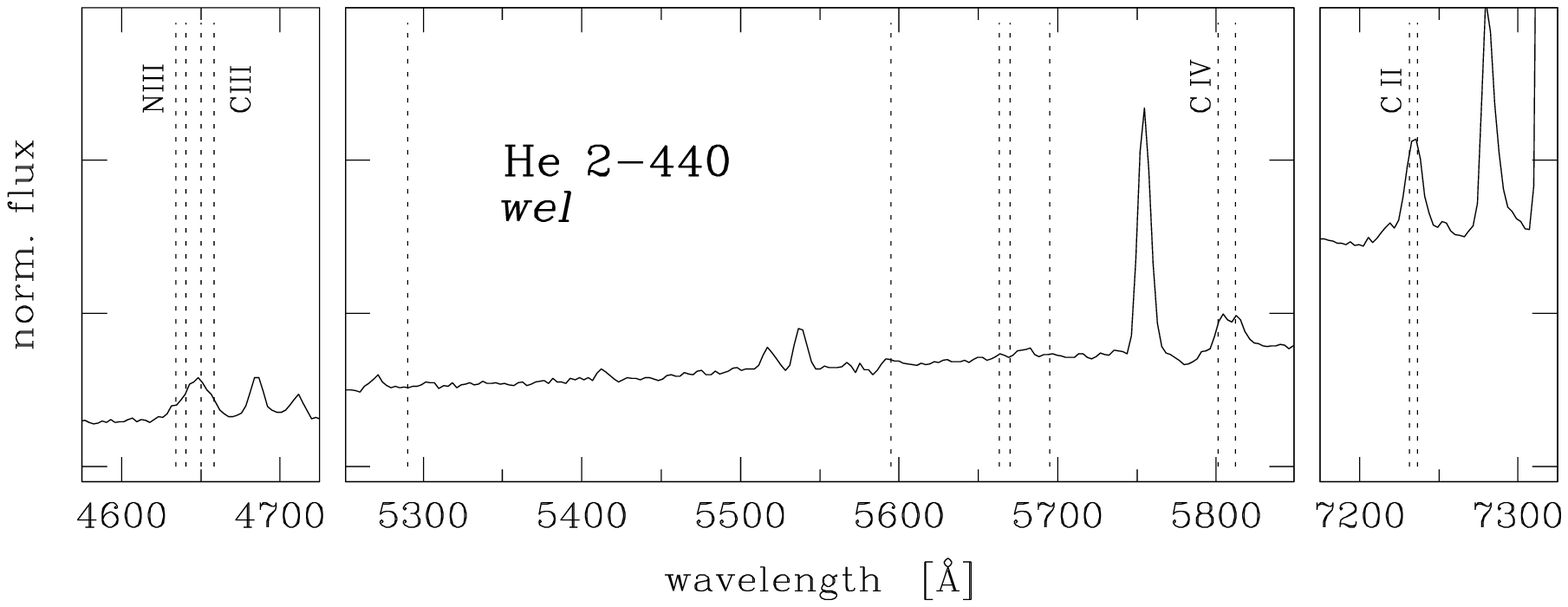}}
\resizebox{\hsize}{!}{\includegraphics{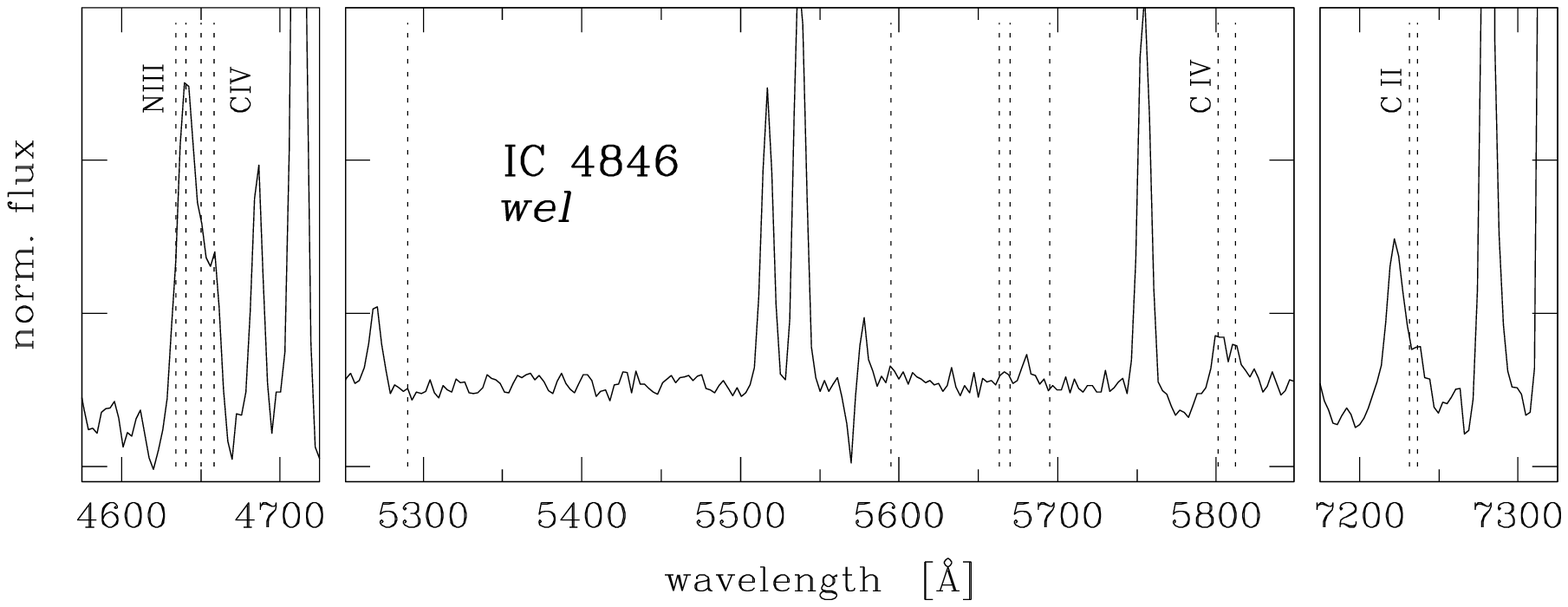}}
\resizebox{\hsize}{!}{\includegraphics{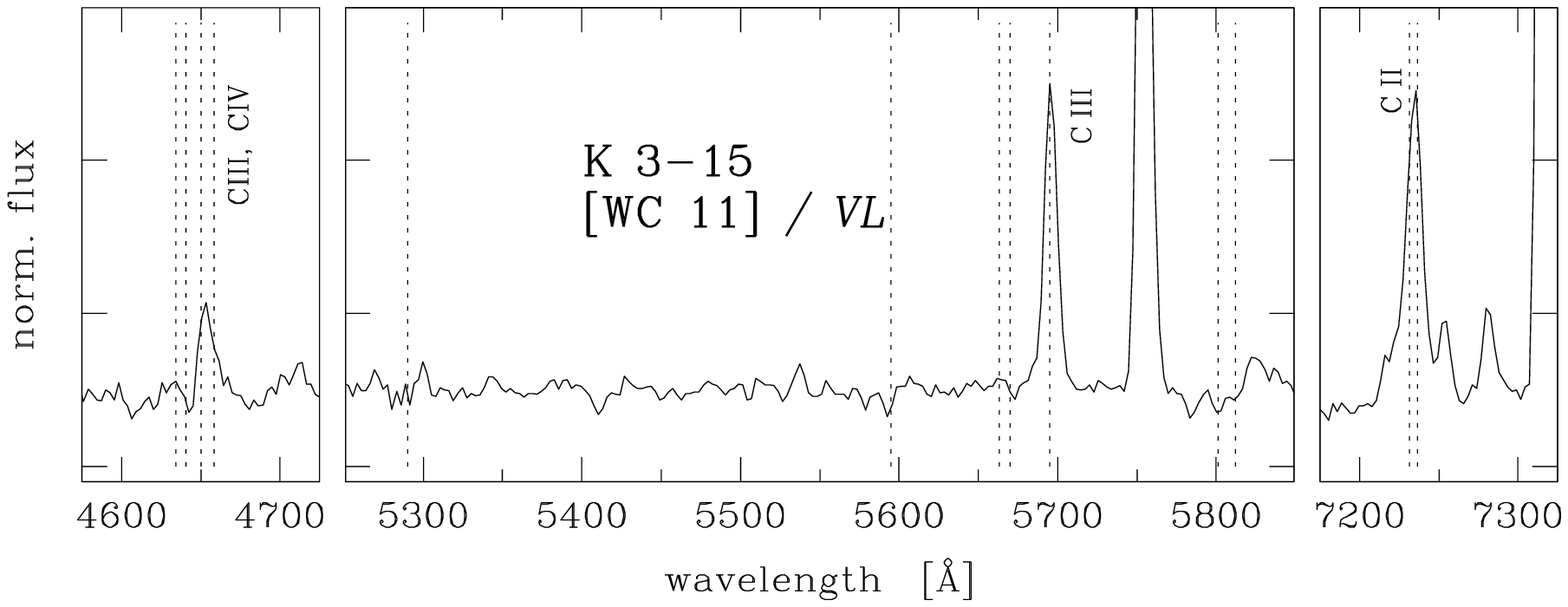}}
\resizebox{\hsize}{!}{\includegraphics{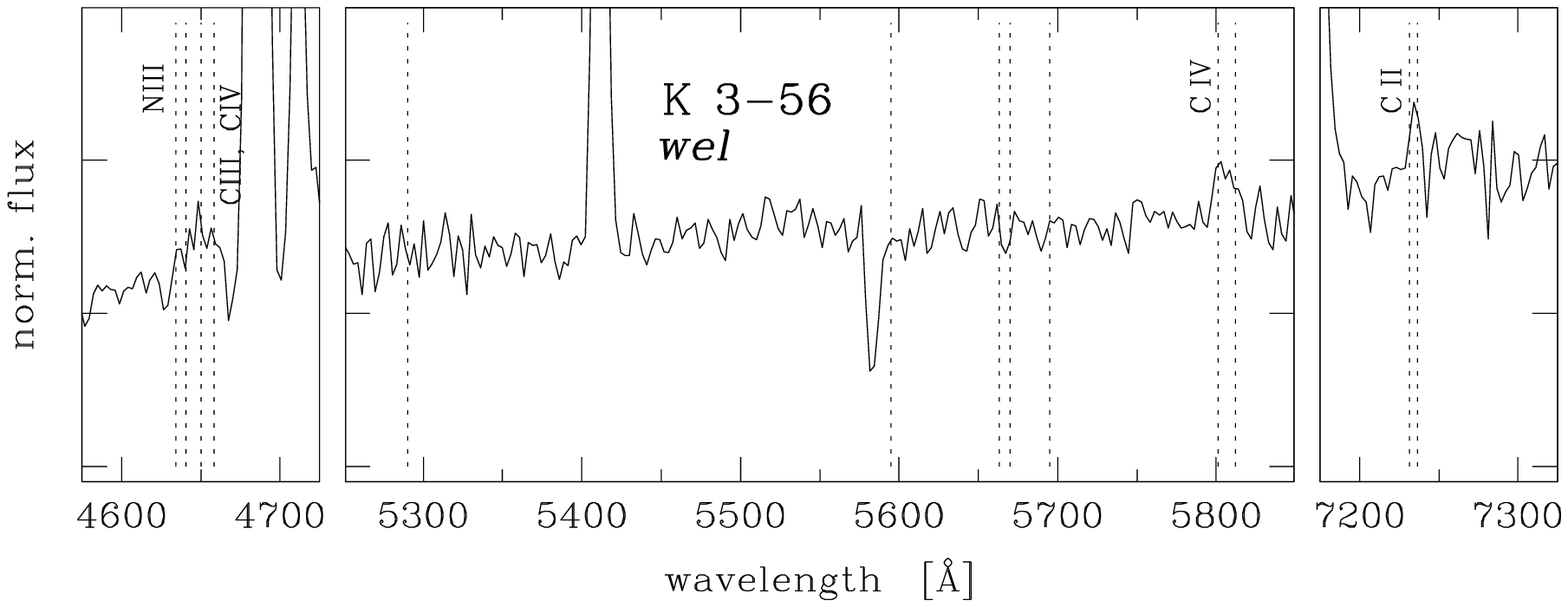}}
\resizebox{\hsize}{!}{\includegraphics{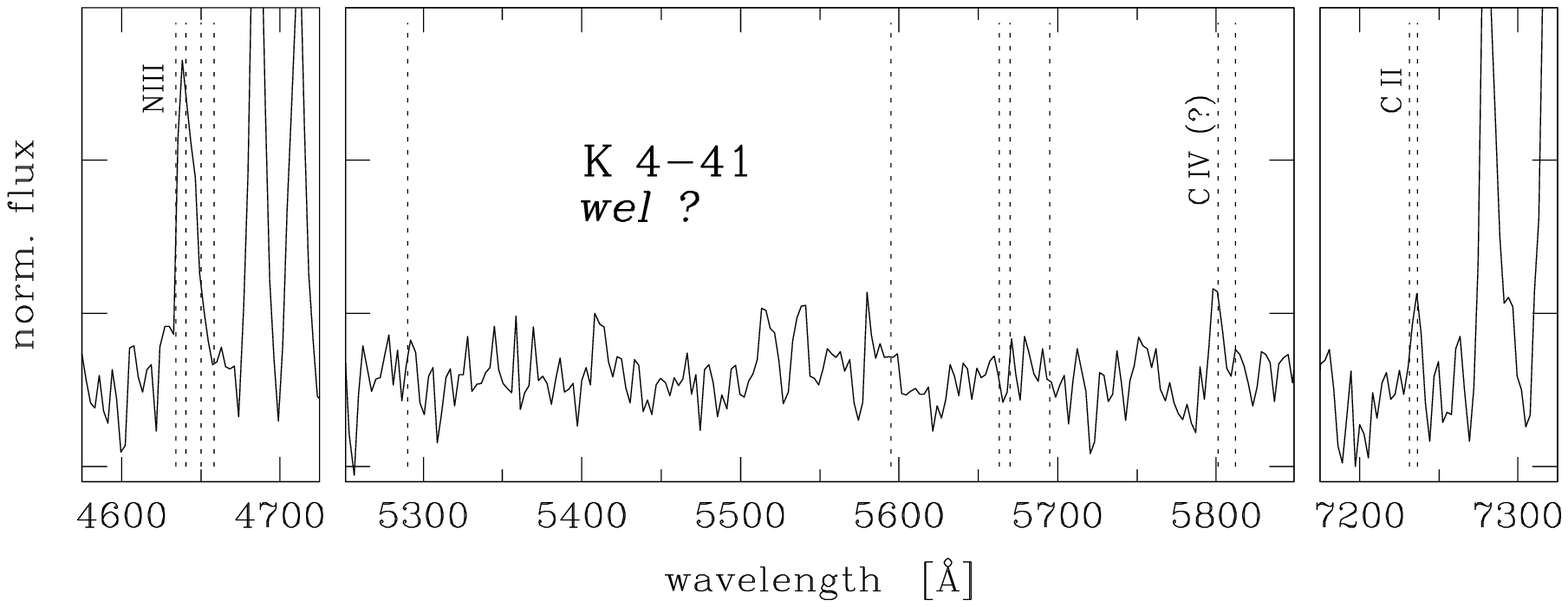}}
\caption[]{
 Spectra of the new PNe with emission-line central stars. Expected
 locations of characteristic features are indicated by dotted lines and
 labelled with the appropriate ion name if present in the spectra.
}
\label{spectra_wel}
\end{figure}

A considerable fraction of the Galactic PNe population has emission-line central
stars of different types \citep{Weidmann2011}. We used our new TNG DOLORES
observations to look for their presence in our sample. We found five objects
(Bl\,2-1, He\,2-440, IC\,4846, K\,3-56, and K\,4-41) to be possible WELs-type
objects \citep[see][]{Tylenda1993}. Only one object (K\,3-15) could tentatively
be classified as a Wolf--Rayet type [WC\,11] central star. Taking the very
low-ionization state of this nebula into account we instead classified it as a
VL-type object - a member of a class of emission-line stars recently introduced
by \cite{Gorny2009}.  The full optical spectrum of K\,3-15 (low-ionization
nebula) and also of K\,3-56 (high-ionization nebula) have been presented in
\figref{spectra_ex}.  In \figref{spectra_wel} we present details of the spectra
of all new emission-line central star candidates in our observed sample. In this
plot the expected locations of the characteristic stellar emission lines are
indicated with dotted lines, and if positively identified they are then marked
with the name of the appropriate ion.

\section{Spectra of the three suspected symbiotic stars}

The full optical spectra of the three suspected symbiotic stars (CTSS 2, K 4-57,
and Th 4-1) are displayed in Figure B.1. Their {\it Spitzer} IR spectra (with
the unusual presence of hot dust emission) are very different from the other
Galactic PNe studied here, and their optical spectra are suggestive of a
possibly symbiotic classification. For example, the 2MASS colours of PN K 4-57
are consistent with a D-type symbiotic PN. A detailed analysis of this small
group of symbiotic stars will be presented elsewhere.

\begin{figure*}
\resizebox{\hsize}{!}{\includegraphics{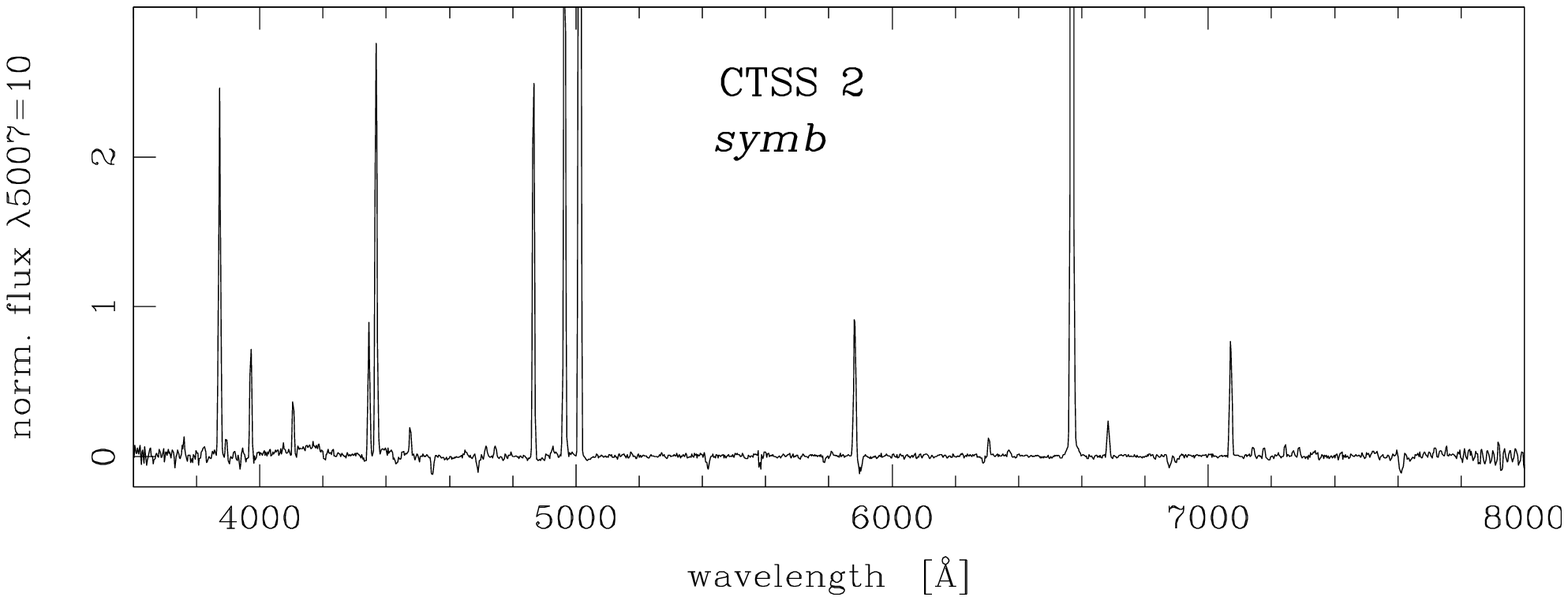}}
\resizebox{\hsize}{!}{\includegraphics{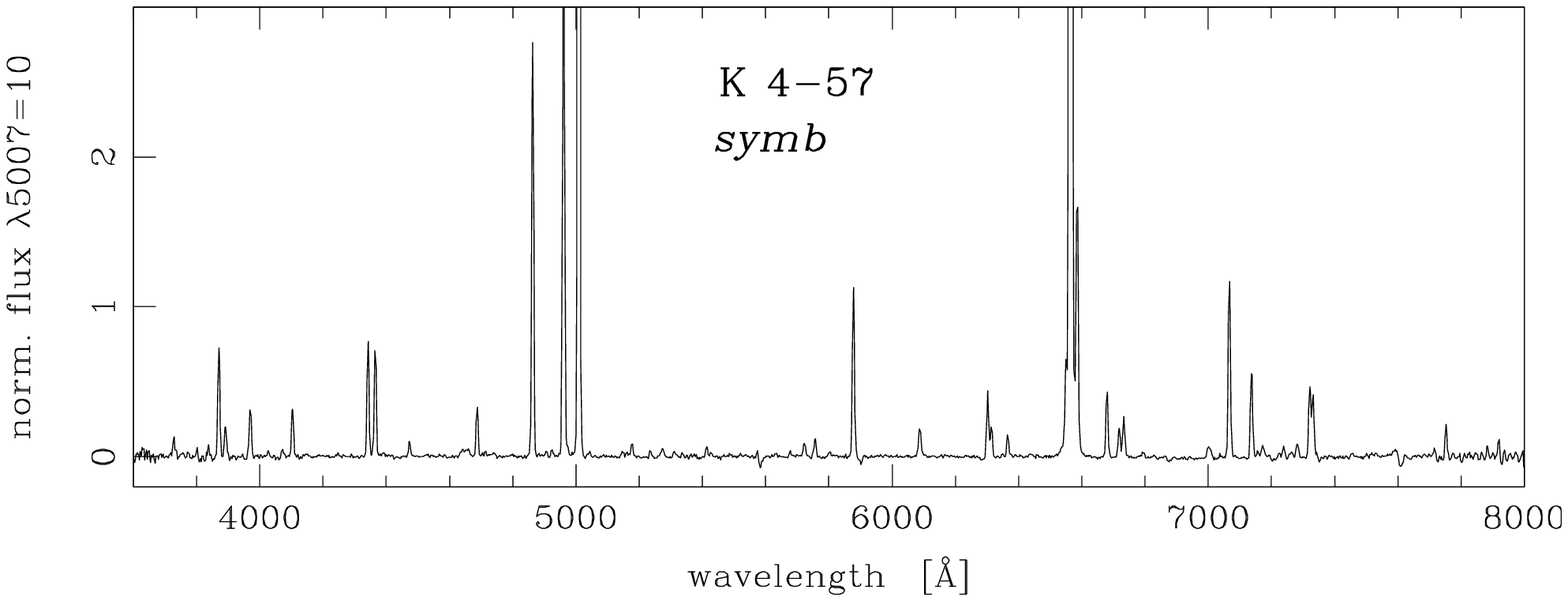}}
\resizebox{\hsize}{!}{\includegraphics{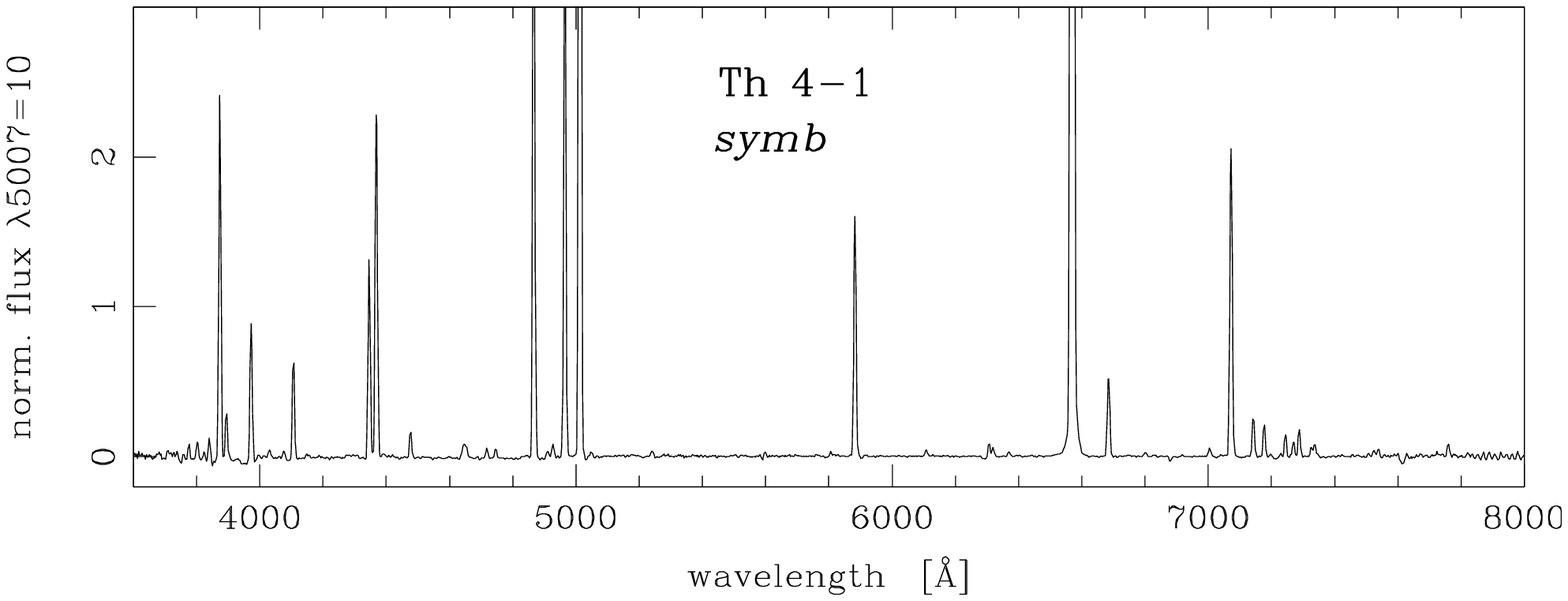}}
\caption[]{
 Spectra of the three suspected symbiotic stars.
}
\label{spectra_symb}
\end{figure*}

\end{appendix}

\clearpage

\setcounter{table}{1}
\begin{table*}
\caption{  
  Observed line fluxes and dereddened intensities on the scale H$\beta$=100 of the main nebular lines. Meaning of the symbols:
  (:) - line error at 20\%; (;) - line error at 40\%;
  (p) - line present but not measurable; (?) - uncertain identification;
  (S) - saturated; (r) - recalculated using doublet pair; 
  (b) - significantly blended with other line; (B) - blend or sum of two lines; 
  (c) - additional reddening correction applied.
}
\begin{tabular}{     l @{\hspace{0.60cm}}
                     l @{\hspace{0.80cm}}
                     r r @{\hspace{0.80cm}} r r @{\hspace{0.80cm}} r r @{\hspace{0.80cm}} r r}
\hline
 ~          &      & \multicolumn{2}{c}{107.4-02.6}& \multicolumn{2}{c}{097.6-02.4}& \multicolumn{2}{c}{095.2+00.7}& \multicolumn{2}{c}{079.9+06.4}\\
 ~          &      & \multicolumn{2}{c}{K 3-87    }& \multicolumn{2}{c}{M 2-50    }& \multicolumn{2}{c}{Bl 2- 1   }& \multicolumn{2}{c}{K 3-56    }\\
            &      &               &               &               &               &               &               &               &               \\
            &      & F($\lambda$)  & I($\lambda$ ) & F($\lambda$)  & I($\lambda$ ) & F($\lambda$)  & I($\lambda$ ) & F($\lambda$)  & I($\lambda$ ) \\
            &      &               &               &               &               &               &               &               &               \\
 3727 & {[O\ II]}  &               &               &    12.34      &    22.72  c   &      9.38  :  &    48.44 :c   &               &               \\
 3869 & {[Ne III]} &     21.88     &    46.79  c   &    66.30      &   113.19  c   &     17.34     &    72.67  c   &    18.79      &    46.30  c   \\
 4068 & {[S II]}   &               &               &     0.33  ;   &     0.46 ;c   &               &               &               &               \\
 4102 &   H I      &     14.36     &    26.23  c   &    17.37      &    26.07  c   &      8.60     &    25.80  c   &    13.59      &    26.29  c   \\
 4340 &   H I      &     29.40     &    47.30  c   &    35.21      &    47.09  c   &     22.67     &    52.03      &    30.44      &    47.38  c   \\
 4363 & {[O III]}  &      7.18     &    11.11      &    10.99      &    14.67      &      2.89     &     6.44      &     7.27      &    11.71      \\
 4471 &   He I     &               &               &     3.59      &     4.50      &      3.15     &     5.75      &               &               \\
 4686 &   He II    &     84.27     &    97.59      &    12.18      &    13.46      &      0.59  ;  &     0.78 ;    &    90.25      &   105.93      \\
 4711 & {[Ar IV]}  &      7.18  :  &     8.17 :    &     4.94      &     4.72      &      1.07  :  &          :    &     7.51      &     8.61      \\
 4725 & {[Ne IV]}  &               &               &               &               &               &               &     1.42  ;b  &     1.59 ;b   \\
 4740 & {[Ar IV]}  &      4.79  :  &     5.32 :    &     3.49      &     3.75      &      0.76  :  &     0.97 :    &     6.09      &     6.83      \\
 4861 &   H I      &    100.00     &   100.00      &   100.00      &   100.00      &    100.00     &   100.00      &   100.00      &   100.00      \\
 4959 & {[O III]}  &    209.23     &   193.41      &   427.13      &   408.92      &    380.56     &   327.03      &   181.74      &   168.30      \\
 5007 & {[O III]}  &    626.15     &   558.00      &  1269.09      &  1190.65      &   1225.81     &   981.25      &   566.31      &   506.08      \\
 5200 & {[N I]}    &               &               &     0.23  ;   &     0.17 ;    &      1.04  ;  &     0.62 ;    &               &               \\
 5515 & {[Cl III]} &               &               &     0.72  :   &     0.54 :    &      0.93  ;  &     0.37 ;    &               &               \\
 5537 & {[Cl III]} &               &               &     0.56  :   &     0.46 :    &      1.66  :  &     0.68 :    &               &               \\
 5755 & {[N II]}   &               &               &     0.47  ;   &     0.35 ;    &      6.35     &     1.84      &               &               \\
 5876 &   He I     &      5.30  ;  &     2.58 ;    &    20.38  :   &    13.68 :    &     68.39     &    16.98      &     3.96  :   &     1.98 :    \\
 6300 & {[O I]}    &               &               &     0.91  ;   &     0.53 ;    &     28.10  ;  &     4.49 ;    &               &               \\
 6312 & {[S III]}  &      3.76  ;  &     1.46 ;    &     2.79  b   &     1.65  b   &             b &           b   &     2.19  ;   &     0.87 ;    \\
 6548 & {[N II]}   &               &               &     4.61  b   &     2.54  b   &     65.69  ;  &     8.30 ;    &               &               \\
 6563 &   H I      &    818.29     &   278.65      &   512.49      &   282.14      &   2312.76     &   288.45      &   794.27      &   277.47      \\
 6583 & {[N II]}   &               &               &    27.25      &    14.89      &    576.90  :b &    70.63 :b   &     4.96  b   &     1.73  b   \\
 6678 &   He I     &      3.25  ;  &     1.04 ;    &     6.75      &     3.64      &     35.60     &     4.03      &     3.55      &     1.16      \\
 6716 & {[S II]}   &               &               &     3.70      &     1.96      &      7.22     &     0.79      &     1.36  ;   &     0.46 ;    \\
 6730 & {[S II]}   &               &               &     5.50      &     2.91      &     27.93     &     3.03      &     1.36  ;   &     0.46 ;    \\
 7006 & {[Ar V]}   &      3.93     &     1.11      &               &               &               &               &     6.68      &     1.97      \\
 7136 & {[Ar III]} &     11.97     &     3.27      &    22.47      &    10.94      &    179.46     &    14.52      &    12.77      &     3.59      \\
 7320 & {[O II]}   &               &               &     1.89  :   &     1.64 :B   &    179.99  B  &    12.69  B   &               &               \\
 7330 & {[O II]}   &               &               &     1.61  :   &     1.64 :B   &    179.99  B  &    12.69  B   &               &               \\
\hline
\end{tabular}
\end{table*}

\setcounter{table}{1}
\begin{table*}
\caption{  
  Continued.
}
\begin{tabular}{     l @{\hspace{0.60cm}}
                     l @{\hspace{0.80cm}}
                     r r @{\hspace{0.80cm}} r r @{\hspace{0.80cm}} r r @{\hspace{0.80cm}} r r}
\hline
            &      & \multicolumn{2}{c}{069.2+02.8}& \multicolumn{2}{c}{068.7+01.9}& \multicolumn{2}{c}{060.5+01.8}& \multicolumn{2}{c}{052.9+02.7}\\
            &      & \multicolumn{2}{c}{K 3-49    }& \multicolumn{2}{c}{K 4-41    }& \multicolumn{2}{c}{He 2-440  }& \multicolumn{2}{c}{K 3-31    }\\
            &      &               &               &               &               &               &               &               &               \\
            &      & F($\lambda$)  & I($\lambda$ ) & F($\lambda$)  & I($\lambda$ ) & F($\lambda$)  & I($\lambda$ ) & F($\lambda$)  & I($\lambda$ ) \\
            &      &               &               &               &               &               &               &               &               \\
 3727 & {[O\ II]}  &     22.76     &    90.22  c   &     3.81  :   &    10.06 :c   &     17.10     &    72.94  c   &     5.26  ;   &    29.46 ;    \\
 3869 & {[Ne III]} &               &               &    35.35      &    81.69  c   &      5.20     &    18.38  c   &    27.11      &   121.65      \\
 4068 & {[S II]}   &      2.20     &     5.42  c   &               &               &      1.29  :  &     3.53 :c   &               &               \\
 4102 &   H I      &     10.88     &    25.59  c   &    14.03      &    25.74  c   &      9.87     &    25.68  c   &     7.10      &    22.41      \\
 4340 &   H I      &     26.75     &    46.45  c   &    30.62      &    46.65  c   &     22.88     &    46.08  c   &    21.05      &    50.25      \\
 4363 & {[O III]}  &               &               &     3.59      &     5.63      &      1.00     &     1.96      &     5.74      &    13.13      \\
 4471 &   He I     &            p  &           p   &     4.79      &     6.78      &      3.39     &     5.72      &     3.11  :   &     5.91 :    \\
 4686 &   He II    &               &               &     4.08      &     4.76      &      1.83     &     2.26      &     1.52  :   &     1.99 :    \\
 4711 & {[Ar IV]}  &               &               &     3.37      &     0.98      &      1.02     &               &     1.83  :   &          :    \\
 4725 & {[Ne IV]}  &               &               &               &               &               &               &               &               \\
 4740 & {[Ar IV]}  &               &               &     2.77      &     3.11      &             ? &           ?   &     2.63  :   &     3.17 :    \\
 4861 &   H I      &    100.00     &   100.00      &   100.00      &   100.00      &    100.00     &   100.00      &   100.00      &   100.00      \\
 4959 & {[O III]}  &      5.68     &     5.18      &   375.12      &   342.22      &    164.24     &   144.49      &   574.88      &   490.54      \\
 5007 & {[O III]}  &     18.09     &    15.72      &  1137.85      &   994.52      &    528.29     &   437.94      &  1850.95      &  1466.56      \\
 5200 & {[N I]}    &            p  &           p   &               &               &      0.50  :  &     0.33 :    &     1.52  ;   &     0.91 ;    \\
 5515 & {[Cl III]} &               &               &     0.52  ;   &     0.29 ;    &      0.70     &     0.33      &     0.80  ;   &     0.32 ;    \\
 5537 & {[Cl III]} &               &               &     0.49  ;   &     0.29 ;    &      1.32     &     0.60      &     1.67  :   &     0.65 :    \\
 5755 & {[N II]}   &     14.24     &     6.51      &            p  &           p   &      8.30     &     2.94      &    11.56      &     3.19      \\
 5876 &   He I     &     10.53     &     4.35      &    38.81      &    16.70      &     55.09     &    17.02      &    82.06      &    19.11      \\
 6300 & {[O I]}    &      5.47     &     1.72      &               &               &      8.98     &     1.92      &    42.90  :   &     6.29 :    \\
 6312 & {[S III]}  &      3.83     &     1.18      &     2.86  :   &     0.95 :    &      7.33  b  &     1.54  b   &    20.41  ;b  &     2.96 ;b   \\
 6548 & {[N II]}   &    206.88     &    55.91      &               &               &    186.11     &    32.55      &   175.52      &    20.16      \\
 6563 &   H I      &   1093.01     &   292.87      &  1019.29      &   289.38      &   1681.72     &   290.82      &  2528.62      &   286.44      \\
 6583 & {[N II]}   &    661.60     &   175.21      &    47.89  b   &    13.45  b   &    766.62     &   130.52      &   721.34      &    80.14      \\
 6678 &   He I     &      4.01  :  &     1.01 :    &    16.73      &     4.47      &     26.85     &     4.28      &    44.02      &     4.50      \\
 6716 & {[S II]}   &      5.45     &     1.36      &     1.88  :   &     0.50 :    &      8.97     &     1.40      &    11.48      &     1.14      \\
 6730 & {[S II]}   &     13.43     &     3.29      &     5.85  :   &     1.51 :    &     22.15     &     3.40      &    34.21      &     3.35      \\
 7006 & {[Ar V]}   &      0.88  ;  &     0.19 ;    &               &               &      0.28  ;  &     0.04 ;    &            p  &           p   \\
 7136 & {[Ar III]} &     16.40     &     3.34      &    69.00      &    15.08      &    140.82     &    16.91      &   279.59      &    20.14      \\
 7320 & {[O II]}   &    118.95     &    41.54  B   &     9.90  B   &     1.99  B   &     60.99  b  &    19.58  B   &    71.21  b   &    16.44  B   \\
 7330 & {[O II]}   &    103.36     &    41.54  B   &     9.90  B   &     1.99  B   &    122.13  b  &    19.58  B   &   192.34  b   &    16.44  B   \\
\hline
\end{tabular}
\end{table*}

\setcounter{table}{1}
\begin{table*}
\caption{  
  Continued.
}
\begin{tabular}{     l @{\hspace{0.60cm}}
                     l @{\hspace{0.80cm}}
                     r r @{\hspace{0.80cm}} r r @{\hspace{0.80cm}} r r @{\hspace{0.80cm}} r r}
\hline
            &      & \multicolumn{2}{c}{051.0+02.8}& \multicolumn{2}{c}{042.9-06.9}& \multicolumn{2}{c}{041.8+04.4}& \multicolumn{2}{c}{038.7-03.3}\\
            &      & \multicolumn{2}{c}{WhMe 1    }& \multicolumn{2}{c}{NGC 6807  }& \multicolumn{2}{c}{K 3-15    }& \multicolumn{2}{c}{M 1-69    }\\
            &      &               &               &               &               &               &               &               &               \\
            &      & F($\lambda$)  & I($\lambda$ ) & F($\lambda$)  & I($\lambda$ ) & F($\lambda$)  & I($\lambda$ ) & F($\lambda$)  & I($\lambda$ ) \\
            &      &               &               &               &               &               &               &               &               \\
 3727 & {[O\ II]}  &               &               &    17.16  :   &     27.77 :c  &     21.33     &    37.54  c   &    18.86      &    40.44  c   \\
 3869 & {[Ne III]} &    201.93  :b &   938.15 :b   &    80.51      &    121.99  c  &               &               &    62.20      &   121.41  c   \\
 4068 & {[S II]}   &               &               &     2.46      &      3.46  c  &      1.47  :  &     2.30 :c   &     2.15  ;   &     3.60 ;c   \\
 4102 &   H I      &               &               &    18.99      &     25.90  c  &     17.11     &    25.81  c   &    15.42      &    25.84  c   \\
 4340 &   H I      &               &               &    37.92      &     46.86  c  &     33.00     &    46.43      &    31.51      &    46.78  c   \\
 4363 & {[O III]}  &               &               &    10.64      &     13.19     &             p &           p   &     4.94      &     7.12      \\
 4471 &   He I     &             ? &           ?   &     4.37      &      5.22     &      1.10  :  &     1.42 :    &     4.25      &     5.61      \\
 4686 &   He II    &     19.44  :  &    25.86 :    &     0.33  :   &      0.32 :   &               &               &    11.72      &    13.27      \\
 4711 & {[Ar IV]}  &     17.67  :  &    22.68 :    &     1.47      &               &               &               &     3.14      &     2.75      \\
 4725 & {[Ne IV]}  &               &               &               &               &               &               &               &               \\
 4740 & {[Ar IV]}  &     15.87  :  &    19.44 :    &     2.01      &      2.11     &               &               &     3.44      &     3.71      \\
 4861 &   H I      &    100.00  :b &   100.00 :b   &   100.00      &    100.00     &    100.00     &   100.00      &   100.00      &   100.00      \\
 4959 & {[O III]}  &    388.31     &   330.11      &   472.25      &    459.77     &      4.82     &     4.51      &   444.71      &   416.16      \\
 5007 & {[O III]}  &   1241.27     &   978.23      &           S   &   1353.14  r  &     14.85     &    13.50      &  1341.19      &  1216.84      \\
 5200 & {[N I]}    &               &               &     0.22  :   &      0.18 :   &               &               &     0.65  ;   &     0.57 ;    \\
 5515 & {[Cl III]} &               &               &     0.30  :   &      0.26 :   &               &               &     0.81  :   &     0.54 :    \\
 5537 & {[Cl III]} &               &               &     0.56  :   &      0.51 :   &      0.21  ;  &     0.14 ;    &     1.33  :   &     0.87 :    \\
 5755 & {[N II]}   &     27.66  :  &     7.40 :    &     1.41      &      1.13     &      8.34     &     5.00      &     2.26      &     1.34      \\
 5876 &   He I     &    104.84  :  &    23.59 :    &    21.30      &     16.67     &      8.56     &     4.85      &    32.30  :   &    17.56 :    \\
 6300 & {[O I]}    &     41.96  b  &     5.90  b   &     6.90  b   &      5.00  b  &      2.55     &     1.17      &     5.65  ;   &     2.56 ;    \\
 6312 & {[S III]}  &     29.35     &     4.08      &     2.90  b   &      2.10  b  &      3.29     &     1.54      &     5.21  b   &     2.32  b   \\
 6548 & {[N II]}   &     28.15     &     3.07      &     6.60  b   &      4.59  b  &     39.56  b  &    16.90  b   &    44.18  b   &    17.89  b   \\
 6563 &   H I      &   2658.51     &   286.43      &   412.31      &    285.88     &    678.23     &   287.90      &   713.82      &   287.24      \\
 6583 & {[N II]}   &    112.45     &    11.88      &    31.76      &     21.98     &    190.27     &    80.17      &   178.92      &    71.41      \\
 6678 &   He I     &     42.41     &     4.11      &     5.77      &      3.95     &      2.84     &     1.14      &    11.26      &     4.36      \\
 6716 & {[S II]}   &      9.05     &     0.85      &     1.50      &      1.02     &      0.76     &     0.32      &    10.49      &     3.99      \\
 6730 & {[S II]}   &     14.05     &     1.31      &     3.48      &      2.37     &      2.42     &     0.96      &    20.60      &     7.80      \\
 7006 & {[Ar V]}   &               &               &               &               &      0.32  :  &     0.11 :    &     0.29      &     0.10      \\
 7136 & {[Ar III]} &    117.95     &     8.00      &    17.34      &     11.12     &      5.65     &     2.02      &    71.48      &    23.81      \\
 7320 & {[O II]}   &     38.39  :  &     6.04 :B   &    12.51      &     13.74  B  &     77.01     &    63.91  B   &     9.57  :   &     5.36 :B   \\
 7330 & {[O II]}   &     64.85  :  &     6.04 :B   &     9.37      &     13.74  B  &    113.39     &    63.91  B   &     7.46  :   &     5.36 :B   \\
\hline
\end{tabular}
\end{table*}

\setcounter{table}{1}
\begin{table*}
\caption{  
  Continued.
}
\begin{tabular}{     l @{\hspace{0.60cm}}
                     l @{\hspace{0.80cm}}
                     r r @{\hspace{0.80cm}} r r @{\hspace{0.80cm}} r r @{\hspace{0.80cm}} r r}
\hline
            &      & \multicolumn{2}{c}{027.6-09.6}& \multicolumn{2}{c}{025.3-04.6}& \multicolumn{2}{c}{011.1+07.0}& \multicolumn{2}{c}{008.6-02.6}\\
            &      & \multicolumn{2}{c}{IC 4846   }& \multicolumn{2}{c}{K 4- 8    }& \multicolumn{2}{c}{Sa 2-237  }& \multicolumn{2}{c}{MaC 1-11  }\\
            &      &               &               &               &               &               &               &               &               \\
            &      & F($\lambda$)  & I($\lambda$ ) & F($\lambda$)  & I($\lambda$ ) & F($\lambda$)  & I($\lambda$ ) & F($\lambda$)  & I($\lambda$ ) \\
            &      &               &               &               &               &               &               &               &               \\
 3727 & {[O\ II]}  &     21.09     &    31.17  c   &    16.06      &     28.43  c  &    291.41     &   688.31  c   &     5.20  :   &    20.69 :    \\
 3869 & {[Ne III]} &     61.13     &    85.88  c   &    66.69      &    109.32  c  &    105.38     &   220.58  c   &    18.53      &    61.98      \\
 4068 & {[S II]}   &      1.08     &     1.44  c   &     1.54      &      2.21  c  &     21.24  :  &    37.12 :c   &               &               \\
 4102 &   H I      &     20.02     &    25.88  c   &    17.93      &     25.93  c  &     15.26  :  &    25.99 :c   &    10.36  :   &    26.24 :    \\
 4340 &   H I      &     38.95     &    46.84  c   &    36.14      &     46.90  c  &     33.98  :  &    46.99 :c   &    24.39      &    49.07      \\
 4363 & {[O III]}  &      6.12     &     7.31      &     7.79      &     10.15     &      7.19  :  &    10.47 :    &     3.85      &     7.64      \\
 4471 &   He I     &      4.47     &     5.18      &     4.16      &      5.16     &      4.54  :  &     6.03 :    &     3.28      &     5.55      \\
 4686 &   He II    &      0.42     &     0.43      &     0.81      &      0.88     &     45.82     &    52.21      &               &               \\
 4711 & {[Ar IV]}  &      1.65     &     0.86      &     1.89      &      1.22     &      2.61  ;  &     2.18 ;    &     1.97  :   &     1.62 :    \\
 4725 & {[Ne IV]}  &               &               &               &               &               &               &               &               \\
 4740 & {[Ar IV]}  &      1.35     &     1.46      &     2.21      &      2.34     &      2.09  ;  &     2.30 ;    &     0.92  ;   &     1.05 ;    \\
 4861 &   H I      &    100.00     &   100.00      &   100.00      &    100.00     &    100.00     &   100.00      &   100.00      &   100.00      \\
 4959 & {[O III]}  &    368.13     &   359.96      &   439.88      &    423.31     &    384.78     &   368.55      &   328.21      &   288.83      \\
 5007 & {[O III]}  &   1081.33     &  1046.40      &  1306.76      &   1235.16     &   1140.33     &  1070.36      &  1048.84      &   869.54      \\
 5200 & {[N I]}    &      0.18  :  &     0.19 :    &     0.28  :   &      0.27 :   &     40.72     &    35.49      &               &               \\
 5515 & {[Cl III]} &      0.44     &     0.35      &     0.39      &      0.32     &      2.65  ;  &     2.02 ;    &     1.22  ;   &     0.57 ;    \\
 5537 & {[Cl III]} &      0.65     &     0.61      &     0.64      &      0.48     &      1.85  ;  &     1.46 ;    &     0.74  ;   &     0.32 ;    \\
 5755 & {[N II]}   &      0.59     &     0.50      &     1.01      &      0.73     &     32.97     &    23.24      &            p  &           p   \\
 5876 &   He I     &     19.34     &    15.72      &    22.08      &     15.53     &     27.31     &    18.37      &    51.05      &    15.77      \\
 6300 & {[O I]}    &      2.45  b  &     1.91  b   &     4.98      &      3.14     &    144.30     &    85.64      &     2.45  :   &     0.53 :    \\
 6312 & {[S III]}  &      1.90  b  &     1.45  b   &     2.31      &      1.44     &      9.72  ;  &     5.74 ;    &     3.89  :   &     0.82 :    \\
 6548 & {[N II]}   &      4.52  b  &     3.32  b   &     8.50  b   &      5.03  b  &   1112.58     &   617.79      &    25.27  :b  &     4.43 :b   \\
 6563 &   H I      &    389.12     &   286.25      &   483.24      &    285.14     &    512.94     &   283.71      &  1643.45      &   284.60      \\
 6583 & {[N II]}   &     23.88     &    17.53      &    37.66      &     22.14     &   3450.20     &  1898.48      &    70.14  :b  &    11.95 :b   \\
 6678 &   He I     &      5.39     &     3.92      &     6.37      &      3.68     &      8.63     &     4.63      &    23.56      &     3.76      \\
 6716 & {[S II]}   &      1.59     &     1.15      &     2.14      &      1.20     &    265.34     &   141.48      &     4.63      &     0.71      \\
 6730 & {[S II]}   &      3.42     &     2.45      &     4.42      &      2.51     &    337.79     &   179.59      &     9.18      &     1.42      \\
 7006 & {[Ar V]}   &               &               &               &               &               &               &               &               \\
 7136 & {[Ar III]} &     14.82     &    10.21      &    19.20      &     10.15     &     91.29     &    44.65      &    58.17      &     7.00      \\
 7320 & {[O II]}   &      3.94  b  &     6.09  B   &     6.30      &      7.30  B  &     14.06  b  &    17.21  B   &    12.41  b   &     3.53  B   \\
 7330 & {[O II]}   &      5.14  b  &     6.09  B   &     8.00      &      7.30  B  &     22.49  b  &    17.21  B   &    20.63  b   &     3.53  B   \\
\hline
\end{tabular}
\end{table*}

\setcounter{table}{1}
\begin{table*}
\caption{  
  Continued.
}
\begin{tabular}{     l @{\hspace{0.60cm}}
                     l @{\hspace{0.80cm}}
                     r r @{\hspace{0.80cm}} r r @{\hspace{0.80cm}} r r}
\hline
            &      & \multicolumn{2}{c}{004.3-02.6}& \multicolumn{2}{c}{000.6-02.3}& \multicolumn{2}{c}{354.9+03.5}\\
            &      & \multicolumn{2}{c}{H 1-53    }& \multicolumn{2}{c}{H 2-32    }& \multicolumn{2}{c}{Th 3- 6   }\\
            &      &               &               &               &               &               &               \\
            &      & F($\lambda$)  & I($\lambda$ ) & F($\lambda$)  & I($\lambda$ ) & F($\lambda$)  & I($\lambda$ ) \\
            &      &               &               &               &               &               &               \\
 3727 & {[O\ II]}  &     12.45  :  &    39.72 :    &     7.05      &     60.68 c   &     34.96  :  &   137.92 :c   \\
 3869 & {[Ne III]} &     23.16     &    64.28      &               &               &               &               \\
 4068 & {[S II]}   &               &               &               &               &               &               \\
 4102 &   H I      &     10.63  :  &    23.12 :    &     6.23:     &     25.32 :c  &      9.49  ;  &    23.81 ;c   \\
 4340 &   H I      &     26.92     &    48.48      &    17.62      &     46.08  c  &     26.24  :  &    52.43 :    \\
 4363 & {[O III]}  &               &               &               &               &               &               \\
 4471 &   He I     &      4.43  :  &     6.82 :    &               &               &               &               \\
 4686 &   He II    &               &               &               &               &               &               \\
 4711 & {[Ar IV]}  &               &               &               &               &               &               \\
 4725 & {[Ne IV]}  &               &               &               &               &               &               \\
 4740 & {[Ar IV]}  &               &               &               &               &               &               \\
 4861 &   H I      &    100.00     &   100.00      &   100.00      &    100.00     &    100.00     &   100.00      \\
 4959 & {[O III]}  &    231.51     &   207.86      &     0.77  ;   &      0.69 ;   &     77.78     &    68.53      \\
 5007 & {[O III]}  &    714.43     &   609.99      &     5.74      &      4.62     &    238.97     &   198.41      \\
 5200 & {[N I]}    &      2.23  ;  &     1.56 ;    &     1.59  :?  &      1.02 :?  &               &               \\
 5515 & {[Cl III]} &      0.58  ;  &     0.32 ;    &            -  &            -  &               &               \\
 5537 & {[Cl III]} &      1.32  ;  &     0.68 ;    &     1.68  ;   &      0.71 ;   &               &               \\
 5755 & {[N II]}   &     14.76     &     6.17      &     8.91  :   &      2.78 :   &      4.79  ;  &     1.71 ;    \\
 5876 &   He I     &     59.43     &    22.09      &     9.98  :   &      2.69 :   &     59.66     &    18.61      \\
 6300 & {[O I]}    &     21.59  :  &     5.87 :    &    17.35      &      3.08     &     10.00  ;b &     2.16 ;b   \\
 6312 & {[S III]}  &      6.99  ;b &     1.89 ;b   &    12.99      &      2.28     &             p &           p   \\
 6548 & {[N II]}   &    465.07     &   107.09      &    34.61  ;b  &      4.92 ;b  &    285.30     &    50.58      \\
 6563 &   H I      &   1316.47     &   300.26      &  2134.37      &    299.43     &   1747.26     &   306.34      \\
 6583 & {[N II]}   &   1623.64     &   365.47      &   129.08  :   &     17.80 :   &   1472.48     &   254.20      \\
 6678 &   He I     &     25.64     &     5.45      &     4.15  :   &      0.54 :   &     33.16     &     5.36      \\
 6716 & {[S II]}   &     14.85     &     3.08      &     7.64      &      0.94     &     44.44     &     6.99      \\
 6730 & {[S II]}   &     40.49     &     8.37      &    11.20      &      1.38     &    137.52     &    21.45      \\
 7006 & {[Ar V]}   &               &               &     2.16  :   &      0.22 :   &               &               \\
 7136 & {[Ar III]} &    188.63     &    31.63      &     7.37      &      0.69     &    120.43     &    14.70      \\
 7320 & {[O II]}   &     30.81     &    11.18  B   &    89.28      &     15.70  B  &      8.12  b  &     4.72  B   \\
 7330 & {[O II]}   &     42.72     &    11.18  B   &   102.42      &     15.70  B  &     35.30  b  &     4.72  B   \\
\hline
\end{tabular}
\end{table*}

\clearpage

\setcounter{table}{2}
\begin{sidewaystable*}
\caption{Derived plasma diagnostics, elemental abundances and interstellar extinction. 
Columns give: 
(1) PN\,G number; 
(2) usual name; 
(3) disc or bulge Galactic location; 
(4) {\it Spitzer} spectra analysed in samples of \cite{Gutenkunst2008},
    \cite{PereaCalderon2009}, or \cite{Stanghellini2012};
(5) dust classification;
(6) references to data on optical lines used in the calculations;
(7) electron density from [S~{\sc ii}] (or assumed 2*10$^3$ if not computable);
(8) and (9) electron temperature from [N~{\sc ii}] and [O~{\sc iii}];
(10) to (16) derived nebular abundance ratios;
(17) logarithmic extinction at H$\beta$ (if not given, original data were already dereddened). 
For columns (7) to (17) there are
three rows for each object: the first row gives the values of parameters
computed from the nominal values of the observational data, whereas the
second and third rows give the upper and lower limits of these parameters
(see text for details).
}
\centering
\label{test_b}
\tiny{
\begin{tabular}{@{\hspace{0.0cm}} rlc @{\hspace{0.10cm}} c @{\hspace{0.10cm}} clcrrcccccccc}
\hline\hline
       PN G & name  & loc.  & sample & type  & ref  & N$_e$ & T$_e$(N~{\sc ii}) & T$_e$(O~{\sc iii}) &   He/H   &    N/H   &    O/H   &   Ne/H   &    S/H   &   Ar/H   &   Cl/H   &    C      \\
\hline
 184.0$-$02.1 & M 1- 5        & d & S & \CCal    & Mi02                     & 5.69E+03 & 12638 & 11177 & 1.06E-01 & 2.80E-05 & 1.21E-04 & 1.65E-05 & 1.32E-06 & 9.32E-07 & 3.33E-07&  1.37E+00 \\
              &               &   &   &          &                          & 7.72E+03 & 13288 & 11542 & 1.12E-01 & 3.26E-05 & 1.42E-04 & 1.93E-05 & 1.57E-06 & 1.06E-06 &4.91E-07&  1.46E+00 \\
              &               &   &   &          &                          & 4.37E+03 & 11730 & 10683 & 9.95E-02 & 2.44E-05 & 1.10E-04 & 1.44E-05 & 1.16E-06 & 8.48E-07 &2.47E-07&  1.25E+00 \\
              &               &   &   &          &                          &          &       &       &          &          &          &          &          &          &          &           \\
 205.8$-$26.7 & MaC 2- 1      & d & S & \CCal    & Cu96                     & 7.90E+03 &       & 11278 & 9.90E-02 & 2.13E-05 & 2.09E-04 &          & 1.09E-06 & 6.74E-07 & 2.20E-06 &  0.00E+00 \\
              &               &   &   &          &                          & 1.00E+05 &       & 11773 & 1.13E-01 & 9.96E-05 & 3.69E-04 &          & 2.66E-06 & 9.43E-07 & 2.99E-06 &  6.61E-02 \\
              &               &   &   &          &                          & 4.63E+02 &       &  9871 & 9.08E-02 & 2.85E-06 & 1.75E-04 &          & 2.97E-07 & 5.86E-07 &     -~~~ &      -~~~ \\
\hline  
\end{tabular}
}
\end{sidewaystable*}

\clearpage

\setcounter{table}{2}
\begin{sidewaystable*}
\caption{Continued}
\centering
\label{test_b}
\tiny{
\begin{tabular}{@{\hspace{0.0cm}} rlc @{\hspace{0.10cm}} c @{\hspace{0.10cm}} clcrrcccccccc}
\hline\hline
       PN G & name  & loc.  & sample & type  & ref  & N$_e$ & T$_e$(N~{\sc ii}) & T$_e$(O~{\sc iii}) &   He/H   &    N/H   &    O/H   &   Ne/H   &    S/H   &   Ar/H   &   Cl/H   &    C      \\
\hline
 295.3$-$09.3 & He 2- 62      & d & S & \OCam    & Go14                     & 4.93E+03 & 16469 & 12945 & 1.03E-01 & 7.38E-05 & 1.41E-04 & 3.81E-05 & 3.18E-06 & 8.37E-07 & 1.00E-06 &  3.26E-01 \\
              &               &   &   &          &                          & 6.64E+03 & 17364 & 13389 & 1.11E-01 & 8.62E-05 & 1.64E-04 & 4.39E-05 & 3.75E-06 & 9.30E-07 & 1.30E-06 &  4.43E-01 \\
              &               &   &   &          &                          & 3.85E+03 & 15468 & 12481 & 9.64E-02 & 5.64E-05 & 1.22E-04 & 3.35E-05 & 2.57E-06 & 7.42E-07 & 6.44E-07 &  2.26E-01 \\
              &               &   &   &          &                          &          &       &       &          &          &          &          &          &          &          &           \\
 296.3$-$03.0 & He 2- 73      & d & S & \DCcr    & Go14                     & 5.05E+03 & 12601 & 11580 & 1.22E-01 & 2.56E-04 & 4.15E-04 & 1.04E-04 & 8.50E-06 & 3.42E-06 & 1.39E-06 &  1.40E+00 \\
              &               &   &   &          &                          & 7.13E+03 & 13090 & 11942 & 1.28E-01 & 3.09E-04 & 4.74E-04 & 1.17E-04 & 9.78E-06 & 3.85E-06 & 1.75E-06 &  1.50E+00 \\
              &               &   &   &          &                          & 4.00E+03 & 11929 & 11241 & 1.16E-01 & 2.12E-04 & 3.59E-04 & 9.25E-05 & 7.16E-06 & 3.04E-06 & 1.02E-06 &  1.30E+00 \\
\hline  
\end{tabular}
}
\end{sidewaystable*}

\clearpage

\setcounter{table}{2}
\begin{sidewaystable*}
\caption{Continued}
\centering
\label{test_b}
\tiny{
\begin{tabular}{@{\hspace{0.0cm}} rlc @{\hspace{0.10cm}} c @{\hspace{0.10cm}} clcrrcccccccc}
\hline\hline
       PN G & name  & loc.  & sample & type  & ref  & N$_e$ & T$_e$(N~{\sc ii}) & T$_e$(O~{\sc iii}) &   He/H   &    N/H   &    O/H   &   Ne/H   &    S/H   &   Ar/H   &   Cl/H   &    C      \\
\hline
 336.3$-$05.6 & He 2-186      & d & S & \CCar    & Ca10                     & 3.16E+03 & 11659 & 13385 & 1.33E-01 & 2.31E-04 & 3.87E-04 & 1.01E-04 & 4.89E-06 & 2.82E-06 &          &      -~~~ \\
              &               &   &   &          &                          & 3.74E+03 & 12060 & 13660 & 1.42E-01 & 2.53E-04 & 4.30E-04 & 1.13E-04 & 5.56E-06 & 3.10E-06 &          &           \\
              &               &   &   &          &                          & 2.50E+03 & 11157 & 13001 & 1.26E-01 & 2.01E-04 & 3.56E-04 & 9.44E-05 & 4.32E-06 & 2.63E-06 &          &           \\
              &               &   &   &          &                          &          &       & 13224 & 1.08E-01 & 1.94E-05 & 1.76E-04 & 4.96E-05 & 2.93E-06 & 1.04E-06 &          &  1.02E+00 \\
 336.9$+$08.3 & StWr 4-10     & d & S & \OCam    & Go14                     &(2.00E+03)&       & 12825 & 1.00E-01 & 1.52E-05 & 1.37E-04 & 4.28E-05 & 2.17E-06 & 8.14E-07 &          &  9.49E-01 \\
              &               &   &   &          &                          &          &       & 13224 & 1.08E-01 & 1.94E-05 & 1.76E-04 & 4.96E-05 & 2.93E-06 & 1.04E-06 &          &  1.02E+00 \\
              &               &   &   &          &                          &          &       & 11893 & 9.39E-02 & 1.21E-05 & 1.21E-04 & 3.67E-05 & 1.96E-06 & 6.62E-07 &          &  8.51E-01 \\
\hline    
\end{tabular}
}
\end{sidewaystable*}

\clearpage

\setcounter{table}{2}
\begin{sidewaystable*}
\caption{Continued}
\centering
\label{test_b}
\tiny{
\begin{tabular}{@{\hspace{0.0cm}} rlc @{\hspace{0.10cm}} c @{\hspace{0.10cm}} clcrrcccccccc}
\hline\hline
       PN G & name  & loc.  & sample & type  & ref  & N$_e$ & T$_e$(N~{\sc ii}) & T$_e$(O~{\sc iii}) &   He/H   &    N/H   &    O/H   &   Ne/H   &    S/H   &   Ar/H   &   Cl/H   &    C      \\
\hline
 354.9$+$03.5 & Th 3- 6       & b & S & \DCcr    & t.w.                     & 1.00E+05 &  4639 &       & 1.17E-01 & 2.25E-03 & 7.86E-02 &          & 1.39E-03 & 2.47E-05 &          &  2.34E+00 \\
              &               &   &   &          &                          & 1.00E+05 &  4990 &       & 1.23E-01 & 4.38E-03 & 2.82E-01 &          & 2.65E-03 & 4.37E-05 &          &  2.43E+00 \\
              &               &   &   &          &                          & 1.00E+05 &  4165 &       & 1.10E-01 & 1.55E-03 & 3.23E-02 &          & 9.91E-04 & 1.82E-05 &          &  2.22E+00 \\
              &               &   &   &          &                          &          &       &       &          &          &          &          &          &          &          &           \\
 355.2$-$02.5 & H 1-29        & b & S & \DCcr    & Cu00,Go04,Gi07           & 4.16E+03 & 11452 & 10172 & 1.21E-01 & 1.43E-04 & 4.16E-04 & 1.00E-04 & 6.30E-06 & 2.09E-06 & 1.72E-06 &  1.61E+00 \\
              &               &   &   &          &                          & 5.80E+03 & 12004 & 11614 & 1.29E-01 & 2.31E-04 & 5.14E-04 & 1.42E-04 & 8.96E-06 & 2.49E-06 & 3.12E-06 &  1.69E+00 \\
              &               &   &   &          &                          & 3.29E+03 & 10759 &  9391 & 1.14E-01 & 7.15E-05 & 2.72E-04 & 5.89E-05 & 2.97E-06 & 1.71E-06 & 3.62E-07 &  1.51E+00 \\
\hline    
\end{tabular}
}
\end{sidewaystable*}

\clearpage

\setcounter{table}{2}
\begin{sidewaystable*}
\caption{Continued}
\centering
\label{test_b}
\tiny{
\begin{tabular}{@{\hspace{0.0cm}} rlc @{\hspace{0.10cm}} c @{\hspace{0.10cm}} clcrrcccccccc}
\hline\hline
       PN G & name  & loc.  & sample & type  & ref  & N$_e$ & T$_e$(N~{\sc ii}) & T$_e$(O~{\sc iii}) &   He/H   &    N/H   &    O/H   &   Ne/H   &    S/H   &   Ar/H   &   Cl/H   &    C      \\
\hline
 358.6$+$01.8 & M 4- 6        & b & S & \DCcr    & Ra97                     & 4.45E+03 & 14396 &       & 1.07E-01 & 3.77E-05 & 7.92E-05 & 5.73E-05 & 1.36E-06 & 9.86E-07 &          &      -~~~ \\
              &               &   &   &          &                          & 5.36E+03 & 14913 &       & 1.15E-01 & 4.25E-05 & 9.41E-05 & 6.86E-05 & 1.59E-06 & 1.11E-06 &          &           \\
              &               &   &   &          &                          & 3.66E+03 & 13616 &       & 9.97E-02 & 3.38E-05 & 7.07E-05 & 5.09E-05 & 1.23E-06 & 9.11E-07 &          &           \\
              &               &   &   &          &                          &          &       &       &          &          &          &          &          &          &          &           \\
 358.7$-$05.2 & H 1-50        & b & P & \DCcr    & WL07                     & 4.87E+03 & 12847 & 11052 & 1.17E-01 & 1.93E-04 & 4.54E-04 & 1.31E-04 & 7.37E-06 & 2.35E-06 & 1.63E-06 &  6.84E-01 \\
              &               &   &   &          &                          & 5.80E+03 & 13386 & 11462 & 1.24E-01 & 2.23E-04 & 5.13E-04 & 1.43E-04 & 8.53E-06 & 2.61E-06 & 2.01E-06 &  7.58E-01 \\
              &               &   &   &          &                          & 3.62E+03 & 12191 & 10829 & 1.08E-01 & 1.52E-04 & 3.86E-04 & 1.08E-04 & 6.04E-06 & 2.06E-06 & 1.16E-06 &  5.86E-01 \\
\hline    
\end{tabular}
}
\end{sidewaystable*}

\clearpage

\setcounter{table}{2}
\begin{sidewaystable*}
\caption{Continued}
\centering
\label{test_b}
\tiny{
\begin{tabular}{@{\hspace{0.0cm}} rlc @{\hspace{0.10cm}} c @{\hspace{0.10cm}} clcrrcccccccc}
\hline\hline
       PN G & name  & loc.  & sample & type  & ref  & N$_e$ & T$_e$(N~{\sc ii}) & T$_e$(O~{\sc iii}) &   He/H   &    N/H   &    O/H   &   Ne/H   &    S/H   &   Ar/H   &   Cl/H   &    C      \\
\hline
 001.4$+$05.3 & H 1-15        & b & G & \OCcr    & Cu00                     & 9.02E+02 &  8982 &       & 1.18E-01 & 2.23E-05 & 2.64E-04 &          & 2.81E-06 & 2.24E-06 &          &  1.45E+0  \\
              &               &   &   &          &                          & 1.78E+03 & 12559 &       & 1.24E-01 & 4.01E-05 & 2.11E-03 &          & 1.22E-05 & 6.22E-06 &          &  1.54E+0  \\
              &               &   &   &          &                          & 4.63E+02 &       &       & 9.92E-02 & 1.33E-05 & 2.69E-05 &          & 3.48E-07 & 5.42E-07 &          &  1.36E+0  \\
              &               &   &   &          &                          &          &       &       &          &          &          &          &          &          &          &           \\
 001.6$-$01.3 & Bl Q          & b & G & \OCcr    & EXs1                     & 2.71E+03 & 12521 &       & 9.62E-02 & 2.79E-04 & 1.84E-04 &          & 8.23E-06 & 1.97E-06 &          &      -~~~ \\
              &               &   &   &          &                          & 3.23E+03 & 13069 &       & 1.03E-01 & 3.58E-04 & 2.06E-04 &          & 1.01E-05 & 2.15E-06 &          &           \\
              &               &   &   &          &                          & 2.12E+03 & 12063 &       & 8.99E-02 & 2.39E-04 & 1.63E-04 &          & 7.26E-06 & 1.71E-06 &          &           \\
\hline    
\end{tabular}
}
\end{sidewaystable*}

\clearpage

\setcounter{table}{2}
\begin{sidewaystable*}
\caption{Continued}
\centering
\label{test_b}
\tiny{
\begin{tabular}{@{\hspace{0.0cm}} rlc @{\hspace{0.10cm}} c @{\hspace{0.10cm}} clcrrcccccccc}
\hline\hline
       PN G & name  & loc.  & sample & type  & ref  & N$_e$ & T$_e$(N~{\sc ii}) & T$_e$(O~{\sc iii}) &   He/H   &    N/H   &    O/H   &   Ne/H   &    S/H   &   Ar/H   &   Cl/H   &    C      \\
\hline
 003.2$-$04.4 & KFL 12        & b & S & \OCam    & Go09                     & 1.00E+05 &       &  8453 & 1.01E-01 & 1.11E-05 & 7.75E-04 & 1.78E-04 & 3.10E-06 & 1.94E-06 &          &  1.17E+00 \\
              &               &   &   &          &                          & 1.00E+05 &       &  9006 & 1.07E-01 & 1.36E-05 & 1.05E-03 & 2.63E-04 & 5.30E-06 & 2.42E-06 &          &  1.27E+00 \\
              &               &   &   &          &                          & 1.00E+05 &       &  7864 & 9.42E-02 & 8.37E-06 & 5.72E-04 & 1.32E-04 & 1.40E-06 & 1.57E-06 &          &  1.07E+00 \\
              &               &   &   &          &                          &          &       &       &          &          &          &          &          &          &          &           \\
 003.6$+$03.1 & M 2-14        & b & P & \DCcr    & Go04                     & 9.38E+03 &  8184 &  7983 & 1.48E-01 & 2.45E-04 & 3.64E-04 & 6.05E-05 & 9.83E-06 & 6.06E-06 & 1.46E-06 &  1.55E+00 \\
              &               &   &   &          &                          & 1.65E+04 &  8622 &  8373 & 1.57E-01 & 3.75E-04 & 5.94E-04 & 1.04E-04 & 2.08E-05 & 8.66E-06 & 2.85E-06 &  1.64E+00 \\
              &               &   &   &          &                          & 6.30E+03 &  7234 &  7149 & 1.38E-01 & 2.07E-04 & 3.01E-04 & 4.46E-05 & 7.73E-06 & 5.12E-06 & 8.74E-07 &  1.43E+00 \\
\hline    
\end{tabular}
}
\end{sidewaystable*}

\clearpage

\setcounter{table}{2}
\begin{sidewaystable*}
\caption{Continued}
\centering
\label{test_b}
\tiny{
\begin{tabular}{@{\hspace{0.0cm}} rlc @{\hspace{0.10cm}} c @{\hspace{0.10cm}} clcrrcccccccc}
\hline\hline
       PN G & name  & loc.  & sample & type  & ref  & N$_e$ & T$_e$(N~{\sc ii}) & T$_e$(O~{\sc iii}) &   He/H   &    N/H   &    O/H   &   Ne/H   &    S/H   &   Ar/H   &   Cl/H   &    C      \\
\hline
 008.1$-$04.7 & M 2-39        & b & S & \OCcr    & WL07                     & 4.94E+03 & 10267 & 26960 & 6.52E-02 & 6.08E-06 & 2.38E-05 & 6.19E-06 & 2.21E-07 & 2.06E-07 & 1.91E-08 &  1.09E+00 \\
              &               &   &   &          &                          & 6.42E+03 & 10478 & 30285 & 6.93E-02 & 6.94E-06 & 2.62E-05 & 6.67E-06 & 2.49E-07 & 2.31E-07 & 2.22E-08 &  1.17E+00 \\
              &               &   &   &          &                          & 4.09E+03 &  9909 & 25945 & 5.87E-02 & 4.85E-06 & 1.96E-05 & 4.99E-06 & 1.80E-07 & 1.79E-07 & 1.40E-08 &  9.58E-01 \\
              &               &   &   &          &                          &          &       &       &          &          &          &          &          &          &          &           \\
 008.2$-$04.8 & M 2-42        & b & S & \F       & WL07                     & 2.15E+03 & 10836 &  8280 & 1.13E-01 & 2.14E-04 & 6.12E-04 & 1.65E-04 & 1.37E-05 & 2.18E-06 & 5.45E-06 &  9.57E-01 \\
              &               &   &   &          &                          & 2.59E+03 & 11246 &  8704 & 1.19E-01 & 2.57E-04 & 6.70E-04 & 1.58E-04 & 1.52E-05 & 2.28E-06 & 6.14E-06 &  1.04E+00 \\
              &               &   &   &          &                          & 1.83E+03 & 10420 &  8219 & 1.05E-01 & 1.58E-04 & 4.72E-04 & 1.24E-04 & 9.97E-06 & 1.85E-06 & 3.30E-06 &  8.59E-01 \\
\hline    
\end{tabular}
}
\end{sidewaystable*}

\clearpage

\setcounter{table}{2}
\begin{sidewaystable*}
\caption{Continued}
\centering
\label{test_b}
\tiny{
\begin{tabular}{@{\hspace{0.0cm}} rlc @{\hspace{0.10cm}} c @{\hspace{0.10cm}} clcrrcccccccc}
\hline\hline
       PN G & name  & loc.  & sample & type  & ref  & N$_e$ & T$_e$(N~{\sc ii}) & T$_e$(O~{\sc iii}) &   He/H   &    N/H   &    O/H   &   Ne/H   &    S/H   &   Ar/H   &   Cl/H   &    C      \\
\hline
 019.2$-$02.2 & M 4-10        & d & S & \OCamcr  & Cu96                     & 5.05E+03 & 15022 & 11853 & 1.06E-01 & 6.01E-05 & 3.77E-04 &          & 3.73E-06 & 1.27E-06 &          &  2.06E+00 \\
              &               &   &   &          &                          & 2.67E+04 & 20565 & 13057 & 1.15E-01 & 1.13E-04 & 6.06E-04 &          & 7.20E-06 & 1.74E-06 &          &  2.15E+00 \\
              &               &   &   &          &                          & 2.30E+03 &  7694 & 10481 & 9.74E-02 & 4.32E-05 & 2.68E-04 &          & 2.45E-06 & 1.06E-06 &          &  1.96E+00 \\
              &               &   &   &          &                          &          &       &       &          &          &          &          &          &          &          &           \\
 019.4$-$05.3 & M 1-61        & d & S & \DCamcr  & Gi07,WL07                & 1.12E+04 & 12880 &  9368 & 1.08E-01 & 1.02E-04 & 4.68E-04 & 1.18E-04 & 5.50E-06 & 2.40E-06 & 1.16E-06 &  1.64E+00 \\
              &               &   &   &          &                          & 2.05E+04 & 13632 &  9588 & 1.14E-01 & 1.17E-04 & 5.84E-04 & 1.43E-04 & 7.19E-06 & 2.75E-06 & 1.62E-06 &  1.74E+00 \\
              &               &   &   &          &                          & 7.29E+03 & 11327 &  9000 & 1.01E-01 & 8.89E-05 & 4.04E-04 & 9.20E-05 & 3.81E-06 & 2.21E-06 & 7.03E-07 &  1.54E+00 \\
\hline    
\end{tabular}
}
\end{sidewaystable*}

\clearpage

\setcounter{table}{2}
\begin{sidewaystable*}
\caption{Continued.}
\centering
\label{test_b}
\tiny{
\begin{tabular}{@{\hspace{0.0cm}} rlc @{\hspace{0.10cm}} c @{\hspace{0.10cm}} clcrrcccccccc}
\hline\hline
       PN G & name  & loc.  & sample & type  & ref  & N$_e$ & T$_e$(N~{\sc ii}) & T$_e$(O~{\sc iii}) &   He/H   &    N/H   &    O/H   &   Ne/H   &    S/H   &   Ar/H   &   Cl/H   &    C      \\
\hline
 060.5$+$01.8 & He 2-440      & d & S & \OCcr    & t.w.                     & 1.00E+05 &  6153 &  8218 & 1.16E-01 & 2.55E-04 & 4.47E-03 & 5.81E-04 & 1.58E-05 & 3.74E-06 & 7.09E-07 &  2.36E+00 \\
              &               &   &   &          &                          & 1.00E+05 &  6585 &  8614 & 1.21E-01 & 2.86E-04 & 5.21E-03 & 6.73E-04 & 1.73E-05 & 4.21E-06 & 9.48E-07 &  2.46E+00 \\
              &               &   &   &          &                          & 1.00E+05 &  6032 &  7979 & 1.09E-01 & 1.88E-04 & 2.67E-03 & 3.01E-04 & 1.20E-05 & 3.20E-06 & 5.49E-07 &  2.27E+00 \\
              &               &   &   &          &                          &          &       &       &          &          &          &          &          &          &          &           \\
 068.7$+$01.9 & K 4-41        & d & S & \OCcr    & t.w.                     & 1.00E+05 &       &  8743 & 1.15E-01 & 1.01E-04 & 7.34E-04 & 1.78E-04 & 7.50E-06 & 3.05E-06 & 1.33E-06 &  1.69E+00 \\
              &               &   &   &          &                          & 1.00E+05 &       &  8968 & 1.22E-01 & 1.30E-04 & 8.33E-04 & 1.84E-04 & 9.34E-06 & 3.39E-06 & 1.95E-06 &  1.77E+00 \\
              &               &   &   &          &                          & 1.00E+05 &       &  8500 & 1.07E-01 & 7.94E-05 & 6.12E-04 & 1.45E-04 & 6.03E-06 & 2.72E-06 & 9.87E-07 &  1.59E+00 \\
\hline\hline  
\end{tabular}
\tablebib{
(FP92)~\citet{FreitasPacheco1992};    
(KB94)~\citet{KingsburghBarlow1994};  
(Cu96)~\citet{Cuisinier1996};         
(Ra97)~\citet{Ratag1997};             
(Cu00)~\citet{Cuisinier2000};         
(Es01)~\citet{Escudero2001};          
(KH01)~\citet{KwitterHenry2001};      
(Pe01)~\citet{Pena2001};              
(Mi02)~\citet{Milingo2002};           
(Es04)~\citet{Escudero2004};          
(Ex04)~\citet{Exter2004};             
(Go04)~\citet{Gorny2004};             
(Gi07)~\citet{Girard2007};            
(WL07)~\citet{WangLiu2007};           
(Go09)~\citet{Gorny2009};             
(Ca10)~\citet{Cavichia2010};          
(He10)~\citet{Henry2010};             
(Go14)~\citet{Gorny2014};             
(t.w.)~this~work.}
}
\end{sidewaystable*}

\clearpage




\end{document}